\documentclass[twocolumn]{aastex62}
\pdfoutput=1 
\usepackage{amsmath,amssymb,amstext}
\usepackage[T1]{fontenc}
\usepackage{apjfonts} 
\usepackage[figure,figure*]{hypcap}
\usepackage{graphicx}
\usepackage[caption = false]{subfig}
\usepackage{rotating}
\usepackage[inline]{enumitem}
\usepackage{comment}

\newcommand{\xmm}{{\sl XMM-Newton }}

\shorttitle{PSRJ0659+1414}
\shortauthors{Arumugasamy et al.}

\begin{document}

\title{POSSIBLE PHASE-DEPENDENT ABSORPTION FEATURE IN THE X-RAY SPECTRUM OF THE MIDDLE-AGED PSR J0659+1414}

\author{Prakash Arumugasamy}
\affiliation{National Centre for Radio Astrophysics, Tata Institute for Fundamental Research, Post Bag 3, Ganeshkhind, Pune 411007, INDIA}
\author{Oleg Kargaltsev}
\affiliation{Department of Physics, The George Washington University, 725 21st Street, NW, Washington, DC 20052, USA}
\affiliation{Astronomy, Physics and Statistics Institute of Sciences (APSIS), The George Washington University, Washington, DC 20052, USA}
\author{Bettina Posselt}
\affiliation{Pennsylvania State University, 525 Davey Lab., University Park, PA 16802, USA}
\author{George G. Pavlov}
\affiliation{Pennsylvania State University, 525 Davey Lab., University Park, PA 16802, USA}
\author{Jeremy Hare}
\affiliation{Department of Physics, The George Washington University, 725 21st Street, NW, Washington, DC 20052, USA}
\affiliation{Astronomy, Physics and Statistics Institute of Sciences (APSIS), The George Washington University, Washington, DC 20052, USA}

\begin{abstract}

We report on the energy-resolved timing and phase-resolved spectral analysis of X-ray emission from PSR J0659+1414 observed with {\sl XMM-Newton} and {\sl NuSTAR}. 
We find that the new data rule out the previously suggested model of the phase-dependent spectrum as a three-component (2 blackbodies + power-law) continuum, which shows large residuals between $0.3-0.7$ keV.
Fitting neutron star atmosphere models or several blackbodies to the spectrum does not provide a better description of the spectrum, and requires spectral model components with unrealistically large emission region sizes.
The fits improve significantly if we add a phase-dependent absorption feature with central energy $0.5-0.6$ keV and equivalent width up to $\approx 50$ eV.
We detected the feature for about half of the pulse cycle.
Energy-resolved pulse profiles support the description of the spectrum with a three-component continuum and an absorption component.
The absorption feature could be interpreted as an electron cyclotron line originating in the pulsar magnetosphere and broadened by the non-uniformity of magnetic field along the line of sight.
The significant phase-variability in the thermal emission from the entire stellar surface may indicate multi-polar magnetic fields and a nonuniform temperature distribution.
The strongly pulsed non-thermal spectral component detected with {\sl NuSTAR} in the 3--20 keV range is well fit by a power-law model with a photon index $\Gamma=1.5\pm0.2$.

\end{abstract}

\keywords{pulsars: individual (PSR J0659+1414 = B0656+14) --- stars: neutron --- X-rays: stars}

\section{Introduction}

Absorption features in X-ray spectra of neutron stars (NS) have been observed since the early days of balloon experiments of X-ray astronomy.
They were first detected from the accreting X-ray pulsar, Hercules X-1 \citep{1978ApJ...219L.105T}.
About 20 such systems have since been detected (\citealt{2012MmSAI..83..230C}; \citealt{2012AIPC.1427...60P}; and references therein).
The high surface magnetic field of these pulsars ($B=10^{12}-10^{13}$ G) causes the charged particles in the accreting plasma to absorb the NS continuum emission at energies between $10-100$ keV via the cyclotron mechanism.
Electron cyclotron features in accreting X-ray pulsars can be used to model the local environmental conditions of the line-forming regions such as the magnetic field strength, electron density, temperature, gravitational redshift, and geometry \citep{2007A&A...472..353S,1999ApJ...517..334A}.
The magnetic field value calculated under the electron cyclotron assumption matches well with the typical surface dipole magnetic fields inferred from spin-down in isolated pulsars ($\sim 10^{12}$ G).
The observations suggest that the absorption occurs in accretion columns near the NS surface.
The absorption features vary with NS spin-phase \citep{Heindl2004}, and these variations have been modeled for various accretion geometries, magnetic field configurations, and viewing angles (e.g., \citealt{Mukherjee2012}).
The hard spectra of the accreting NSs --- due to high temperatures in the accretion column --- allow detection of multiple cyclotron harmonics in some cases (e.g., \citealt{1999ApJ...523L..85S}).

Absorption features in non-accreting NS, with comparatively lower luminosities, turned out to be much harder to detect, requiring a new generation of sensitive X-ray spectral instruments.
First such features were detected in the spectrum of 1E 1207--5209 \citep{Sanwal2002}, which belongs to a class of young NSs in supernova remnants called Central Compact Objects (CCO; \citealt{Pavlov2004, 2008deLuca, DeLuca2017}).
The absorption features were attributed to electron cyclotron absorption in the NS atmosphere (\citealt{Bignami2003,Suleimanov2010,Suleimanov2012}).
Absorption features in phase-integrated and phase-resolved spectra have also been reported for 6 of the 7 known high-B Isolated Neutron Stars (INSs; \citealt{2007Haberl,2017Borghese}), and a couple of magnetars \citep{Tiengo2013}.
INSs and magnetars possess magnetic fields orders of magnitude higher than those of rotation-powered pulsars (RPPs), ranging from $10^{13}$ G to $10^{15}$ G. 
The features, detected at energies between 0.2 and 1.5 keV could, therefore, be  attributed to proton cyclotron absorption in the strong global magnetic dipole fields or even stronger local magnetic loops close to the surface \citep{Tiengo2013}.
However, absorption by atoms in strong magnetic fields (\citealt{Ruder1994, Mori2002}) or, in some cases, strong localized non-uniformities in the surface temperature distributions \citep{Vigano2014} are also possible.

The detection of absorption features such as those obtained from INSs and CCOs require small distances, as in the case of INSs, or high luminosities, as in the case of most CCOs.
The detections are further helped by the simple, single dominant thermal continuum model that seems to fit most INS and CCO spectra.
In very young RPPs the X-ray spectra are dominated by non-thermal emission while older RPPs, where thermal emission from the NS becomes visible, are faint.
The surface emission from the bulk of the NS surface shifts to the ultraviolet regime for old (characteristic age $\tau_c \gtrsim 1$ Myr) pulsars (e.g., \citealt{Pavlov2017}), while their X-ray emission becomes substantially weaker.
Therefore, RPPs in the narrow range of intermediate ages (10 kyr $\lesssim \tau_c \lesssim$ 1 Myr) offer the best opportunity for detecting spectral features in soft X-rays.
However, their $0.1-10$ keV spectra typically require multiple continuum components: thermal emission from the bulk surface, polar cap region or hot spot, and often non-thermal emission from the magnetosphere.
Hence, the detection of absorption features requires high signal-to-noise ratio (S/N) spectra and limits the target choices to several bright and nearby sources.

First indications of absorption features in RRPs were found only recently.
The nearby, 340 kyr old Geminga pulsar shows narrow, unmodeled residuals around 0.5 keV in continuum model fit to its phase-resolved spectra \citep{Jackson2005}.
The 100 kyr old pulsar, PSR J1740+1000 provided stronger evidence for phase-dependent absorption features with $E=0.55-0.65$ keV which become strongest near the soft X-ray pulse minimum \citep{Kargaltsev2012}.
These properties indicate either absorption by elements other than H near the NS surface or electron cyclotron absorption in the NS magnetosphere. 

J0659+1414 (=B0656+14, hereafter B0656) is a 110 kyr old pulsar with a period of 385 ms, spin-down power of $\dot{E} = 3.8\times10^{34}$ erg s$^{-1}$, and at a parallax distance of $d = 288^{+33}_{-27}$ pc \citep{brisken2003distance}.
This nearby and bright pulsar was first detected in the radio by \cite{Manchester1978}.
It has since been detected at multiple radio frequencies, in the infrared, optical and ultraviolet (\citealt{Caraveo1994,Pavlov1996,Pavlov1997,Koptsevich2001,Kargaltsev2007,Durant2011}), X-rays (\citealt{Cordova1989,Pavlov2002nsps,DeLuca2005}), and $\gamma$-rays (\citealt{Ramanamurthy1996, Weltevrede2010}).
Pulsations have also been found at all observed wavelengths.
Due to the proximity and its age, B0656 is one of the few optimal targets to look for possible spectral features in isolated pulsars. 

The last update on B0656's X-ray spectrum and pulsations was obtained from a 41 ks \xmm  observation.
\cite{DeLuca2005} found that the $0.2-8$ keV spectrum can be satisfactorily described by two blackbodies, with temperatures of about $T_{\rm BB,c} = 0.65$ MK and $T_{\rm BB,h} = 1.25$ MK from $R_{\rm BB,c}\sim 20$ km and $R_{\rm BB,h}\sim 1.8$ km effective radii, respectively, and a power-law component with photon index $\Gamma\approx 2$, confirming earlier {\sl Chandra} results \citep{Pavlov2002nsps}.
They also reported that the phase-resolved spectrum can be described by a single model with varying normalizations for the three components, while the temperatures and photon index were kept constant.
To investigate the phase dependence of the X-ray continuum more accurately and look for spectral features, we carried out a deeper \xmm observation of B0656, supplemented by a simultaneous {\sl NuSTAR} observation.
In this paper we report on spectral and timing analysis of  the data obtained in those observations, focusing on a possible absorption spectral feature around $0.5-0.6$ keV seen for about half of the pulsar period.

\begin{table}
\centering
\caption{Pulsar J0659+1414 Parameters Summary.\label{tbl-1_summary}}
\begin{tabular}{cl}
\tableline\tableline
Parameter & Value \\
\tableline
Right ascension (J2000){\def\hfill{\hskip 60pt plus 1fill}\dotfill}	&	$06^{\rm h} 59^{\rm m} 48\fs1472(7)$ \\
Declination (J2000) \dotfill	&     $+14^\circ 14\arcmin 21\farcs160(10)$ \\
Position epoch (MJD) \dotfill		&	51544	\\
Galactic longitude/latitude ($l/b$)\dotfill		&	$201\fdg11$ / $8\fdg26$	\\
Period ($P$) \dotfill						&	384.9189079(1) ms	\\
Period derivative ($\dot{P}$) \dotfill			&	$5.49586(9) \times 10^{-14}\;{\rm s\;s}^{-1}$	\\
Frequency ($\nu$) \dotfill					&	2.59794980081(64) Hz	\\
Frequency derivative ($\dot\nu$) \dotfill		&	$-3.709345(58) \times 10^{-13}\;{\rm s}^{-2}$	\\
Epoch of timing solution (MJD) \dotfill		&	55555	\\
Dispersion measure (DM)\dotfill				&	13.977(13) cm$^{-3}$ pc	\\
Distance ($d$)\dotfill		&	$288_{-27}^{+33}$ pc	\\
Characteristic age ($\tau_c$) \dotfill		&	$110$ kyr	\\
Spin-down power ($\dot{E}$) \dotfill			&	$3.8 \times 10^{34}$ erg ${\rm s}^{-1}$	\\
Surface magnetic field ($B_{\rm surf}$) \dotfill	&	$4.7 \times 10^{12}$ G	\\
\tableline
\end{tabular}
\renewcommand{\thefootnote}{\fnsymbol{footnote}}\tablecomments{The parameters are taken from the DE405 ephemerides files\footnote{\url{https://confluence.slac.stanford.edu/display/GLAMCOG/LAT+Gamma-ray+Pulsar+Timing+Models}} based on \cite{Ray2011}.
The parallax distance was measured by \cite{brisken2003distance}}
\end{table}

\section{Observation and Data Analysis}
\subsection{XMM-Newton}
\begin{figure}
\centering
\includegraphics[width=0.48\textwidth]{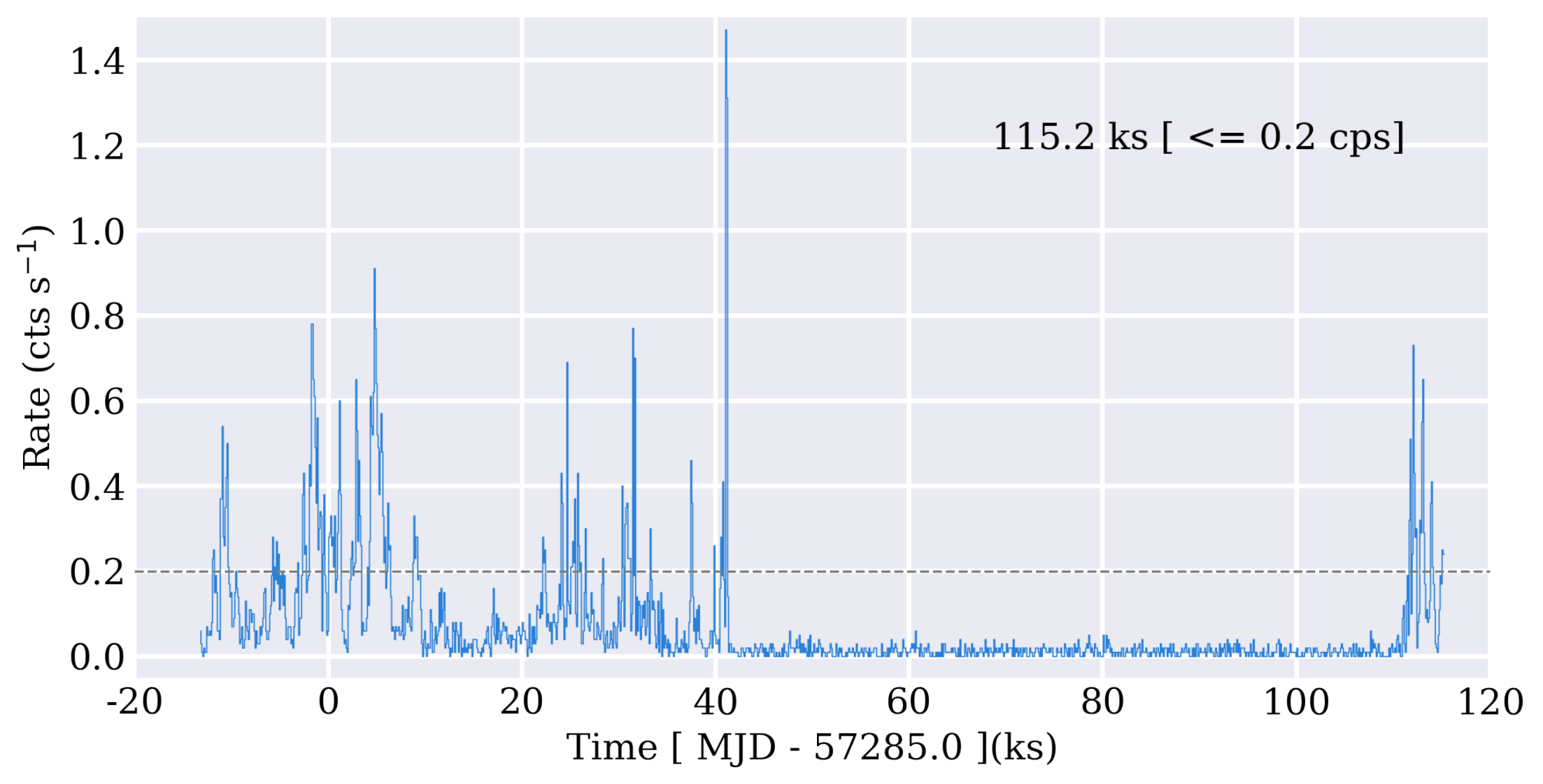}
\caption{\label{fig:fig2_flaring}Background flaring light curve obtained from EPIC-pn (entire SW region) for the full observation duration and energies $>10$\,keV. The net exposure time is 115 ks after a flaring cut at $\leq0.2$ counts s$^{-1}$ level.}
\end{figure}

B0656 was observed with the European Photon Imaging Camera (EPIC) of the \xmm observatory (obsid 0762890101) on 2015 September 19 (MJD\,57285) for 130 ks.
The EPIC-pn was operated in Small Window (SW) mode which uses a $4\farcm37\times4\farcm37$ window to achieve a 5.7 ms time-resolution.
The EPIC data processing was done with the \xmm Science Analysis System (SAS) ver.\ 15.0.0\footnote{\url{http://xmm.esac.esa.int/sas}} \citep{Gabriel2004}, applying standard filtering tasks.
The flaring background extracted from the full EPIC-pn SW area is shown in Figure \ref{fig:fig2_flaring} using the $>10$ keV light curve.
We remove events observed in the time intervals corresponding to flaring count rates greater than 0.2 counts s$^{-1}$ from the data.
This ensures high S/N data by retaining $\approx 115$ ks of the good time intervals (GTI) while eliminating periods of very high flaring background.
Further standard filtering restricts the events to pattern $\leq 4$ (singles and doubles) and  rejects events close to the chip gaps and bad pixels through flag=0.

EPIC-MOS1 and EPIC-MOS2 were operated in timing mode which restricts the window size to $\approx 1\farcm83\times11\arcmin$, to achieve a time-resolution of 1.75 ms.
The timing mode MOS1 data is unsuitable for scientific use due to severe electronic cross-talk from the physical damage it sustained from a micro-meteoroid impact in Rev. 961\footnote{\url{http://xmm2.esac.esa.int/docs/documents/CAL-TN-0018.pdf}}.
The MOS2 had only 27 ks of useful data available, due to an unresolved technical problem that failed to generate the complete observation files.
Hence, we refrain from using the MOS1 and MOS2 data.

\begin{figure}
\centering
\includegraphics[width=0.5\textwidth]{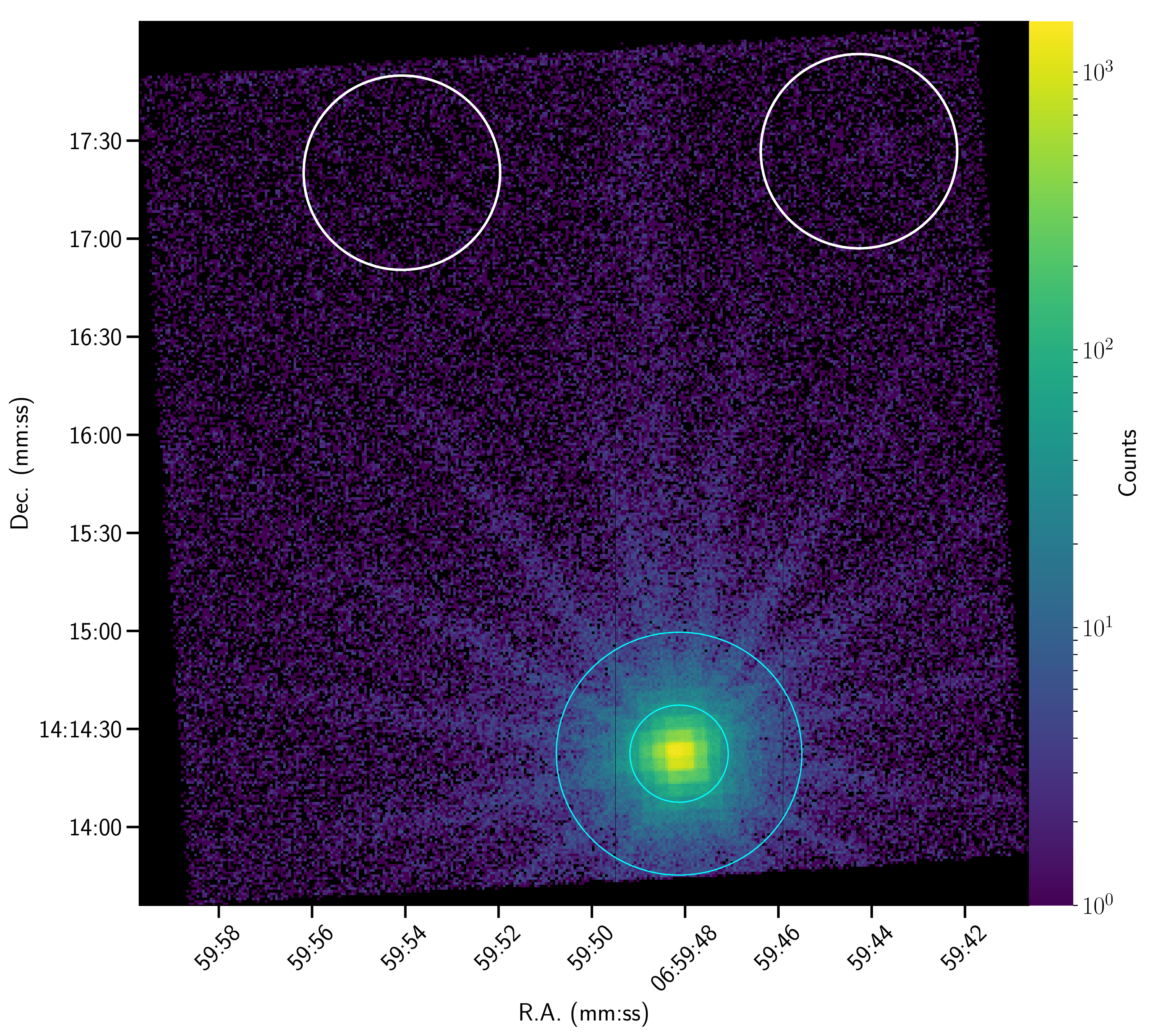}
\caption{\label{fig:fig1_cmap}The $0.3-7$ keV binned count-map for 115 ks exposure with EPIC-pn.
The cyan $15\arcsec$ and $37\farcs5$ radius circles show the source regions used for phase-integrated and phase-resolved spectral analyses, respectively.
Two white $30\arcsec$ radius circles show the background regions.
}
\end{figure}

\subsection{NuSTAR}

{\sl NuSTAR} \citep{2013ApJ...770..103H}  observed B0656 simultaneously with {\sl XMM-Newton} on 2015 September 19 (MJD 57285) for 288 ks. We reduced the {\sl NuSTAR} data using NUSTARDAS v1.7.1 with the calibration database (CALDB) v20170614. The data were filtered using the {\tt nupipeline} task with {\tt saacalc=2}, {\tt saamode=optimized}, and {\tt tentacle=yes}, which reduces the background contribution due to  {\sl NuStar's} passage through the South Atlantic Anomaly. This left 127 and 126 ks of livetime-corrected exposure time for the FPMA and FPMB detectors, respectively.  
We limited the {\sl NuSTAR} data analysis to the 3-20 keV energy range to minimize the background contribution, which dominates at higher energies. All photon arrival times were  corrected to the Solar system barycenter, including  the clock offset corrections, to achieve a timing accuracy of $\sim 2$ ms\footnote{\url{http://www.srl.caltech.edu/NuSTAR_Public/NuSTAROperationSite/clockfile.php}}. 

\begin{figure}
\centering
\includegraphics[width=0.5\textwidth]{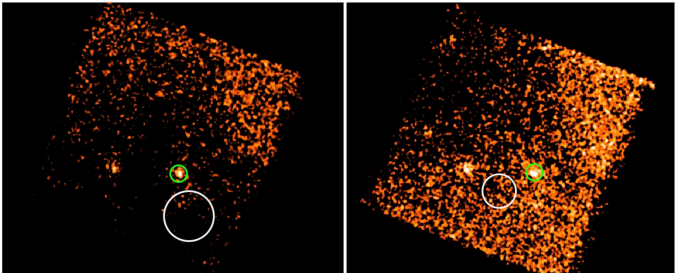}
\vspace{-2.0\baselineskip}\caption{\label{fig:figN_image}{\sl NuSTAR} FPMA (left) and FPMB (right) images in the 3-20 keV energy range of PSR J0659+1414 smoothed with a Gaussian kernel of $3''$ width. The green ($r=30''$) and white ($r_A=90''$ and $r_B=61''$) circles show the source and background extraction regions, respectively.}
\end{figure}

\section{Phase-Integrated Spectral Analysis}
\subsection{Spectral Extraction}
We extracted the phase-integrated EPIC-pn spectrum from a $15\arcsec$\explain{The hard X-rays ($\gtrsim 1$ keV) psf and optimal extraction region are close to $15\arcsec$. The PL $\Gamma$ is determined almost exclusively from the hard X-ray tail in which this extraction region produces the highest S/N spectrum. In fact, larger apertures lead to background counts higher than source counts at high energies and hence biased (softer) PL fits. This is why DeLuca fit a softer PL.} circular aperture in the $0.3-7$ keV range.
The lower energy cut-off is chosen at 0.3 keV, below which the nominal accuracy of the energy and the calibration of the effective area are unreliable$^2$.
The higher energy cut-off is chosen at 7 keV, above which the background contribution exceeds the source flux.
As shown in Figure \ref{fig:fig1_cmap}, we selected background regions sufficiently far away from the bright pulsar to avoid contamination from the point spread function (PSF) wings.
Our relatively small aperture ($15\arcsec$, optimized for the $2-7$ keV range) reduces the background contribution in the spectrum at energies $\gtrsim 2$ keV (34\% for $15\arcsec$ as opposed to 55\% for $37\farcs5$, which is the largest size allowed by the proximity of the chip edge), and hence allows high S/N at high energies.
As opposed to the larger $r = 37\farcs5$ aperture, the S/N ratio decreases by $\approx 10\%$ from 567 in the $0.3-2$ keV band and increases by $\approx 20\%$ from 17.6 in the $2-7$ keV band, when using the $15\arcsec$ extraction aperture.
The spectrum was extracted using the XMM-SAS task \texttt{especget}, and grouped using \texttt{specgroup} by fixing a minimum S/N of 4 per spectral bin while oversampling the detector's energy resolution by a factor $\leq3$\footnote{For energies <1 keV, the energy resolution of EPIC-pn is known to be $\approx 100$ eV (full width at half maximum for spectra considering single and double events; XMM UHB, 3.3.4., Fig26)}.
The $0.3-7$ keV net count rate is 3.78 counts per second, and the spectrum is constructed using 276299 events extracted from the $15\arcsec$ source region.

For {\sl NuSTAR}'s FPMA and FPMB detectors, the source photons were extracted from circular regions with a radius of 30$''$ centered on B0656.
The background photons were extracted from source-free regions away from the pulsar (see Figure \ref{fig:figN_image}).

\subsection{Spectral Fitting}\label{phintSpectral}

We used PyXspec ver.\ 2.0.0 interface with XSPEC ver.\ 12.9.0n \citep{Arnaud1996} for X-ray spectral modeling.
We modeled absorption by the interstellar medium (ISM) using the T\"{u}bingen-Boulder model \citep{2000ApJ...542..914W} through its XSPEC implementation \texttt{tbabs}, setting the abundance table to \texttt{wilm} \citep{2000ApJ...542..914W} and photoelectric cross-section table to \texttt{bcmc} \citep{1992ApJ...400..699B}, with updated He photo-ionization cross-section based on \citet{1998ApJ...496.1044Y}.
We implemented a Bayesian model fitting and parameter estimation routine using \texttt{emcee} \citep{2013PASP..125..306F}, a pure-Python implementation of Goodman and Weare's Affine Invariant Markov chain Monte Carlo (MCMC) Ensemble sampler, for posterior sampling.
The routines library also implements Bayesian evidence integral calculations using the so-called thermodynamic integration method \citep{Goggans2004}.

To reduce the computational overhead associated with our Bayesian analysis, we first ruled out the simpler single- and two-component models by checking the goodness of fit using the $\chi^2$ statistic and looking for localized systematic deviations in the residuals.
We tried the power-law (PL; \texttt{powerlaw} in XSPEC), blackbody (BB; \texttt{bbodyrad} in XSPEC), neutron star atmosphere (NSA --- \texttt{nsa} in XSPEC --- \citealt{Zavlin1996,Pavlov1995}; NSMAXG --- \texttt{nsmaxg} in XSPEC --- \citealt{2008ApJS..178..102H}; \citealt{2007MNRAS.377..905M}) models\footnote{NSA models the spectrum of a magnetized hydrogen atmosphere in hydrostatic and radiative equilibrium, with allowance for general relativistic effects.
XSPEC provides the NSA models for three magnetic field values, B = 0 ($< 10^8 - 10^9$ G), $10^{12}$, and $10^{13}$ G.
The neutron star mass and radius are typically fixed at $1.4\;M_\odot$ and 10 km, respectively,
and the parameters $\log T_{\rm eff}$ (temperature at the NS surface) and normalization $= 1/d^2$ (where d is the distance to the pulsar in parsecs) are free to vary.
The NSMAXG models hydrogen as well as other mid-Z (C, O, Ne) element plasma atmospheres.
It also includes a range of discrete magnetic field values between $10^{10}-10^{13}$ G, with specific magnetic inclinations and observing angles.
NSA and NSMAXG model the emission radius through the model normalization, norm $\propto R_{\rm e}^2/R_{\rm NS}^2$.
}.
The main difference between the NSA and NSMAXG is that the former assumes full ionization in the atmosphere while the latter takes into account the presence of not fully ionized atoms and allows the atmosphere to consist of elements other than Hydrogen (e.g., Carbon).

Single-component BB, NSA, NSMAXG, and PL models produce unacceptable fits with large systematic residuals at higher energies.
Two-component blackbody or atmosphere models do not fit the spectrum at higher energies either ($\chi_\nu^2 \approx 12$ for $\nu = 63$ degrees of freedom [dof]).
The flattening of the spectrum at energies $\gtrsim 2$ keV can be described by a PL with a photon index $\Gamma \sim 1.7$.
So, as the next step, we tried two-component models combining PL with BB, NSA, and NSMAXG.
These models fail to fit the spectrum too ($\chi_\nu^2 \gtrsim 4$) due to large systematic residuals at low energies, $\lesssim 1$ keV.

\begin{figure*}[ht]
\captionsetup[subfigure]{labelformat=empty}
\subfloat[]{\label{fig:J0659Sa}\includegraphics[width = 0.5\textwidth]{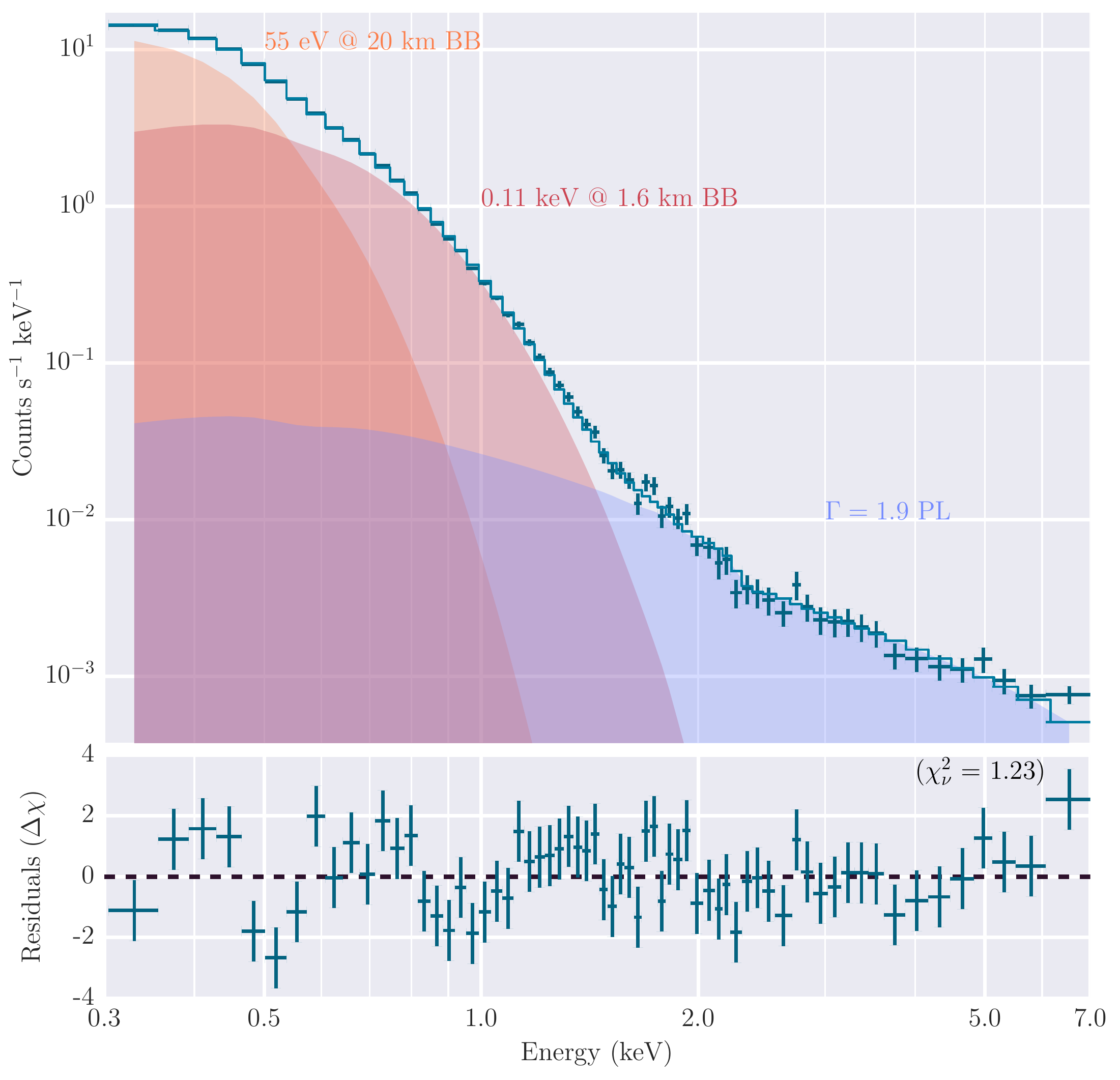}}
\subfloat[]{\label{fig:J0659Sb}\includegraphics[width = 0.51\textwidth]{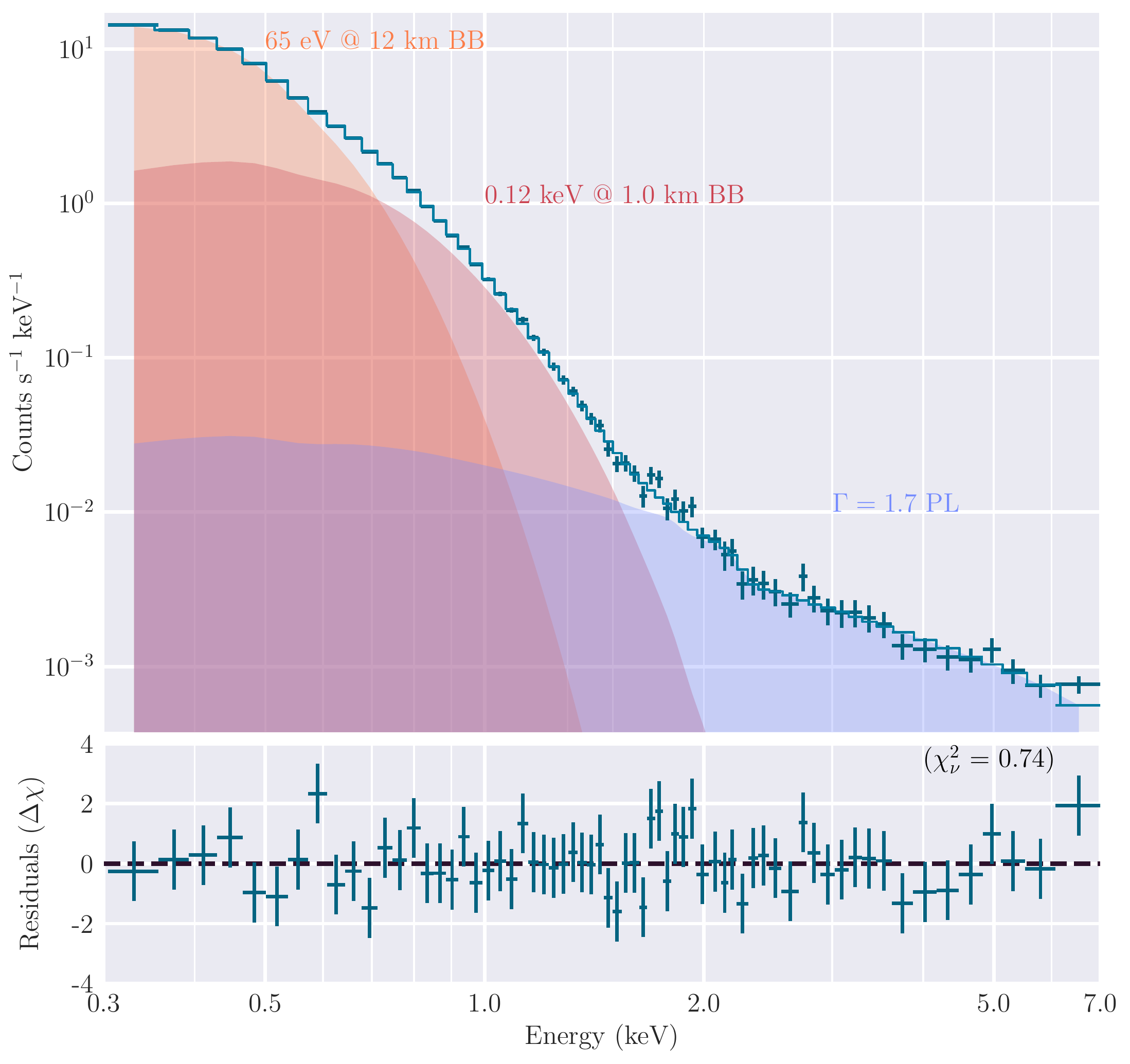}}
\caption{2BBPL model, without (left) and with (right) \texttt{Gabs} absorption component, fit to the phase-integrated spectrum of J0659+1414. The model parameters correspond to maximum posterior probability, which in the case of G2BBPL are different from the distribution medians listed in Table \ref{tbl-2_intspectra}.}
\label{J0659spectra}
\end{figure*}

After firmly ruling out the one- and two-component models, we fit a three-component model similar to \cite{Pavlov2002nsps} and \cite{DeLuca2005}, combining two BB with a PL (2BBPL).
The overall quality of the fit with the 2BBPL model is marginally acceptable ($\chi_\nu^2=1.2$ for $\nu=59$).
Figure \ref{J0659spectra} shows the spectra, the best-fit 2BBPL model, the contributions from individual components, and the residuals of the fit.
However, the fit produces systematic residuals around 0.5 and 1 keV.
To better accommodate these features, we tried the two-temperature atmosphere models and additional blackbody or absorption line components.

A double NSMAXG plus PL model gives a fit quality comparable to that of 2BBPL, but with unrealistic emission radius ($R_{e}$) and distance parameters.
For this fit, we froze the NS mass and radius at $M_{\rm NS}=1.4$ M$_\odot$ and $R_{\rm NS}=10$ km, respectively, and fixed $B=10^{12}$ G (the models with $B=0$ give poorer fits).
The hot component has an effective temperature $T_{\rm eff,h} = 0.53-0.58$ MK, emitted from a region of radius $R_{\rm e,h} \sim 5$ km, at $d=288$ pc distance, which is much larger than the conventional polar cap radius, $R_{\rm pc} = \sqrt{2\pi\,R_{NS}^3/cP} \approx 230$ m.
The cold component has $T_{\rm eff,c} = 0.16-0.22$ MK, emitted from an unrealistically large sphere, $R_{\rm e,c} \sim 1300$ km, for the same distance.
A double NSMAXG plus PL model with the model switch 1260 (H atmosphere and $B=4\times10^{12}$ G) provides a fit better than the other ones, but still with unrealistic emission size parameters.
The hotter and colder NSMAXG components imply emission regions $\sim 2$ and $\sim 30$ times the typically assumed NS radius, respectively. 
Therefore, we conclude that the NSA and NSMAXG models do not provide realistic parameters for the pulsar's thermal emission.

We also tried to fit a triple blackbody plus power-law model (3BBPL).
The best fit parameters at 90\% confidence, using $d=288$ pc for estimating radii, are as follows:
$kT_{\rm BB} = 37-49$ eV, $74-98$ eV, and $120-170$ eV from $R_{e} = 32-93$ km, $3-7$ km, and $0.2-1$ km, respectively, and $\Gamma = 1.4-1.9$.
The overall fit quality is good, and the residuals are well-behaved\footnote{However, this model still fails to adequately describe the phase-resolved spectrum in certain phase bins (see section \ref{phresolved}).}, but the cool BB radius substantially exceeds a plausible NS radius rendering this fit unphysical.

Next, we attempted to eliminate the residuals, seen in the 2BBPL fit around 0.5 keV by adding an absorption line modeled as a Gaussian profile (\texttt{Gabs} in XSPEC).
We denote this model by G2BBPL.
The Gabs component is a multiplicative energy-dependent factor,
\begin{align}
F(E) = \exp (-\tau) \label{eq:FGabs} ;\qquad
\tau = \tau_0 \, \exp\left[-\frac{(E-E_c)^2}{2\sigma^2}\right],
\end{align}
where the optical depth, $\tau_0 = s/\sqrt{2\pi\sigma^2}$, at line's central energy $E_c$ is related to the strength ($s$) and Gaussian width ($\sigma$) of the line.

When comparing nested models that become identical when one of the parameters goes to zero, the use of likelihood ratio tests or F-test leads to biased  estimates \citep{Protassov2002}.
A Bayesian model comparison by either computing posterior predictive probability values or evidence-based Bayes factors is, however, free of such biases.
Below, we compare the 2BBPL and G2BBPL models using their Bayes factors in order to assess the significance of the absorption component.

For Bayesian spectral analysis, we construct a Gaussian likelihood function to fit the above-mentioned models to the background-subtracted spectrum.
The log likelihood function is,
\begin{align}
\ln P(D\vert M)=-\frac{1}{2}\sum_n\left[ \frac{(D_n-M_n)^2}{\sigma_n^2} \right]
\end{align}
where $D_n$, with uncertainties $\sigma_n$, represent the observed spectral flux in $n$-th energy bin, and $M_n$ represent the corresponding model spectrum.
We choose non-informative priors for the model parameters, uniform over linear (uniform prior) or log parameter space (Jeffreys prior).
\begin{align}
P(\theta\vert M) &= \frac{1}{\theta_{\rm max}-\theta_{\rm min}} \quad[{\rm uniform\;prior}] \\
P(\ln\theta\vert M) &= \frac{1}{\ln\theta_{\rm max}-\ln\theta_{\rm min}} \quad[{\rm Jeffreys\;prior}]
\end{align}
The posterior distribution is obtained by taking the product of the likelihood function and the priors, $P(M\vert D) \propto P(D\vert M)\cdot P(\theta\vert M)$.
As seen in Table \ref{tbl-2_intspectra}, the prior range is sufficiently large, significantly exceeding the range of the resulting posterior distribution.
Non-informative priors contribute a constant factor to the posterior probability within the allowed parameter range and zero outside of it.
Hence, the posterior is only affected by the data and the likelihood function, and large parameter ranges in the priors have no effect on the posterior.
The posterior distribution is sampled with the \texttt{emcee} sampler using 72 walkers\footnote{http://dan.iel.fm/emcee/current/user/faq/\#walkers} (parallel Markov chains), taking between $12000-16000$ steps.
The initial parameter guesses for the walkers are determined using the $\chi^2$ fit, which ensures a faster convergence.
The stability of the converged posterior sampling is tested out to 100,000 steps.
After eliminating the initial $\sim 10\%$ burn-in steps during which the walkers stabilize, we have over 720,000 individual samples representing the posterior distribution.

The distributions of parameter values are more meaningful than a set of `best-fit' values.
A best fit can be rigorously defined in Bayesian analysis through a maximum-a-posteriori (MAP) estimate, which is the set of parameters corresponding to the statistical mode of the sampled posterior.
Determination of mode of the posterior distribution, however, is often computationally difficult, hence it is customary to give the median values of the marginalized parameter distributions.
In some cases the parameters producing the maximum posterior probability are a better representation of the MAP than the median.

\begin{figure*}[ht]
\captionsetup[subfigure]{}
\subfloat[]{\label{fig:J0659M1a}\includegraphics[width = 0.5\textwidth]{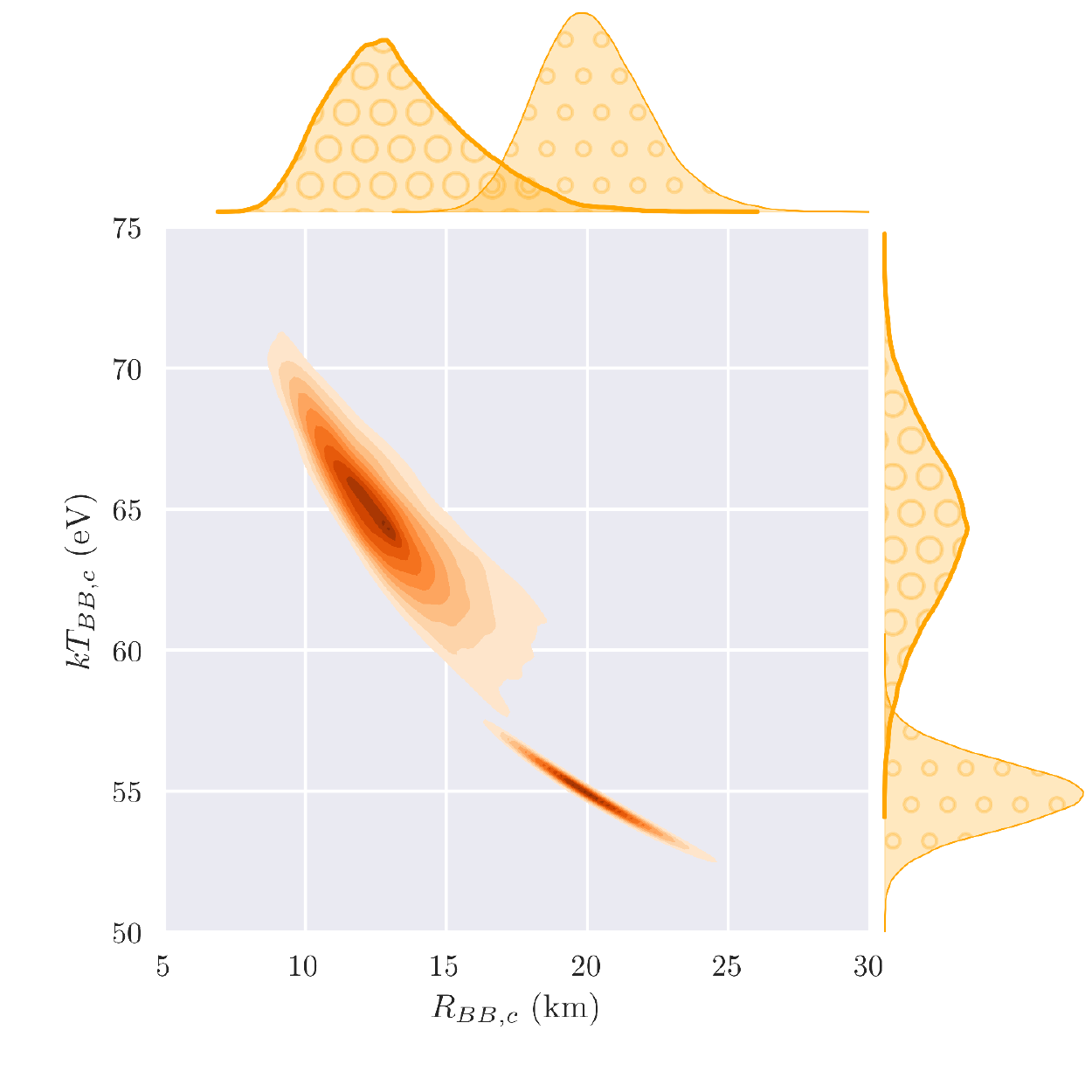}} 
\subfloat[]{\label{fig:J0659M1b}\includegraphics[width = 0.5\textwidth]{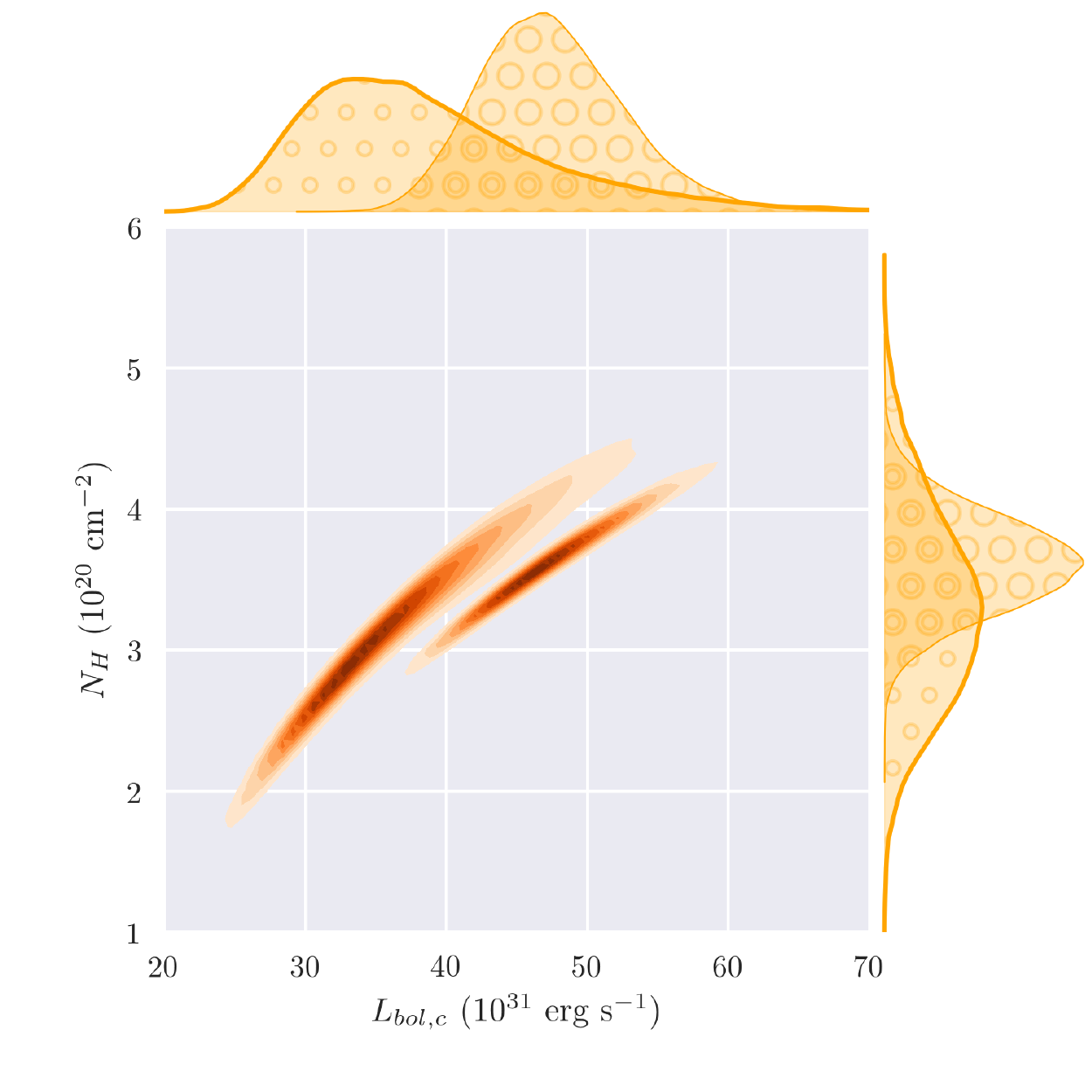}} \\
\subfloat[]{\label{fig:J0659M1c}\includegraphics[width = 0.5\textwidth]{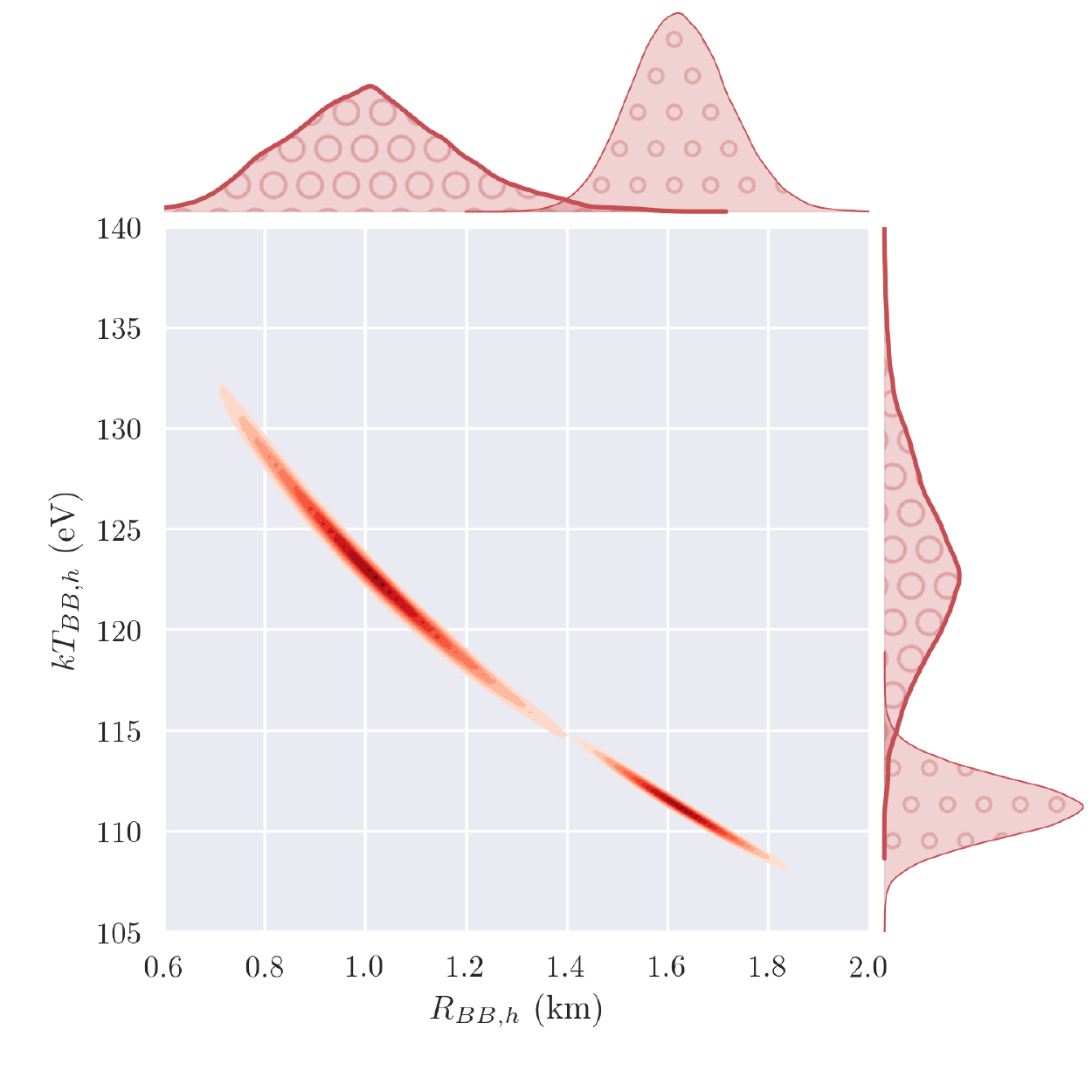}}
\subfloat[]{\label{fig:J0659M1d}\includegraphics[width = 0.5\textwidth]{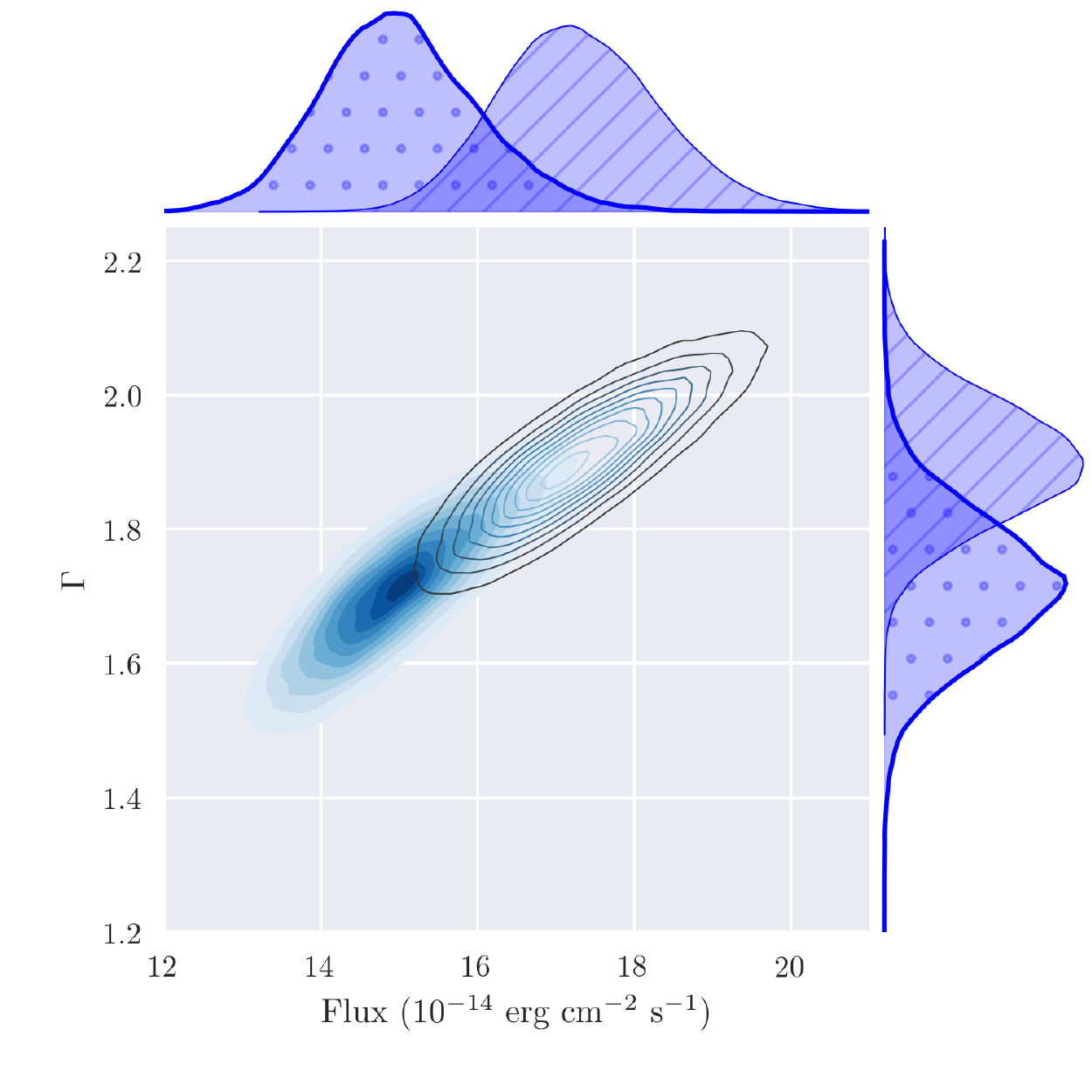}}
\caption{Two-dimensional marginalized joint plots with $10-90$ percentile contours (in increments of 10\%) for 2BBPL and G2BBPL model parameters: (a) cold BB temperature $kT_{\rm BB,c}$ versus effective BB radius $R_{\rm BB,c}$; (b) absorbing, equivalent hydrogen column density $N_{\rm H}$ versus cold BB bolometric luminosity $L_{\rm Bol,c}$; (c) hot BB $kT_{\rm BB,h}$ versus corresponding $R_{\rm BB, h}$; (d) PL photon index $\Gamma$ versus unabsorbed PL flux in 0.3-7 keV.
The one-dimensional marginalized parameter distributions are plotted on the axes; G2BBPL distributions are plotted with thicker lines.}
\label{J0659marginals1}
\end{figure*}

We use Bayesian parameter estimation to obtain the parameter distributions for a 2BBPL model fit.
Our posterior probable parameter ranges are compatible with the previous results of \cite{DeLuca2005}.
The most probable 2BBPL model parameters, with 90\% credible limits, include a $kT_{\rm BB,c}=53-57$ eV `cold' BB component emitted from a region of $R_{\rm BB,c}=18-23$ km effective radius, a $kT_{\rm BB,h}=109-113$ eV `hot' BB from a region of $R_{\rm BB,h}=1.5-1.7$ km, and a PL component with $\Gamma = 1.8-2.0$ and an unabsorbed flux $F_{\rm PL}^{\rm unab}=(1.6-1.9)\times10^{-13}$ erg cm$^{-2}$ s$^{-1}$ in $0.3-7$ keV.
For the G2BBPL model, the probable range of continuum parameters with 90\% credible limits are as follows: $kT_{\rm BB,c} = 60-68$ eV and $R_{\rm BB,c} = 10-17$ km, $kT_{\rm BB,h} = 0.12-0.13$ keV and $R_{\rm BB,h} = 0.8-1.2$ km, and $\Gamma = 1.6-1.8$ (Table \ref{tbl-2_intspectra}).
The quoted $R_{\rm BB}$ and $T_{\rm BB}$ values are as observed at infinity \added{(i.e., not corrected for the gravitational redshift and photon trajectory bending)}.
The median and 90 percentile limits of model parameters and derived quantities, obtained from marginalized distributions, are listed in Table \ref{tbl-2_intspectra}.
The model fits, with individual component contributions and residuals, are shown in Figure \ref{J0659spectra}.
We use the maximum posterior probability parameters as the `best-fit' model for the spectral plot, since this model produces smaller residuals than that using median values.
In the joint-distribution plots (Figure \ref{J0659marginals1}), we compare the marginalized joint posterior distributions of the continuum parameters, and show the systematic parameter offsets between the 2BBPL and G2BBPL models.

\begin{figure*}[ht]
\centering
\includegraphics[width=0.8\textwidth]{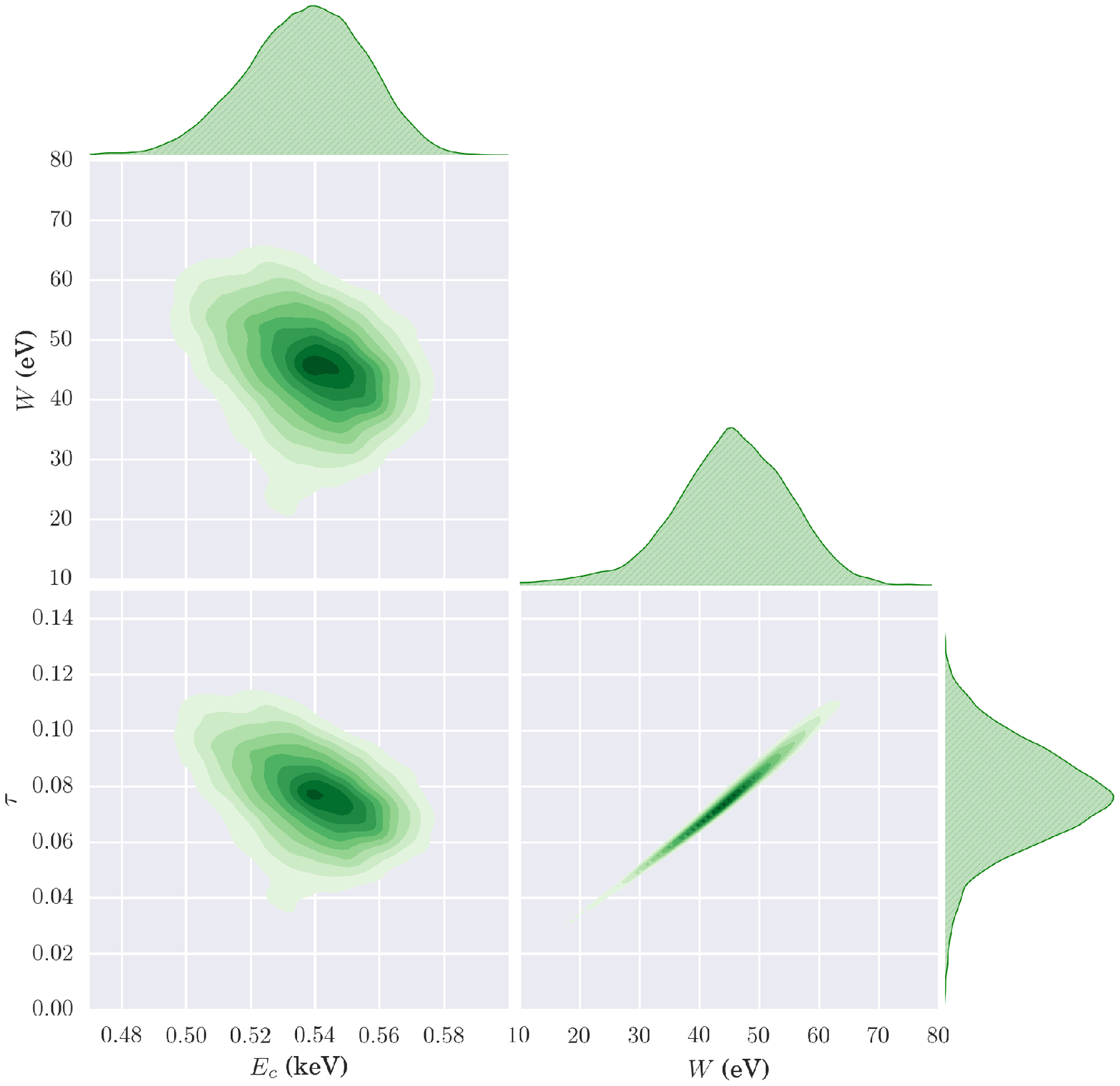}
\caption{\label{J0659marginals2}Two-dimensional marginalized pair plots with $10-90$ percentile contours (in increments of 10\%) for \texttt{Gabs} component parameters from the G2BBPL model: absorption line energy $E_{\rm c}$ versus equivalent width $W$ (top); $E_{\rm c}$ versus optical depth $\tau$ (bottom-left); $W$ versus $\tau$ (bottom-right). The one-dimensional marginalized parameter distributions are plotted opposite to the axes.}
\end{figure*}

The joint-distribution plots for the \texttt{Gabs} absorption line energy, optical depth, and equivalent width are plotted in Figure \ref{J0659marginals2}.
The equivalent width is defined as
\begin{align}
W = \int \left[ 1 - F(E) \right] dE,
\end{align}
where $F(E)$ is given by equation \eqref{eq:FGabs}.
The $90\%$ probable ranges for central line energy is $0.51-0.57$ keV, for optical depth $\tau_0=0.055-0.099$ and for line width $\sigma = 0.07-0.13$ keV.
The equivalent width of the absorption line is between $33-58$ eV.

We compare the two models, 2BBPL and G2BBPL, using the Bayesian Information Criteria (BIC; \citealt{schwarz1978estimating}).
The median BIC values are listed in Table \ref{tbl-2_intspectra} with smaller (preferred) values corresponding to the G2BBPL model and $\Delta$BIC = 56, between the two models in question.
$\Delta$BIC > 10 implies very strong evidence against the model with higher BIC value \citep{kass1995bayes}.
We also evaluate the Bayesian evidence ($Z$) and calculate the Bayes factor (\citealt{GoodmanBayes1}; \citealt{GoodmanBayes2}) for G2BBPL over 2BBPL, $\Delta\ln Z = \ln(Z_{\rm G2BBPL}/Z_{\rm 2BBPL}) = 8.7$.
The evidence is defined as the integral of the unnormalized posterior over the entire parameter space, $Z\equiv\int d\theta P(D\vert M)\cdot P(\theta\vert M)$.
Assuming equal prior odds for the two models, $P(M_2)/P(M_1)$ = 1, where, $M_2$ represents G2BBPL and $M_1$ represents 2BBPL, the posterior odds-ratio for model G2BBPL over 2BBPL is $O_{21} = \frac{Z(M_2)}{Z(M_2)} \frac{P(M_2)}{P(M_1)} = 6000$.
Hence, the G2BBPL model is 6000 times more probable than the 2BBPL model.

\begin{table}
\centering
\caption{Marginalized parameter distributions represented by medians with $10-90$ percentile intervals for the phase-integrated {\sl XMM-Newton} spectrum.\label{tbl-2_intspectra}}
\begin{tabular}{lccc}
\tableline\tableline
Parameter (Units) & priors (range) & 2BBPL & G2BBPL \\
\tableline
$N_{\rm H}$ ($10^{20}$ cm$^{-2}$) & $10^{-3} - 100$ & $3.6^{+0.5}_{-0.4}$  & $3.0^{+0.7}_{-0.9}$  \\
$E_{\rm c}$ (keV)	& $0.3-0.9$ &  		---						& $0.54^{+0.02}_{-0.03}$ \\
$\sigma$ (keV)	& $10^{-3}-1$ &  		---						& $0.10^{+0.03}_{-0.03}$  \\
$\tau_0$\tablenotemark{a} & --- & 		---		 	& $0.077^{+0.022}_{-0.022}$ \\
$W$\tablenotemark{b} (eV) & --- & 		---		 	& $46^{+12}_{-13}$ \\[1.25Ex]
$kT_{\rm BB,c}$ (eV) & $30 - 100$ & $55^{+2}_{-2}$ 					& 64$^{+4}_{-4}$ \\
BB norm ($10^{5}$)\tablenotemark{c} & $10^{-5}-10^{5}$ & $4.86^{+1.37}_{-1.03}$	& $2.04^{+1.30}_{-0.72}$  \\
$R_{\rm BB,c}$ (km)  & --- & $20^{+3}_{-2}$ 				& $13^{+4}_{-3}$ \\
$L_{\rm bol,c}$ ($10^{32}$ erg s$^{-1})$\tablenotemark{d} & --- & $4.7^{+0.7}_{-0.6}$	& $3.7^{+1.3}_{-0.8}$  \\[1.25Ex]

$kT_{\rm BB,h}$ (eV) & $90-500$ & $111^{+2}_{-2}$ 					& $123^{+6}_{-5}$ \\
BB norm \tablenotemark{c} & $1-10^{6}$ & $3173^{+523}_{-457}$		& $1216^{+613}_{-449}$  \\
$R_{\rm BB,h}$ (km)  & --- & $1.6^{+0.1}_{-0.1}$ 				& $1.0^{+0.2}_{-0.2}$ \\
$L_{\rm bol,h}$ ($10^{31}$ erg s$^{-1})$\tablenotemark{d} & --- & $5.2^{+0.5}_{-0.4}$	& $3.0^{+0.8}_{-0.7}$  \\[1.25Ex]

$\Gamma$ & $0.5-4$ & $1.9^{+0.1}_{-0.1}$ 					& $1.7^{+0.1}_{-0.1}$  \\
PL norm $(10^{-5})$\tablenotemark{e} & $10^{-2}-10^{2}$ & $3.29^{+0.42}_{-0.38}$	& $2.58^{+0.44}_{-0.39}$  \\
$F^{\rm unabs}_{0.3 - 7\,{\rm keV}}$ ($10^{-13}$ erg cm$^{-2}$ s$^{-1}$)\tablenotemark{f} &  & $1.73^{+0.12}_{-0.14}$	& $1.50^{+0.12}_{-0.14}$  \\
BIC\tablenotemark{g} & --- & 666								& 610 \\
Evidence\tablenotemark{g} ($\log Z$) & --- & -307				& -299 \\
\tableline
\end{tabular}
\tablecomments{Parameters without prior value ranges are derived quantities.}
\tablenotemark{a}{Optical depth $\tau_0 = s/\sqrt{2\pi\sigma^2}$, where $s$ and $\sigma$ are the absorption line strength and standard deviation of the Gaussian line profile, respectively.}
\tablenotemark{b}{Equivalent width of the absorption line.}
\tablenotemark{c}{BB normalization $= R_{\rm BB}^2/d_{10\, {\rm kpc}}^2$, where $R_{\rm BB}$ is the effective radius of BB emission region in km, and $d_{10\,{\rm kpc}}$ (= 0.0288 for B0656) is the distance in units of 10 kpc.}
\tablenotemark{d}{Blackbody bolometric luminosity.}
\tablenotemark{e}{PL normalization in units of $10^{-5}$ photons cm$^{-2}$ s$^{-1}$ keV$^{-1}$ at 1 keV.}
\tablenotemark{f}{Unabsorbed PL flux in the $0.3-7$ keV range.}
\tablenotemark{g}{BIC is the Bayesian Information Criterion. Evidence values quoted here are the un-normalized Bayesian log evidence.}
\end{table}

Thus, we have shown that the systematic residuals present near 0.5 and 1 keV can be eliminated by adding an absorption feature modeled with Gabs, which not only produces a better overall fit ($\chi_\nu^2=0.74$) with no significant local systematic residuals, but also results in more realistic radii for the cold BB.

NS radii (as observed by a distant observer) are in the range $12-18$ km \citep{Steiner2010}, while the conventional polar cap radius, for a dipole B-field and period $P=385$ ms of B0656, is $R_{\rm pc} \approx 230$ m. 
For the 2BBPL fit, the cold BB effective emission region radii are in the range, $R_{\rm BB, c} = 16-24$ km (with 90\% probability), after accounting for the distance uncertainty.
The $R_{\rm BB,c}$ range, obtained from the fits, has some overlap with the expected NS radius, but the $R_{\rm BB,h} = 1.3-1.9$ km range is significantly larger than the conventional polar cap radius.
The G2BBPL model fit gives $R_{\rm BB,c}=9-17$ km, consistent with the expected NS radius, and $R_{\rm BB,h}=0.7-1.3$ km closer to the expected polar cap radius.
These derived parameters only marginally favor the G2BBPL model over the 2BBPL model, but the Bayes factor and odds ratio calculations show that the G2BBPL model is $\sim 6000$ times more probable than the 2BBPL model.

We use the {\sl NuSTAR} data to better constrain the  PL component.
We chose the 3 keV lower energy cut off to minimize the contamination by thermal emission from hotter parts of the NS surface while that 20 keV upper energy cut off is dictated by the growing background contribution.
In the 3-20 keV energy band there are 165 and 133 net counts detected by the FPMA and FPMB detectors, respectively. The source and background spectra were binned to have 1 count per bin and fit with W statistic\footnote{ https://heasarc.gsfc.nasa.gov/xanadu/xspec/manual/XSappendixStatistics.html}.
The Hydrogen absorption column density was frozen to the best-fit {\sl XMM-Newton} EPIC-pn spectrum value ($N_{\rm H}=3\times10^{20}$ cm$^{-2}$), however, the spectrum $>$ 3 keV is not expected to be affected by absorption for any plausible $N_{\rm H}$ value.
The PL model provides a good fit with $\Gamma=1.5\pm0.2$ consistent with the index found in the {\sl XMM-Newton} spectrum (see Table 2). 
We also find a normalization difference between the FPMA and FPMB detectors, with the FPMB PL normalization being $0.89\pm0.15$ of that of FPMA.

The ratios of the observed best-fit $3-7$ keV flux values (in units of $10^{-14}$ erg cm$^{-2}$ s$^{-1}$), $F_{\rm FPMA}/F_{\rm PN} = (4.99\pm0.92)/(5.45\pm0.63) = 0.92\pm0.20$ and $F_{\rm FPMB}/F_{\rm PN} = (4.41\pm0.94)/(5.45\pm0.63) = 0.81\pm0.20$, are lower than the cross-calibration values $F_{\rm FPMA}/F_{\rm PN} = 1.10\pm0.02$ and $F_{\rm FPMB}/F_{\rm PN} = 1.14\pm0.02$ quoted in \cite{Madsen2017}. However, the differences are within $2\sigma$ and $3\sigma$, respectively.

\begin{figure}
\centering
\includegraphics[width=0.5\textwidth]{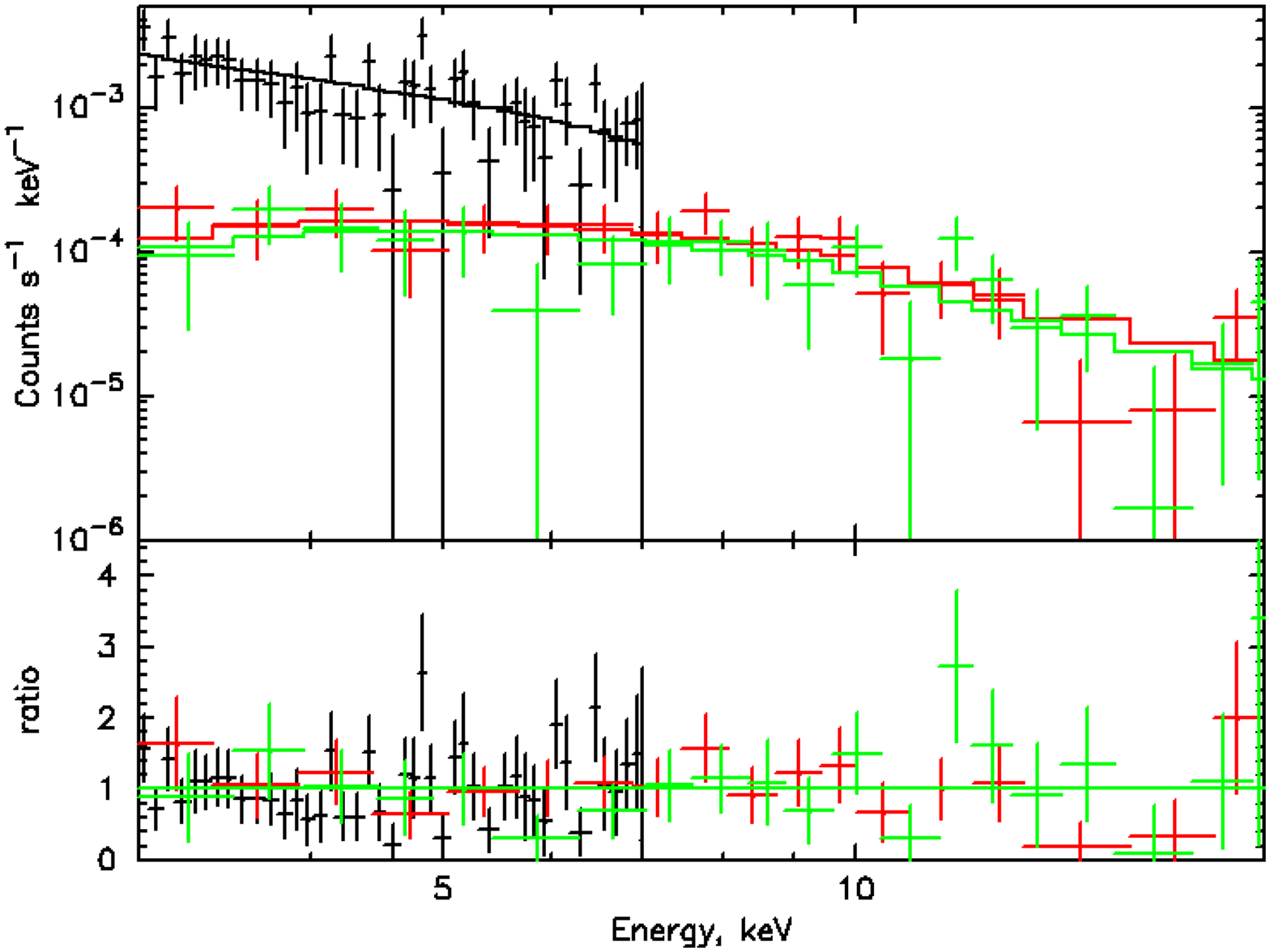}
\caption{\label{figN_spec}{\sl XMM-Newton} EPIC-pn (black; 3-7 keV) and {\sl NuSTAR} FPMA (red; 3-20 keV) and FPMB (green; 3-20 keV) spectra and the best-fit absorbed PL model ($\Gamma=1.46\pm0.15$) are shown. 
The fit is performed with unbinned data (using C-statistics). The  spectra shown are binned for visualization purposes only.}
\end{figure}

We have also simultaneously fit the {\sl XMM-Newton} EPIC-pn (379 net counts in the 3-7 keV band, extracted from the region described in Section 3.1) and {\sl NuSTAR} FPMA and FPMB spectra (in 3-20 keV) with an absorbed PL model, keeping the hydrogen absorption column frozen to the same value and allowing for different normalizations.
The best-fit photon index $\Gamma=1.46\pm0.15$ is consistent with the photon indices found by both the joint {\sl NuSTAR} FPMA/FPMB spectra and in the fit to the {\sl XMM-Newton} spectrum. The PL normalizations for the  FPMA and FPMB spectra are  0.77$^{+0.12}_{-0.11}$ and 0.69$^{+0.12}_{-0.10}$ of the EPIC-pn normalization, respectively (see Figure \ref{figN_spec}). 
Freezing the normalizations  at their best-fit values results in $\Gamma=1.46\pm0.12$.

\section{Timing Analysis}
\subsection{Energy-Integrated Timing with XMM-Newton}\label{eItime}
\begin{figure}
\captionsetup[subfigure]{labelformat=empty}
\subfloat[]{\label{fig:z22test}\includegraphics[width = 0.49\textwidth]{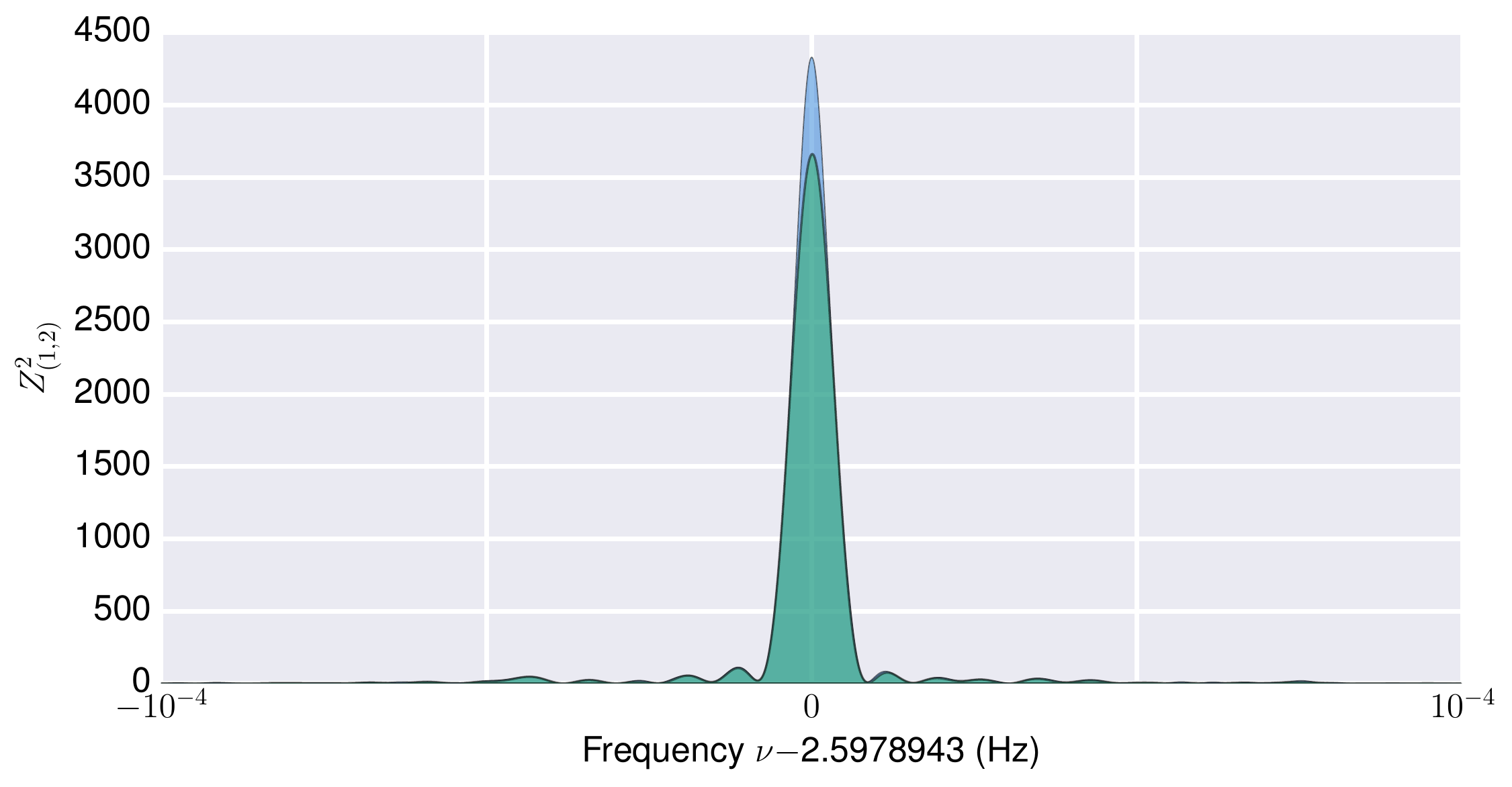}}
\vspace{-2\baselineskip}\\
\subfloat[]{\label{fig:J0659Tb}\includegraphics[width = 0.49\textwidth]{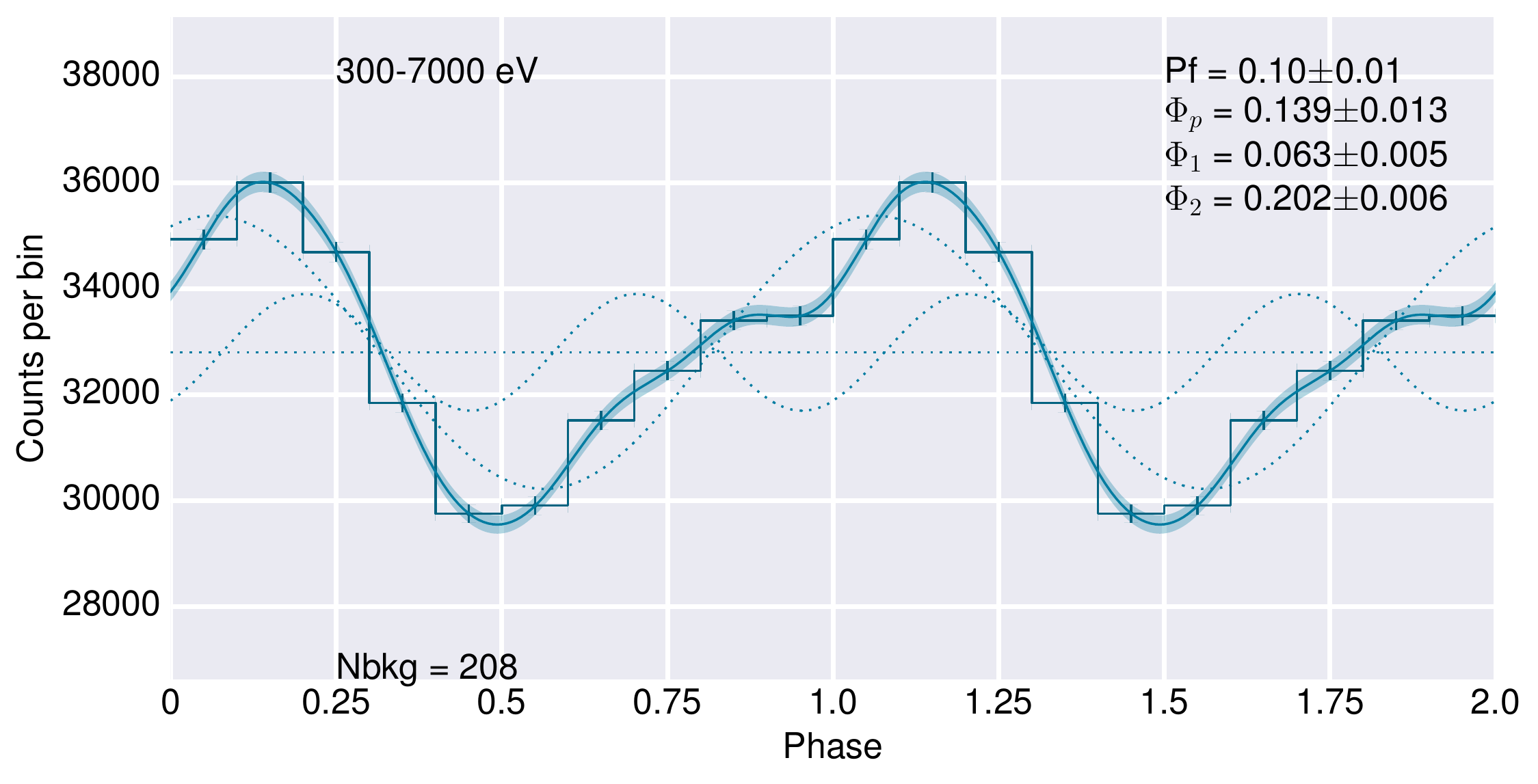}}
\caption{Top: $Z_{\rm n}^2$ statistics for two significant harmonics, $Z_1^2$ (green) and $Z_2^2$ (blue), obtained for $0.3-7$ keV events extracted from EPIC-pn source region.
Bottom: Pulse profile from phase-folded (phase $\phi=0$ corresponds to $T_{\rm ref,0}=57285.585338$) and binned ($N_{\rm bins}=10$) $0.3-7$ keV events.
The continuous line is the Fourier smoothed profile with shaded area representing the Poisson uncertainty.
The dotted lines show the average level and first and second significant harmonics that represent the curve.
The plot is annotated with the observed pulsed-fraction (Pf), background counts per bin (Nbkg), and phases of the maxima of the full profile ($\Phi_p$) and the first two significant harmonic components ($\Phi_1, \Phi_2$).}
\label{J0659timing}
\end{figure}

We extracted 328,002 events from an energy-dependent source aperture applied to the GTI filtered, $0.3-7$ keV energy-restricted\explain{I had earlier included photons below 0.3 keV for timing, but we decided all the different energy ranges are confusing and look inconsistent. Anyway, no new information is added by expanding the energy range here.}, barycentered (SAS task \texttt{barycen}) data.
The extraction aperture varies from $37\farcs5$ below 0.8 keV to $12\farcs5$ above 4.5 keV (over 10 equally-spaced energy intervals in log-scale), ensuring high S/N extraction at all energy ranges.
We used the $Z_n^2$ test \citep{Buccheri1983} for $n = 1-4$, and the H-test \citep{1989A&A...221..180D} and found statistically significant contributions only up to $n=2$ harmonics. 
For the reference time corresponding to the middle of the observation, $T_{\rm ref} = $ MJD 57285.58708124, we found $Z_{\rm 2,max}^2 =4327$ at the frequency $\nu = 2.59789433(5)$ Hz, where the $1\sigma$ uncertainty of the last significant digit is given in parentheses (Figure \ref{J0659timing}).
The detected frequency is consistent with the frequency predicted from the $\gamma$-ray ephemeris (Table \ref{tbl-1_summary}) extrapolated to the \xmm observation epoch, $\nu_{\gamma} = 2.597894339(1)$ Hz.
Within the GTI time span of 115 ks, $Z_n^2$ is not sensitive to frequency derivative values $|\dot{\nu}| \lesssim 10^{-11}$ Hz s$^{-1}$ (significantly larger than the pulsar's $\dot{\nu} = -3.7 \times 10^{-13}\;{\rm s}^{-2}$).

The binned pulse profile obtained by phase-folding with the above-determined frequency is shown in Figure \ref{J0659timing}.
All the folded profiles have the phase zero at $T_{\rm ref,0} = $ MJD 57285.585338.
The pulsed fraction defined as the ratio of the area enclosed in the pulse profile above the light curve minimum to the total profile area, is $10\%\pm1\%$.
Due to the negligible background contribution ($\approx0.6\%$ in the 0.3-7 keV), the intrinsic and observed pulsed fractions are virtually identical. 

We study the harmonic behavior of the profile by taking the discrete Fourier transform of the binned profile as implemented in the Python library SciPy \citep{scipy2001}.
A smoothed profile is obtained by using the Fourier shift theorem to obtain interpolated values within a bin.
The uncertainties in the pulsed fraction and phase of profile and harmonic component maxima are obtained through bootstrap re-sampling \citep{FeigelsonBabu2012}, assuming independent Poisson distribution of counts in each profile bin.
The smoothed $0.3-7$ keV pulse profile along with Poisson uncertainty bounds, the zero, first, and second significant harmonic contributions are shown in Figure \ref{J0659timing}.

The pulsations significance and the broad profile shape in the $0.3-7$ keV range are dominated by the soft photons ($\leq 0.7$ keV) from the cold BB part of the spectrum.
If the cold BB is emitted from the entire stellar surface, as suggested by the fit BB emission area sizes, the pulsations imply a non-uniform temperature distribution within the area attributed to cold component.
The profile cannot be approximated by a simple sine curve, but requires at least two phase-shifted harmonic sine curves, as seen from the Fourier components.

\subsection{Energy-Resolved Timing with XMM-Newton} \label{eVPhiTiming}
\begin{figure}
\centering
\includegraphics[width=0.49\textwidth]{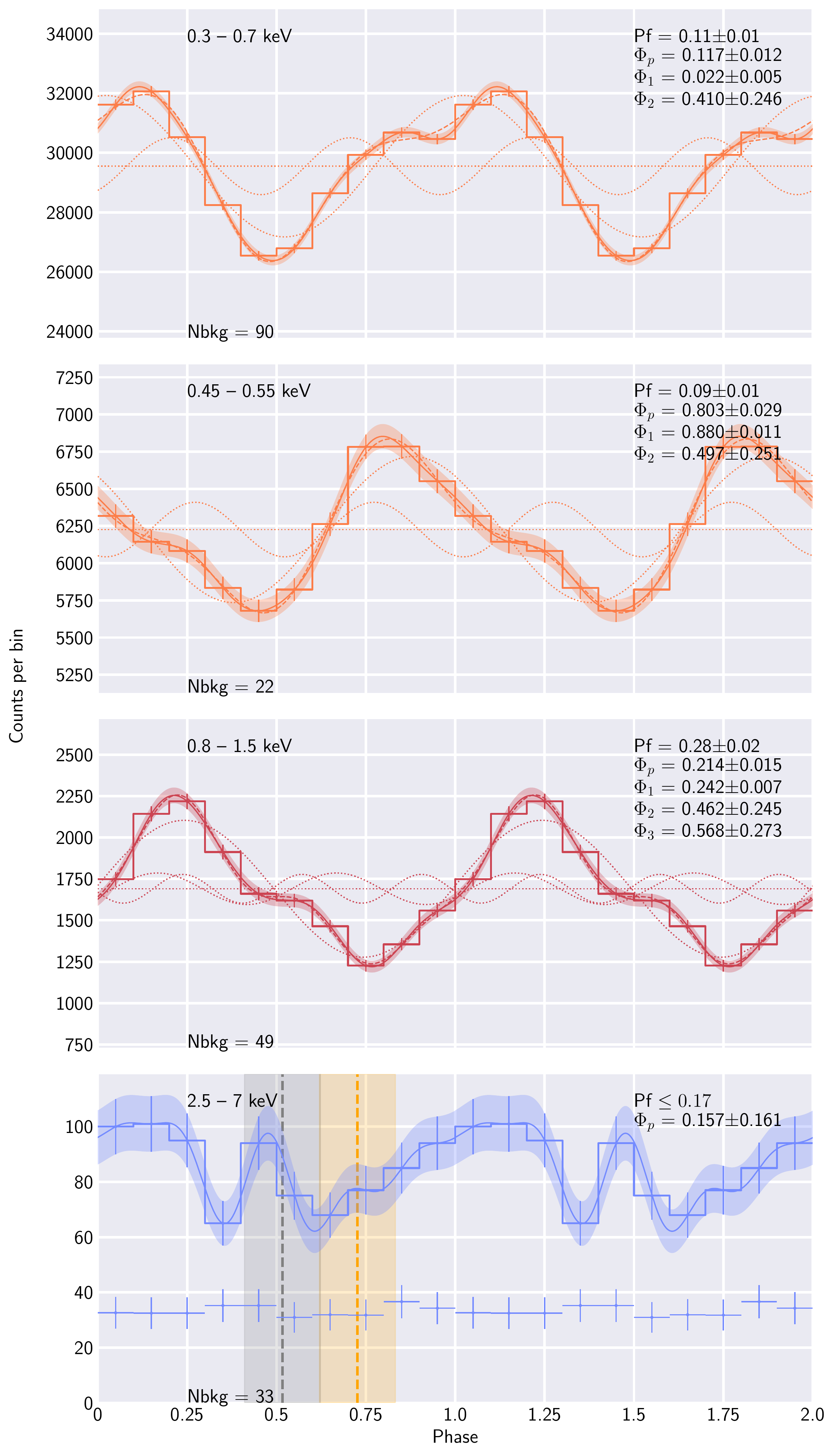}
\caption{\label{fig:stackedprofiles} Top to bottom: {\sl XMM-Newton} pulse-profiles from 295298 counts in $0.3-0.7$ keV, 131598 counts in $0.4-0.6$ keV, 16899 counts in $0.8-1.5$, and 854 counts in $2.5-7$ keV range.
The Fourier smoothed profiles with the uncertainty ranges are over-plotted. 
The dotted lines show the contributions from the zero and the first few Fourier components.
The plot is annotated with the observed pulsed-fraction (Pf), background counts per bin (Nbkg), and phase of maxima of the total profile ($\Phi_p$) and the Fourier components ($\Phi_n$).
For the $2.5-7$ keV profile, the background light curve is shown but the Fourier components are omitted because there are not enough counts for a robust decomposition.
In the bottom panel, we also show the phases for the 1.4 GHz radio (grey) and $\gamma$-ray (yellow) peaks (dashed) with uncertainties (shaded regions).
The radio and $\gamma$-ray peak phase uncertainties shown here are due to extrapolation of the $\gamma$-ray ephemeris from MJD 55677.
The radio-$\gamma$ offset, $\phi_\gamma - \phi_{\rm rad} = 0.21\pm0.01$ \citep{Weltevrede2010}, is not subject to this uncertainty.}
\end{figure}

We use the phase-integrated spectral fits to approximately determine the energy ranges within which each of the model components dominates.
From the G2BBPL fit in Figure \ref{J0659spectra}, it is apparent that the cold BB is dominant in  $\lesssim 0.7$ keV range (which includes the 0.4-0.6 keV absorption feature region), hot BB in $\approx 0.8-1.5$ keV, and the PL at energies $\gtrsim 2$ keV.
The pulse profiles in these energy ranges are shown in Figure \ref{fig:stackedprofiles}.
The pulsations have very high significance in the $0.3-0.7$ keV and $0.8-1.5$ keV ranges, whereas the significance is only marginal in the $2.5-7$ keV band.
Additionally, we show pulsations in the $0.4-0.6$ keV range, where the absorption feature is prominent.
These pulse profiles show large (statistically significant) differences in the 4 energy bands, which reflects different origins of the dominating emission.

The cold BB dominant energy range, $0.3-0.7$ keV, contains 90\% of the events extracted in the $0.3-7$ keV, hence the profile shape is very similar to the energy-integrated profile in Figure \ref{J0659timing}.
However, in the $0.4-0.6$ keV range, where the absorption feature is prominent, the profile shape changes, and the peak is significantly offset ($\Delta\phi=0.72\pm0.03$ in phase) compared to the peak of the $0.3-0.7$ keV profile.
This statistically significant profile change indicates likely phase-variability of the absorption feature seen in the phase-integrated spectrum.
The profile shape changes in the $0.8-1.5$ keV energy range with a $\Delta\phi = 0.10\pm 0.02$ offset in peak and a higher pulsed fraction, when compared to the $0.3-0.7$ keV profile.
The profile in this hot BB dominant energy range is broadly described by a sine curve, but second and third harmonics are needed to better model the profile extrema.
As expected, the hot BB component emitted from a hotspot covering only a fraction of the stellar surface has a higher pulsed fraction than the cold BB component from the entire surface.
There are too few source counts in the non-thermal component to reliably infer its pulse profile.

We phase-connect our X-ray profiles with the $\gamma$-ray profile using the timing solution provided by \cite{Ray2011}\footnote{https://confluence.slac.stanford.edu/display/GLAMCOG/LAT+Gamma-ray+Pulsar+Timing+Models} and show the phase of $\gamma$-ray peak (bottom panel of Figure \ref{fig:stackedprofiles}).
We also show the 1.4 GHz radio peak using the radio -- $\gamma$-ray peak offset reported by \cite{Weltevrede2010}.

\begin{figure}
\centering
\includegraphics[width=0.49\textwidth]{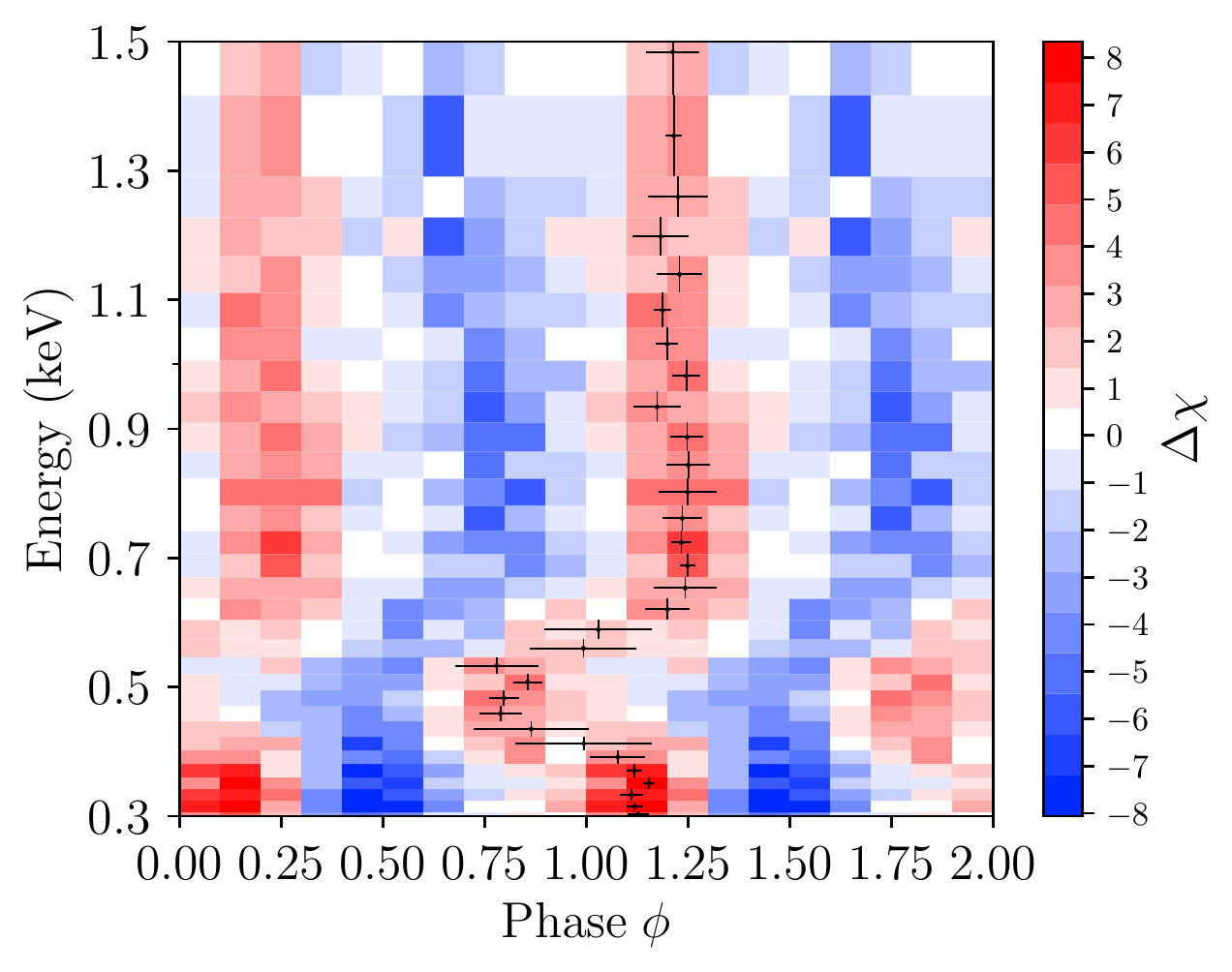}
\caption{\label{fig:eVphaseplot}Binned phase--energy plot for {\sl XMM-Newton} events extracted from the source region. The bin color represents $\Delta\chi_{i,j}$ (red - positive, blue - negative) values defined in equation (\ref{eq:delchi}). The overlaid error bars represent the peak of the pulse profile in each energy bin.}
\end{figure}

The large number of counts allows us to explore a more finely-resolved pulsation behavior through a pulsation significance map (Figure \ref{fig:eVphaseplot}).
This diagram visualizes the significance of pulsations in the two-dimensional phase-energy space.
A conventional phase-energy diagram typically shows the normalized count contribution in the different phase-energy bins, but it does not provide the significance of the variations.
To visualize the significance of the variations, we evaluated the deviations of the counts in the phase bin $j$ and energy bin $i$ from the phase-averaged value:
\begin{equation}\label{eq:delchi}
\Delta\chi_{i, j} = \frac{N_{i, j} -  \bar{N}_{i}}{\sqrt{N_{i, j}}},
\end{equation}
where $N_{i, j}$ is the number of counts in the $j^{\rm th}$ phase bin and $i^{\rm th}$ energy bin, and $\bar{N}_{i}$ is the phase-averaged counts in the $i^{\rm th}$ energy bin.
We binned the events in the phase-energy space, using 20 equal-sized phase bins ($N_\phi$), and choosing variable size energy bins to maintain $\gtrsim 30$ counts per phase-energy bin.
The phase-energy plot is restricted to the $0.3-1.5$ keV energy range, where we have enough source counts.
The pulse peaks between $\phi=0.0-0.2$ for energies $\lesssim 0.4$ keV, and between $\phi\sim0.15-0.35$ for energy bins $\gtrsim 0.6$ keV, as the dominant thermal component changes.
In the intermediate energy range $0.4-0.6$ keV, the peak shifts to $\phi\approx0.8$.
This non-monotonic behavior is inconsistent with a hotter surface becoming more visible.

A simple two blackbody interpretation is inadequate to explain this variation of profile peak with energy.
We expect a smooth shifting of the pulse peak from $\phi\approx0.12$ to $\phi\approx0.22$ as the dominant emission changes from cold BB to hot BB.
On the other hand, an absorption feature around 0.5 keV can explain the profile changes if the absorption is phase-dependent, affecting the cold BB photons in the $\phi\approx0.0-0.3$ phase range.
Hence, the observed energy-dependent pulse profile evolution, showing strong variability in a relatively narrow energy range, further supports the phase-dependent absorption interpretation.

\begin{figure}
\centering
\includegraphics[width=0.49\textwidth]{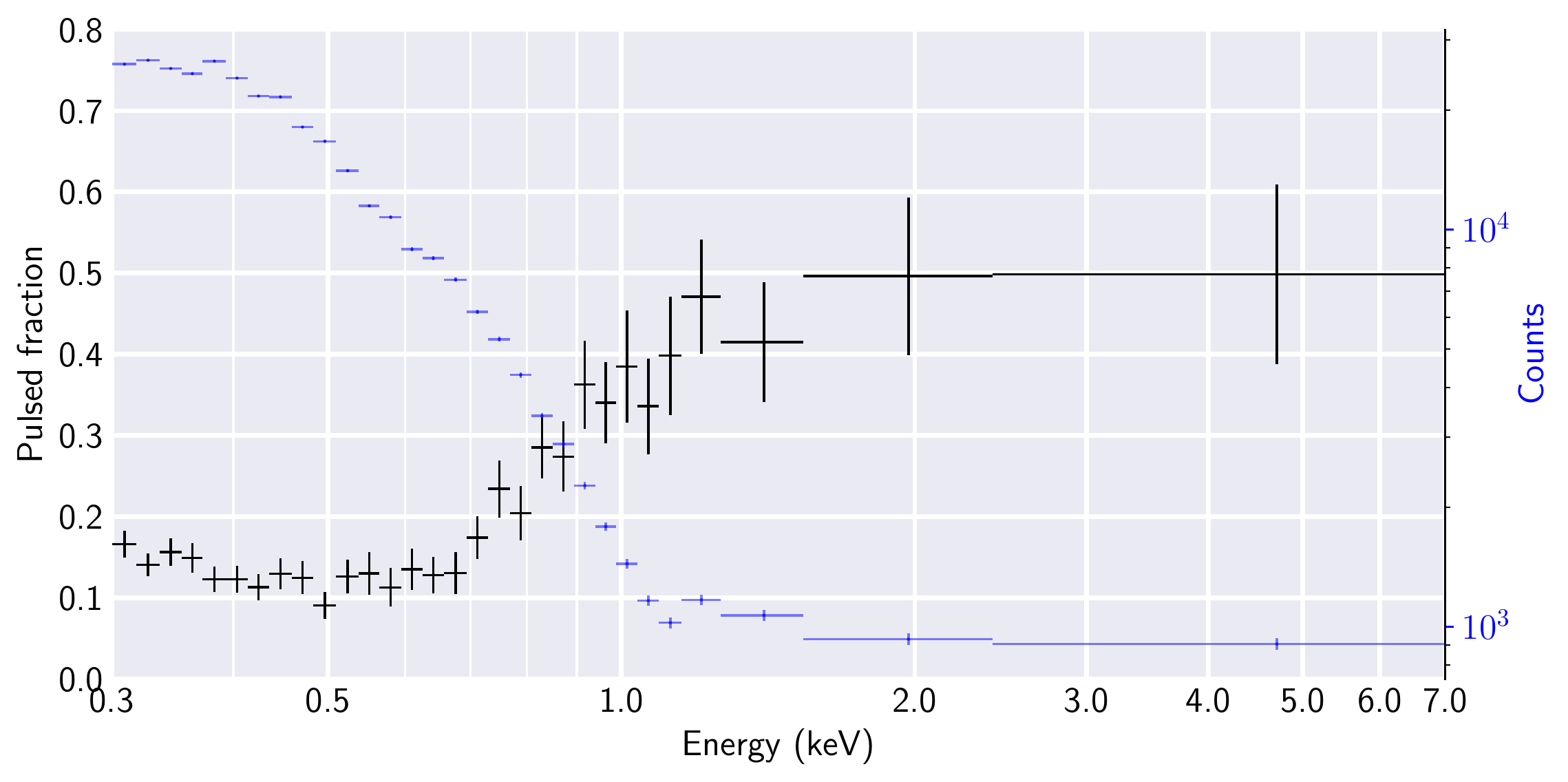}
\caption{\label{fig:Pfplot}Intrinsic pulsed fraction as a function of energy using EPIC-pn data. The number of counts in each bin is also plotted, in blue.}
\end{figure}

In Figure \ref{fig:Pfplot}, we show the variation in intrinsic pulsed fraction (corrected for the background) of the X-ray profile with energy.
The pulsed fraction shows a minimum near 0.5 keV, corresponding to the central energy of the absorption-like feature.
The steep rise in pulsed fraction above 0.7 keV (corresponding to the energies at which the hot BB emission dominates) is due to strong surface temperature inhomogeneity caused by a localized hotspot.
At the energies where the non-thermal emission dominates, the Pf does not show significant variation with energy, but the number of counts is too small there to conclude this with certainty.
The energy dependence of pulsed fraction is broadly consistent with the model describing the phase-integrated spectrum.

\subsection{Timing of the NuSTAR data} 

\begin{figure}
\centering
\includegraphics[width=0.49\textwidth]{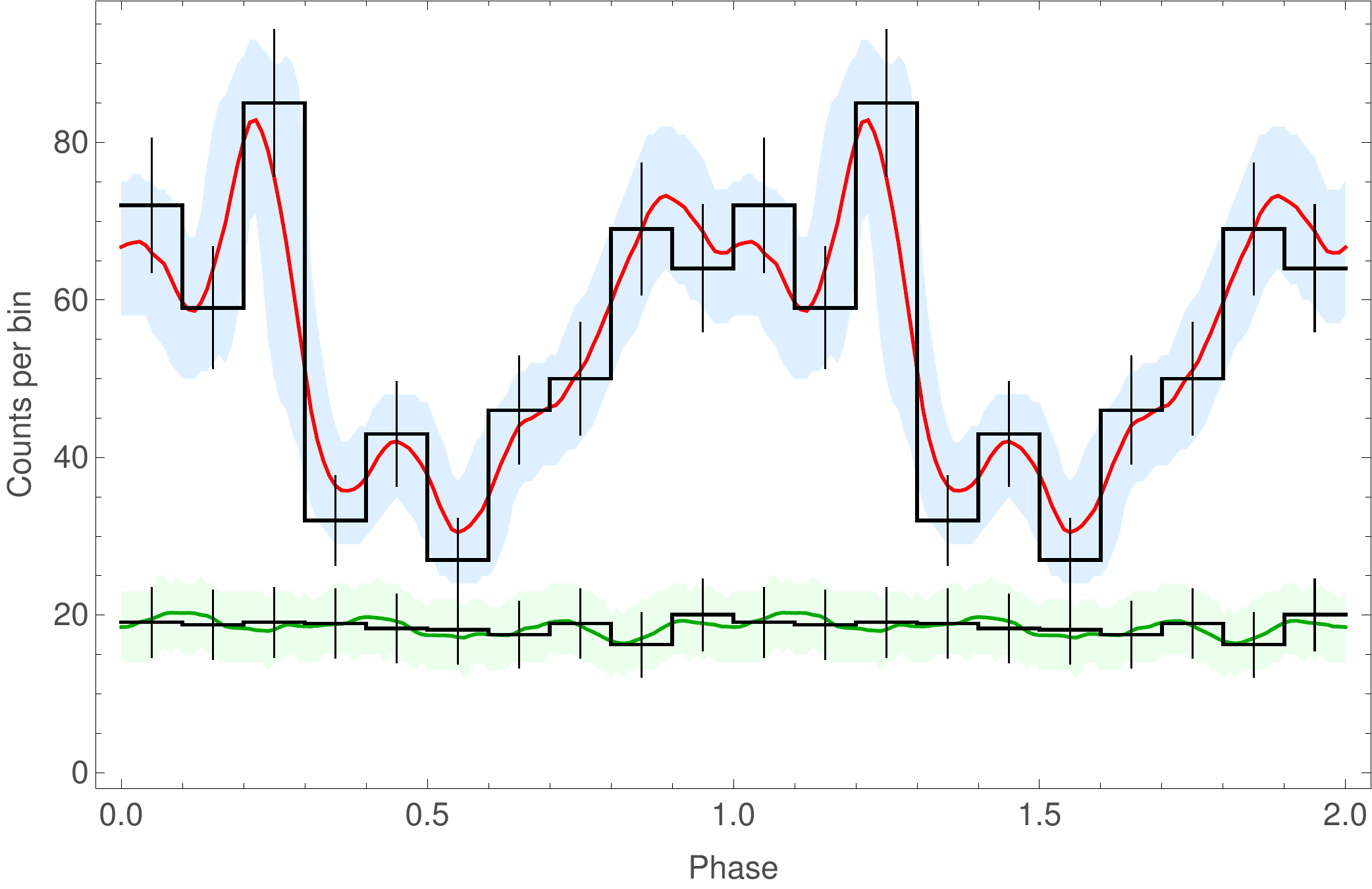}
\caption{\label{figN:fold_lc}  Pulse profile of B0656 as seen by {\sl NuSTAR} in the 3-20 keV energy range. 
Also shown are the smoothed pulse profile (red curve), the background,  and the $1\sigma$ uncertainties of the source and background pulse profiles (shaded blue and green areas, respectively).}
\end{figure}

We have also searched for pulsations in the {\sl NuSTAR} data in the 3-20 keV energy band 
which contains 547 total counts extracted from the source region for the combined FPMA and FPMB detectors. 
The GTI  span for the {\sl NuSTAR} observations is $T_{\rm span}=287,693$ s.
Since we know the pulsation frequency with a small uncertainty  from the simultaneous {\sl XMM-Newton} observation, there is no need to search for period.
To check that the {\sl NuSTAR} and {\sl XMM-Newton} pulsation frequencies are consistent with each other and estimate the {\sl NuSTAR} pulsation significance, we calculated the $Z_1^2$ and $Z_2^2$ statistics within a narrow region around the best-fit {\sl XMM-Newton} frequency, $\nu_{\rm XMM}=2.59789433$ Hz, in steps of $10^{-8}$ Hz.
We found peak values $Z_{\rm 1,max}^2=33.0$ and $Z_{\rm 2,max}^2=36.0$, corresponding to $5.4\sigma$ and $5.1\sigma$ significance, respectively, both occurring at $\nu = 2.5978943(3)$ Hz, in excellent agreement with $\nu_{\rm XMM}$.
The pulse profile, folded with the above-determined frequency and the same definition of reference phase as for the EPIC-pn, is shown in Figure \ref{figN:fold_lc}.
The corresponding pulsed fraction is ${\rm Pf} = 48\%\pm 9\%$ at the 68\% confidence level, which corresponds to the intrinsic pulsed fraction of $71^{+14}_{-13}\%$.

\section{Phase-Resolved Spectral Analysis}\label{phresolved}

We divided the source events into 10 equal-sized phase bins and extracted spectra using filtering and binning criteria identical to those in the phase-integrated analysis.
However, we chose to use the larger extraction radius, $r = 37\farcs5$ instead of $15\arcsec$, because below we will be mostly concerned with the analysis of the thermal part of the spectrum where background contribution is negligible even in the larger aperture.
We require a minimum S/N $\gtrsim 4$ per bin for spectral binning, subtract the background, and fit models using $\chi^2$ minimization in XSPEC.
After obtaining the best-fit model for the spectrum extracted from each phase bin, we use Bayesian analysis to obtain the posterior distribution and credible ranges for the parameters, similar to the phase-integrated analysis.

\begin{figure*}[ht]
\centering
\includegraphics[width=\textwidth]{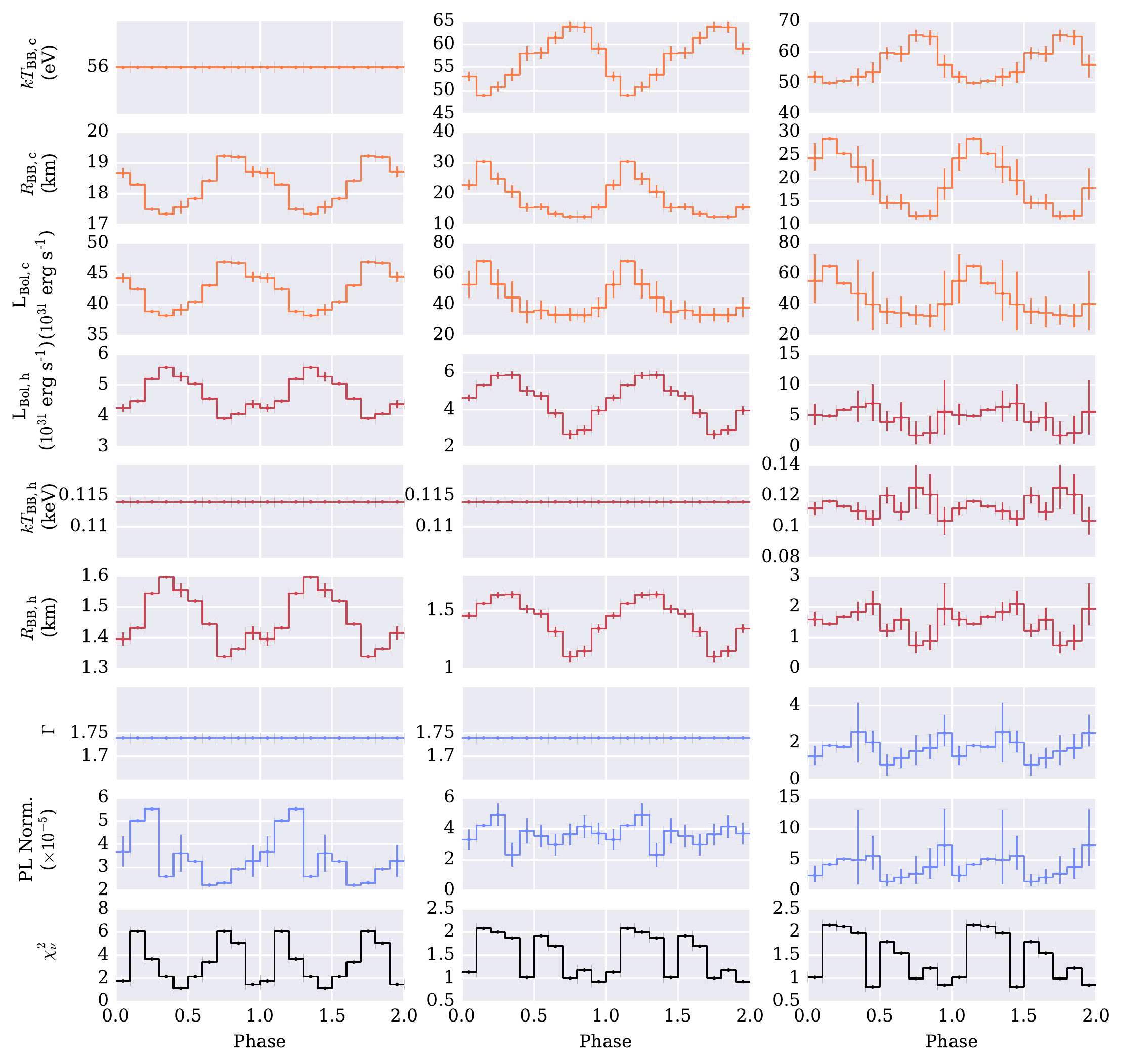}
\caption{\label{fig:Parmods2BBPL}
\replaced{For the 2BBPL we test different phase-resolved fits:
values fixed to those of deLuca (left column), cold BB is allowed to vary (middle column), all parameters (except $N_H$) are allowed to vary (right column).}{Fitting parameters in 10 phase bins for the 2BBPL model. The left column shows parameter variations for fixed $N_H$, $kT$, and $\Gamma$, similar to \cite{DeLuca2005}, the cold BB temperature is allowed to vary in the middle column, and all continuum parameters are allowed to vary in the right column.} 
The parameters are \replaced{cold BB (orange) temperature ($kT_{c}$), effective-radii ($R_{c}$) and bolometric luminosities ($L_{\rm Bol, c}$), hot BB (red) luminosities, temperatures, and radii, and PL (blue) photon index ($\Gamma$), and normalizations, and best-fit reduced chi-square values ($\chi_\nu^2$)}{cold BB (orange) and hot BB (red) temperatures, radii and luminosities,  PL (blue) photon index and normalization, and $\chi_\nu^2$ values.}
Parameter uncertainties ($90\%$ confidence) are calculated only for fits with $\chi_\nu^2 < 2$.
Frozen parameters are shown in the plots with horizontal straight lines.}
\end{figure*}

We start our analysis from fitting 2BBPL model to the phase-resolved data. Firstly, following the \cite{DeLuca2005} analysis, we froze the X-ray absorption ($N_H=3.03\times10^{20}$ cm$^{-2}$), the temperatures ($kT_{\rm BB,c} = 56$ eV and $kT_{\rm BB,h} = 114$ eV), and the photon index ($\Gamma = 1.74$) to their phase-integrated median values from our fits (see Table \ref{tbl-2_intspectra}) and fitted 3 remaining parameters (BB and PL normalizations) in each of the phase bins. The result  is shown in the left column of Figure \ref{fig:Parmods2BBPL}. We see that the fit is rather poor ($\chi_\nu^2 \gtrsim 2$), with large residuals in all phase bins.
Relaxing the constraints by freeing up $T_{\rm BB,c}$ improves the fits considerably (see $\chi_\nu^2$ values in Figure \ref{fig:Parmods2BBPL}).
However, the fit quality still remains poor for the phase bins between $\phi=0.1$ and $\phi=0.7$.
We then perform fits where all the parameters, with the exception of $N_H = 3.03\times10^{20}$ cm$^{-2}$, are allowed to vary.
This results in only marginal improvements to the fit in some of the phase bins ($\phi = 0.0-0.1$ and $0.4-1.0$).
The phase dependences of the cold and hot BB temperatures, radii, bolometric luminosities, power-law photon indexes, normalizations, and $\chi_\nu^2$ values of the best-fit models, are plotted in the right column of Figure \ref{fig:Parmods2BBPL}.
The exercise above shows that the 2BBPL model does not describe the spectrum well for all phases.

Although 2BBPL fit is formally acceptable, in terms of $\chi_\nu^2$, it results in systematic residuals throughout the $0.3-7$ keV range  (see Figure \ref{J0659phspectra3}, left column) and is unable to fit the phase-resolved spectra at all phase ranges.
Unlike \cite{DeLuca2005}, we cannot satisfactorily fit the phase-resolved spectra with 2BBPL by fixing the temperatures and $\Gamma$ across all phases at the values obtained from the phase-integrated fits.
Allowing the BB temperatures and $\Gamma$ to vary with phase, in addition to the normalizations, improves the fits only in some phase bins.
The fits still exhibit large systematic residuals around 0.5 keV in the $0.0-0.6$ phase range (Figures \ref{fig:Parmods2BBPL} and \ref{J0659phspectra3}).
Only the spectra in the range $\phi=0.6-1.0$ can be satisfactorily described by the 2BBPL continuum model.

The poor 2BBPL fits in the $\phi = 0.1-0.7$ phase range are due to unmodeled residuals at energies $\lesssim 1$ keV, where the thermal emission dominates.
To simplify spectral fitting for the purposes of exploring absorption-like feature in the $0.4-0.6$ keV range, we truncate the spectra above 1.3 keV.
This effectively removes the spectral bins where the PL component makes any noticeable contribution.
The PL component contributes $\sim 5\%$ to the flux below 1.3 keV in the phase-integrated spectrum.
The removal of the bins above 1.3 keV and the PL component does not affect the strength of the absorption feature substantially.
On truncation of the PL part of the spectra, the cold BB normalizations change by $10\%-40\%$ and the hot BB normalizations change by $5\%-30\%$, which are within $1.5\sigma$ and $1\sigma$, respectively, of their statistical uncertainties.
Therefore, below we will fit only double-blackbody models with (G2BB) and without (2BB) the absorption-line component.
In Figure \ref{J0659phspectra3} we show the result of these fits side-by-side for each of the 10 phase ranges.
The G2BB plots (right panels) show residuals of the best-fit model as well as residuals obtained by removing the \texttt{Gabs} absorption component from the same model {\sl without re-fitting}, to highlight the shape and magnitude of the modeled absorption feature.
In the $\phi = 0.6-1.0$ phase range, the \texttt{Gabs} component, although fit, is not required, and the 2BB model is sufficient, while outside this range the \texttt{Gabs} component is required.
We use Bayesian fitting to obtain the posterior distributions of the model parameters.
The box and whiskers plot in Figure \ref{fig:Parmods} shows the phase variation of the distribution of fitting parameters: the absorption line energy and width of the Gaussian profile, cold and hot BB temperatures and effective radii/areas, and derived quantities such as line equivalent width and bolometric luminosities for the BB components.

The cold BB parameters exhibit anti-correlated temperature and radius changes which cannot be reduced to variability of a single parameter by fixing the other.
The derived bolometric luminosity follows the approximate trend observed in the $0.3-0.7$ keV pulse profile.
Anti-correlation is also observed for the hot BB parameters but the trend in phase variability of the bolometric luminosity is not recovered.

The central energy of the absorption feature shifts from $\sim 0.5$ keV to $\sim 0.6$ keV as the phase increases from 0.0 to 0.6, in agreement with what is seen in Figure \ref{fig:eVphaseplot}.
The optical depth and equivalent width reach a maximum in the $\phi=0.2-0.3$ bin, where the feature is the strongest.
Therefore, we use the spectrum extracted in the $0.2-0.3$ phase range ($N_{\rm src}/N_{\rm bkg}$ = 37491/182) to compare the 2BB and G2BB models using Bayesian evidence and odds ratio.
Again, assuming equal prior odds for the two models, $P(M_2)/P(M_1)$ = 1, where $M_2$ represents G2BB and $M_1$ represents 2BB, the posterior odds-ratio for model G2BB over 2BB is $O_{21} = 11856$.

We use the same $0.2-0.3$ phase range spectrum to also test the suitability of models other than 2BB and G2BB.
The single and double continuum models with BB, NSA, and NSMAXG produce fits with significant residuals below $\sim 0.7$ keV.
Among these models, the best fit is given by the 2BB model with $\chi_\nu^2 = 2.5$ for $\nu = 23$, and parameters $kT_{\rm BB} = 51$ eV and 0.12 keV with effective radii $R_{\rm BB} = 25$ km and 1.5 m, respectively.
We attempted to add one or two BB components to the 2BB model but failed to obtain a statistically better fit

A single BB with two \texttt{Gabs} components (2GBB) provides a marginally acceptable fit ($\chi_\nu^2 = 1.3$ for $\nu = 20$) with residuals comparable to those of the G2BB model.
In this fit, one of the \texttt{Gabs} components, with $E_c=0.80$ keV and $\sigma=0.30$ keV, behaves like a very broad absorption feature in the $kT=95$ eV BB spectrum emitted from an $R_{\rm BB}=7.3$ km region.
The parameters of the second \texttt{Gabs} component, $E_c=0.53$ keV and $\sigma=0.10$ keV, are virtually identical to those in the best-fit G2BB model.

\begin{figure*}[ht]
\captionsetup[subfigure]{labelformat=empty}
\subfloat[]{\label{fig:J0659p11}\includegraphics[width = 0.45\textwidth]{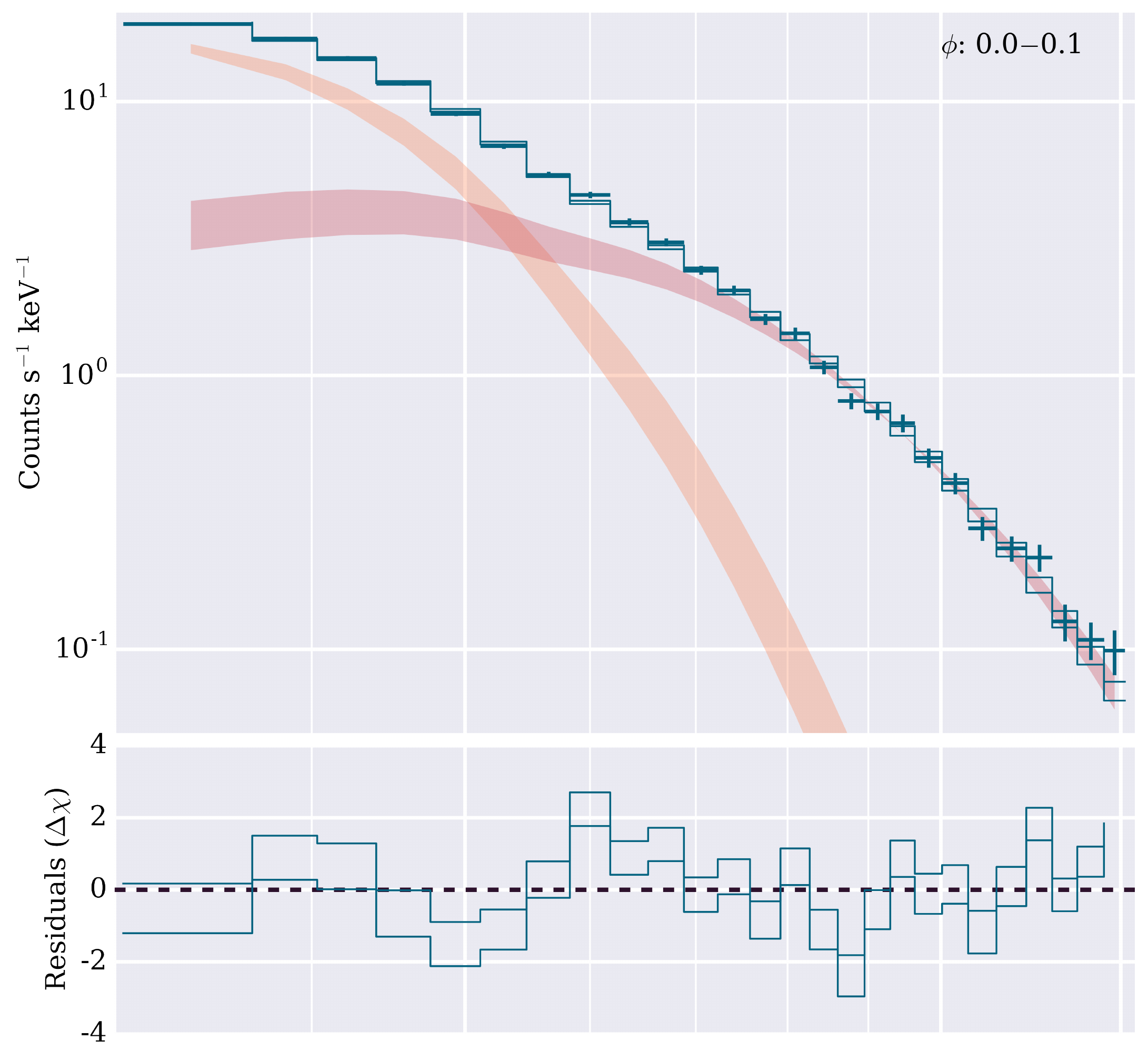}}
\subfloat[]{\label{fig:J0659p12}\includegraphics[width = 0.45\textwidth]{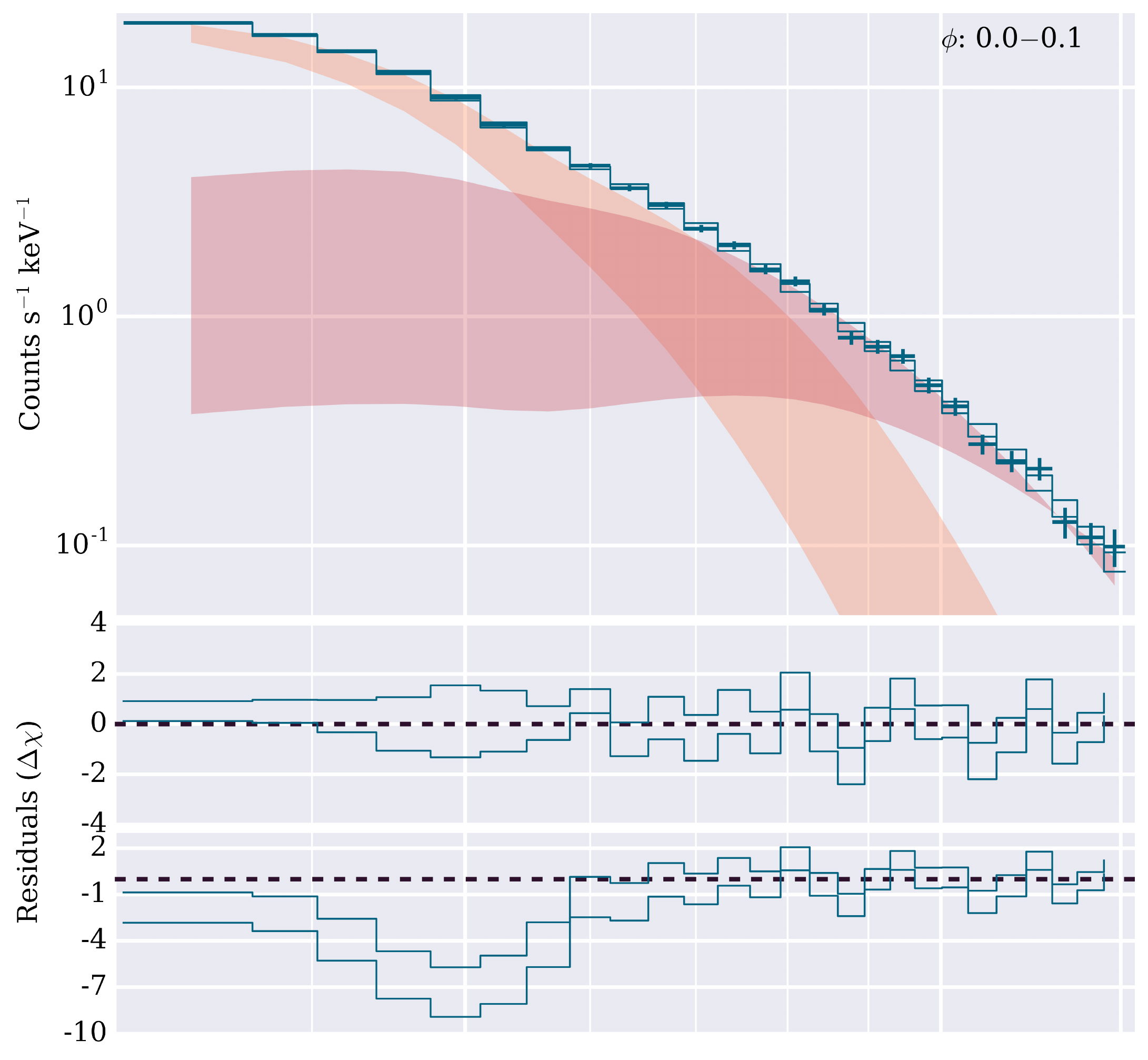}} 
\vspace{-1.75\baselineskip}\\
\subfloat[]{\label{fig:J0659p21}\includegraphics[width = 0.45\textwidth]{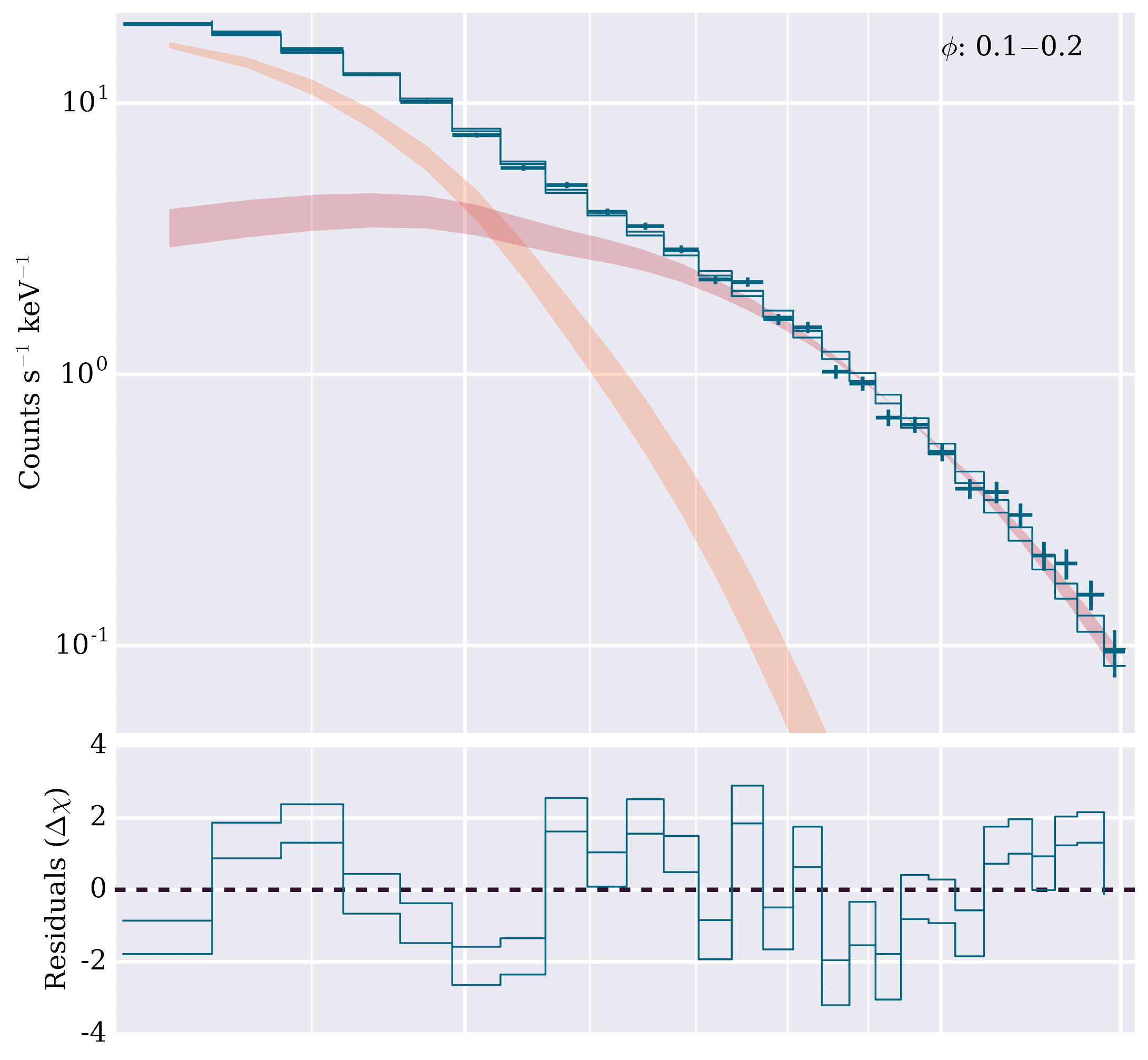}}
\subfloat[]{\label{fig:J0659p22}\includegraphics[width = 0.45\textwidth]{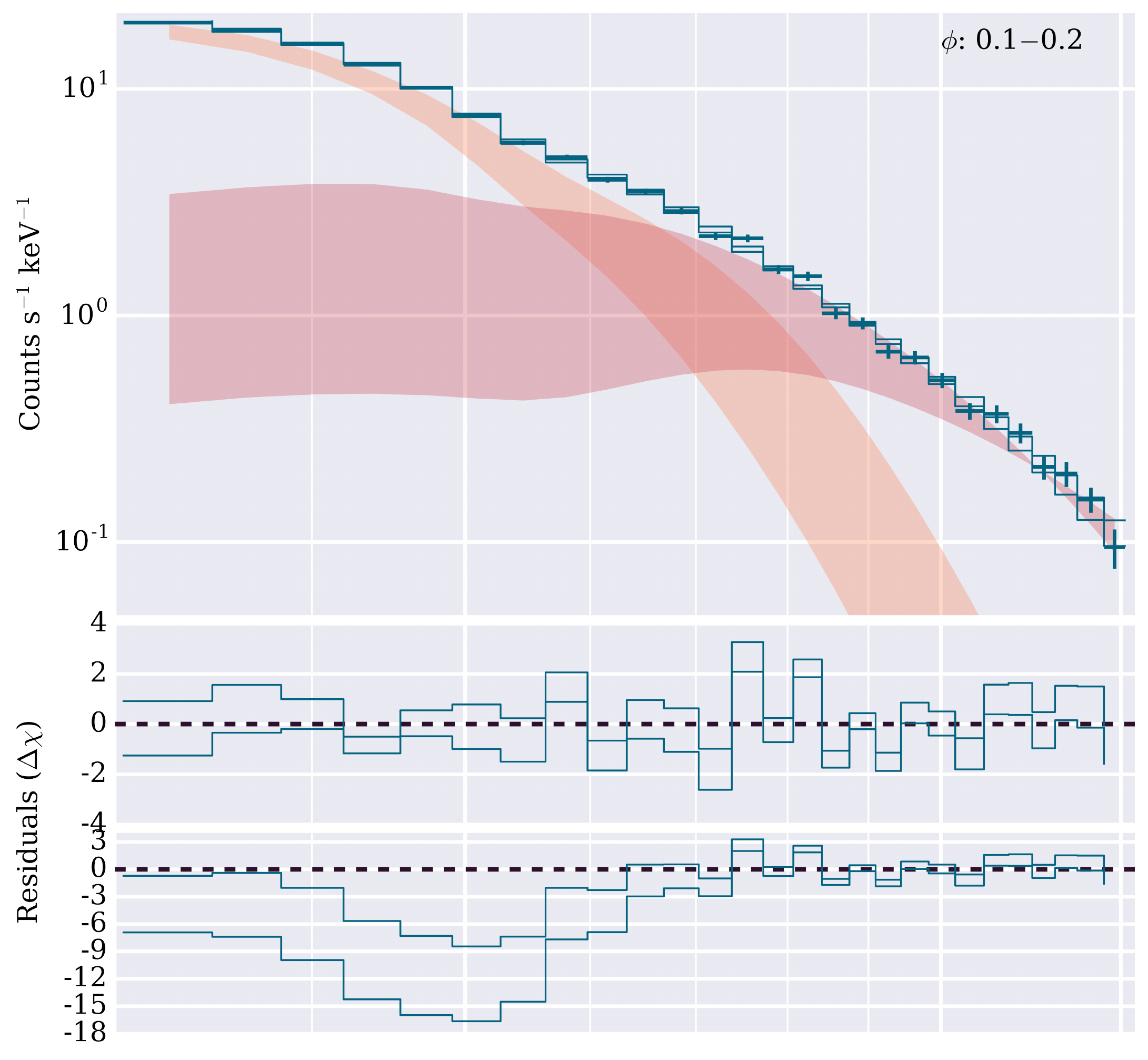}} 
\vspace{-1.75\baselineskip}\\
\subfloat[]{\label{fig:J0659p31}\includegraphics[width = 0.45\textwidth]{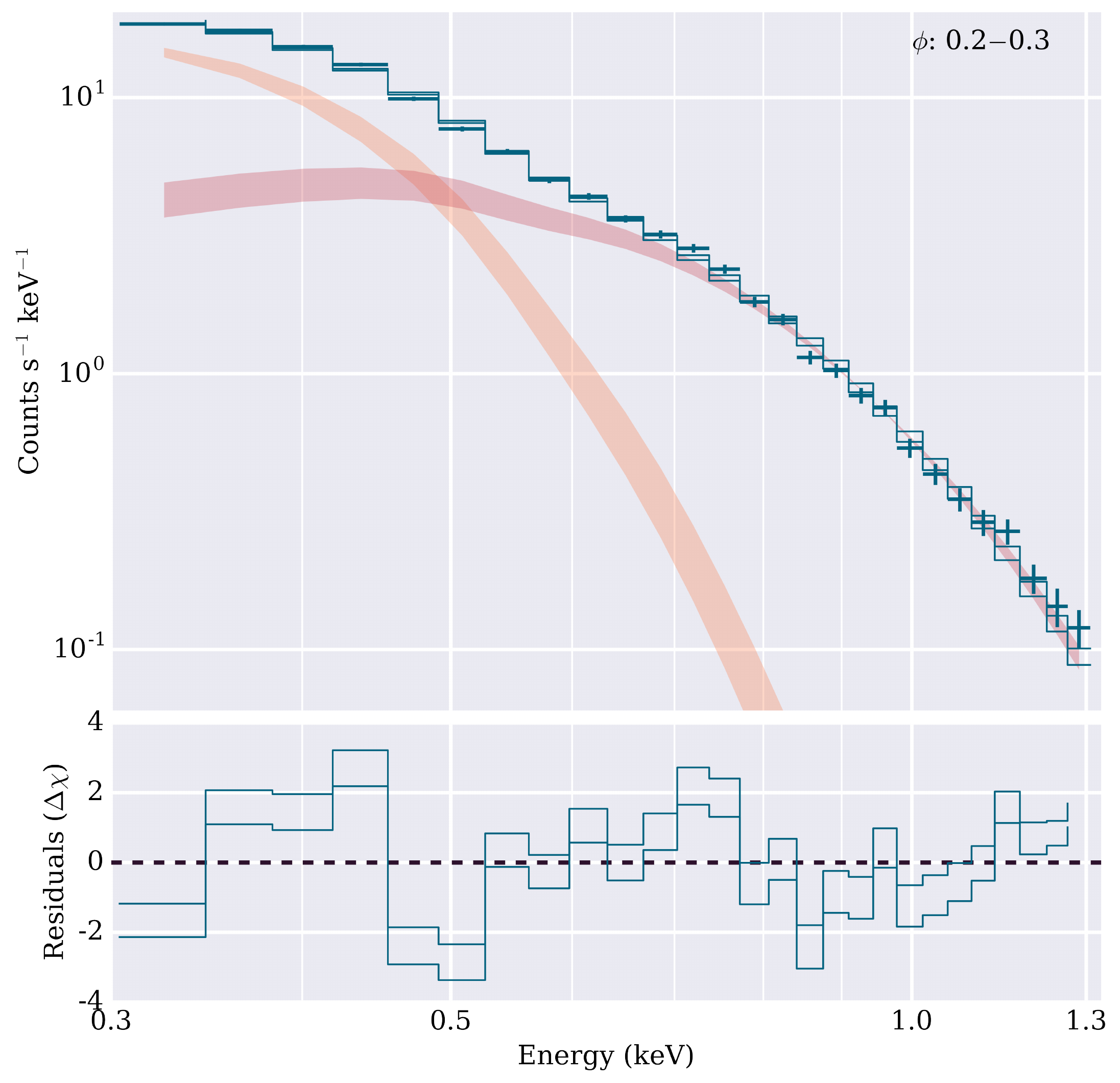}}
\subfloat[]{\label{fig:J0659p32}\includegraphics[width = 0.45\textwidth]{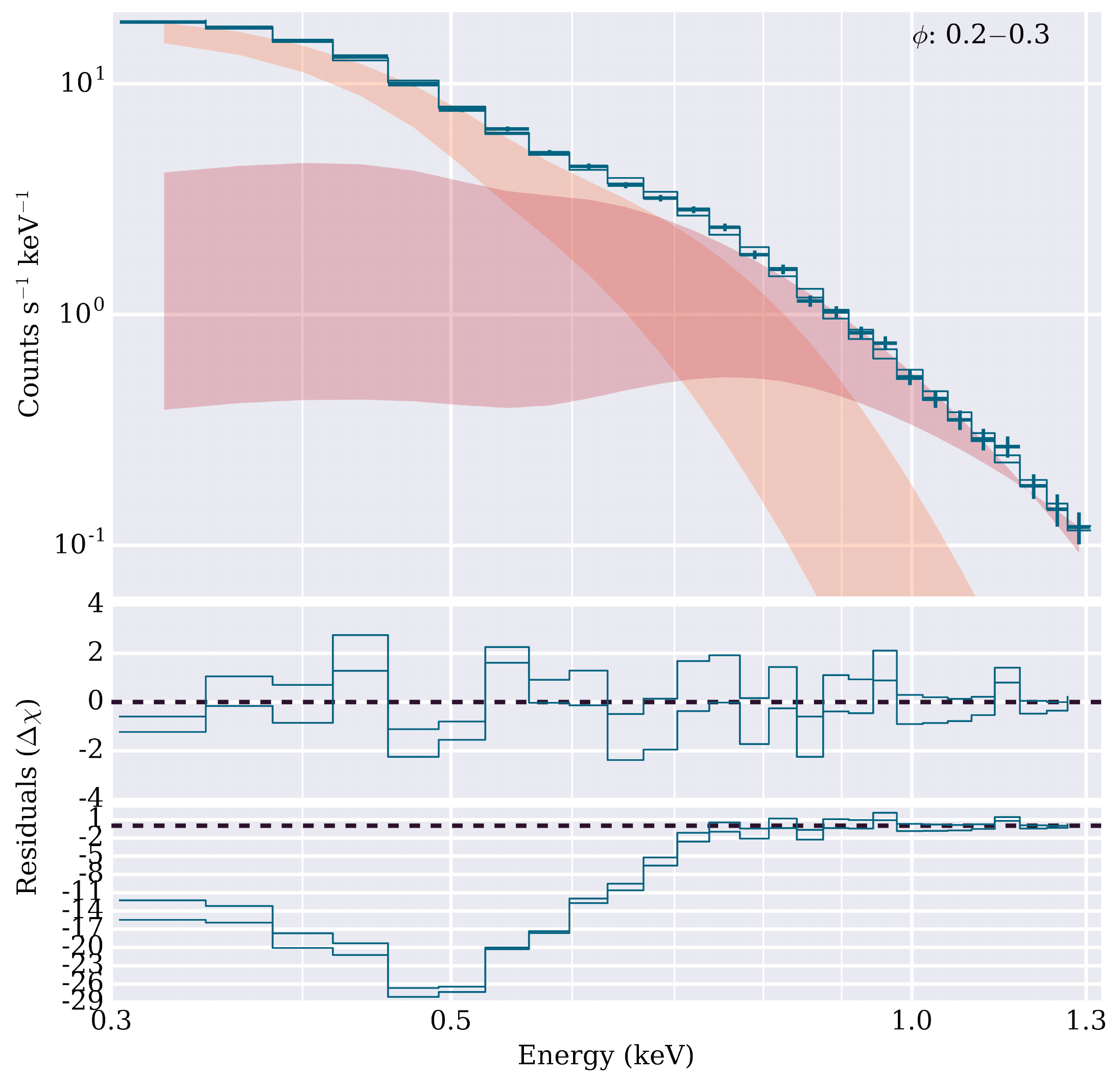}}
\caption{}
\label{J0659phspectra0}
\end{figure*}

\begin{figure*}\ContinuedFloat
\captionsetup[subfigure]{labelformat=empty}
\subfloat[]{\label{fig:J0659p41}\includegraphics[width = 0.45\textwidth]{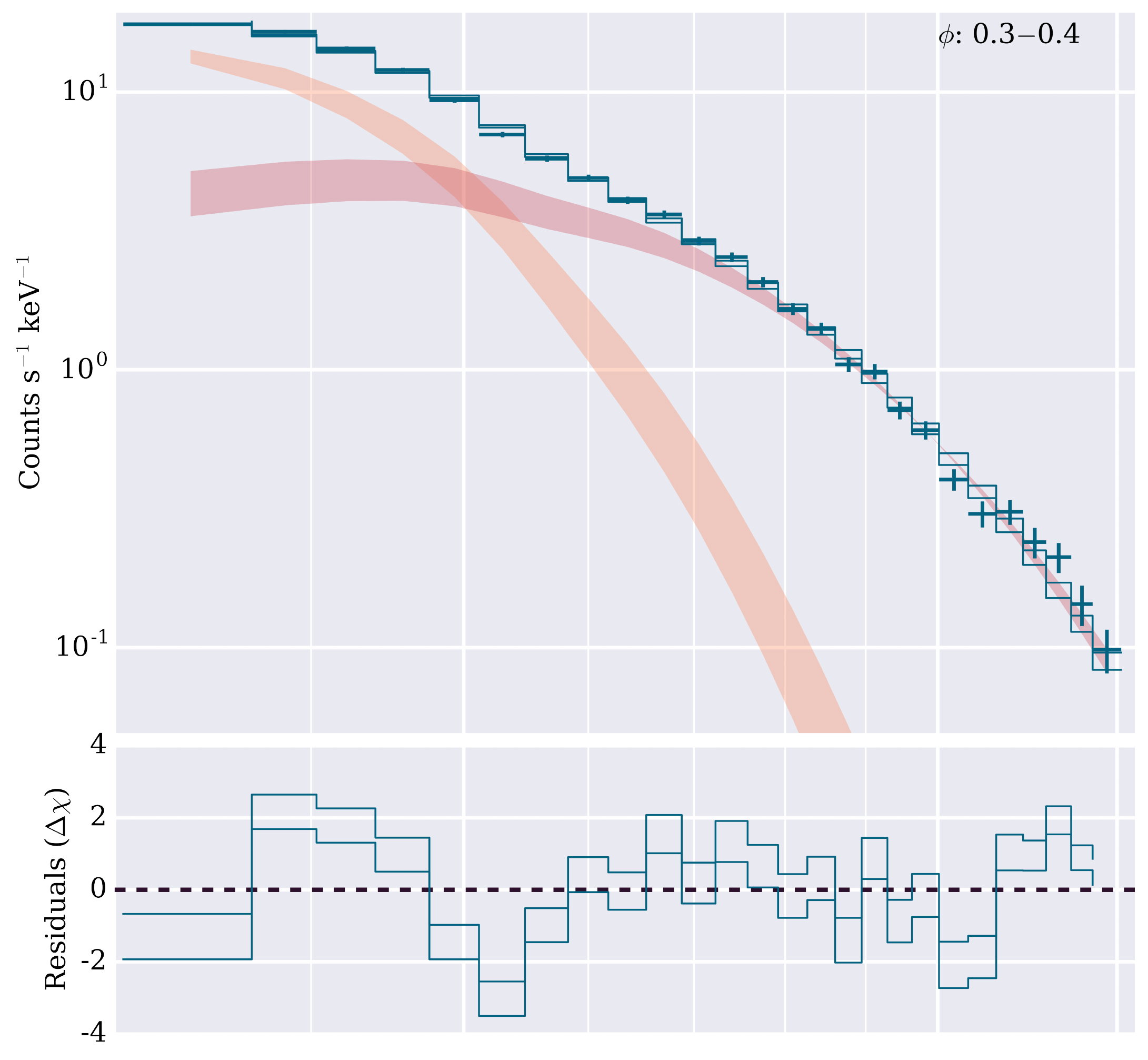}}
\subfloat[]{\label{fig:J0659p42}\includegraphics[width = 0.45\textwidth]{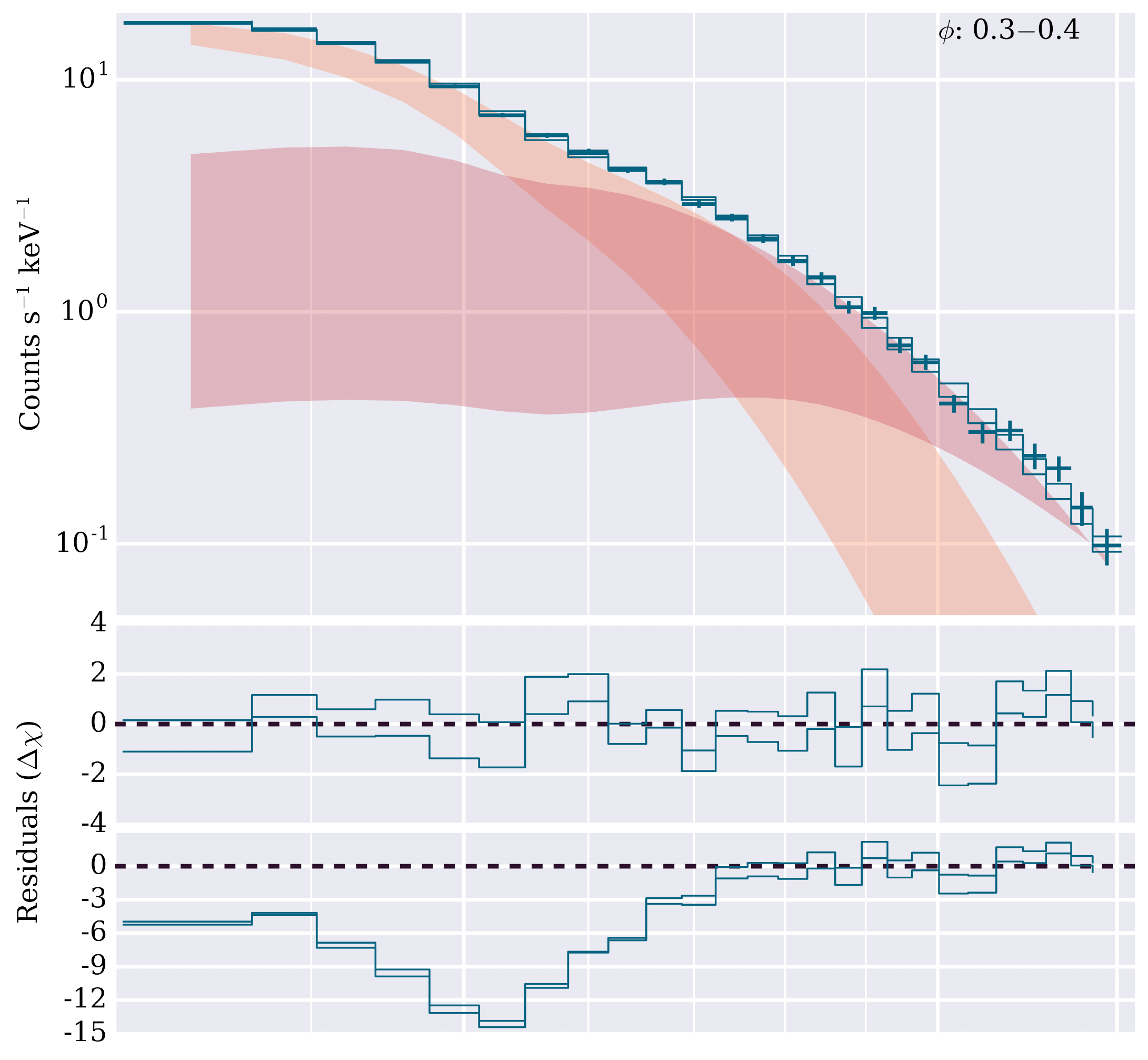}}
\vspace{-1.75\baselineskip}\\
\subfloat[]{\label{fig:J0659p51}\includegraphics[width = 0.45\textwidth]{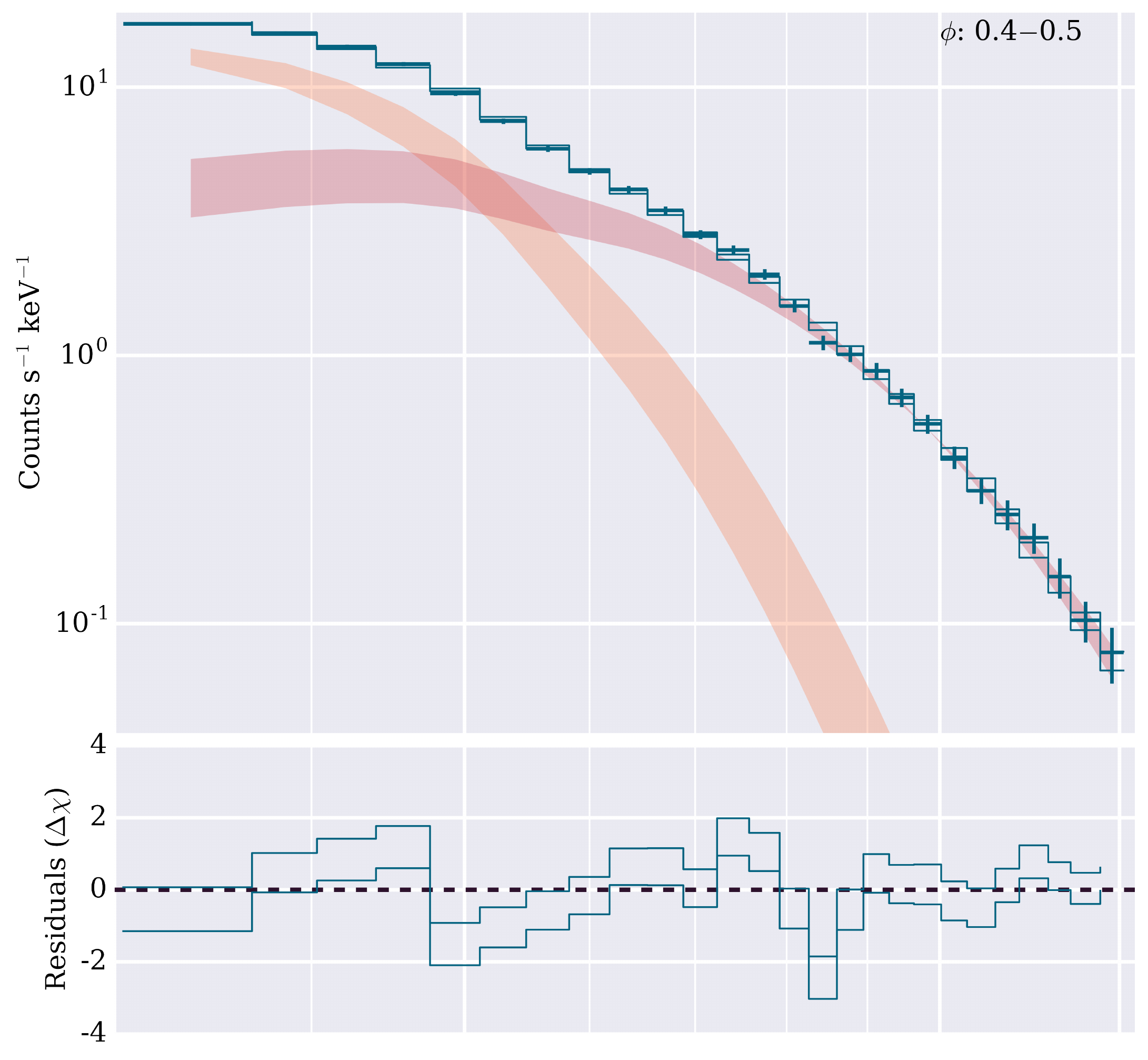}}
\subfloat[]{\label{fig:J0659p52}\includegraphics[width = 0.45\textwidth]{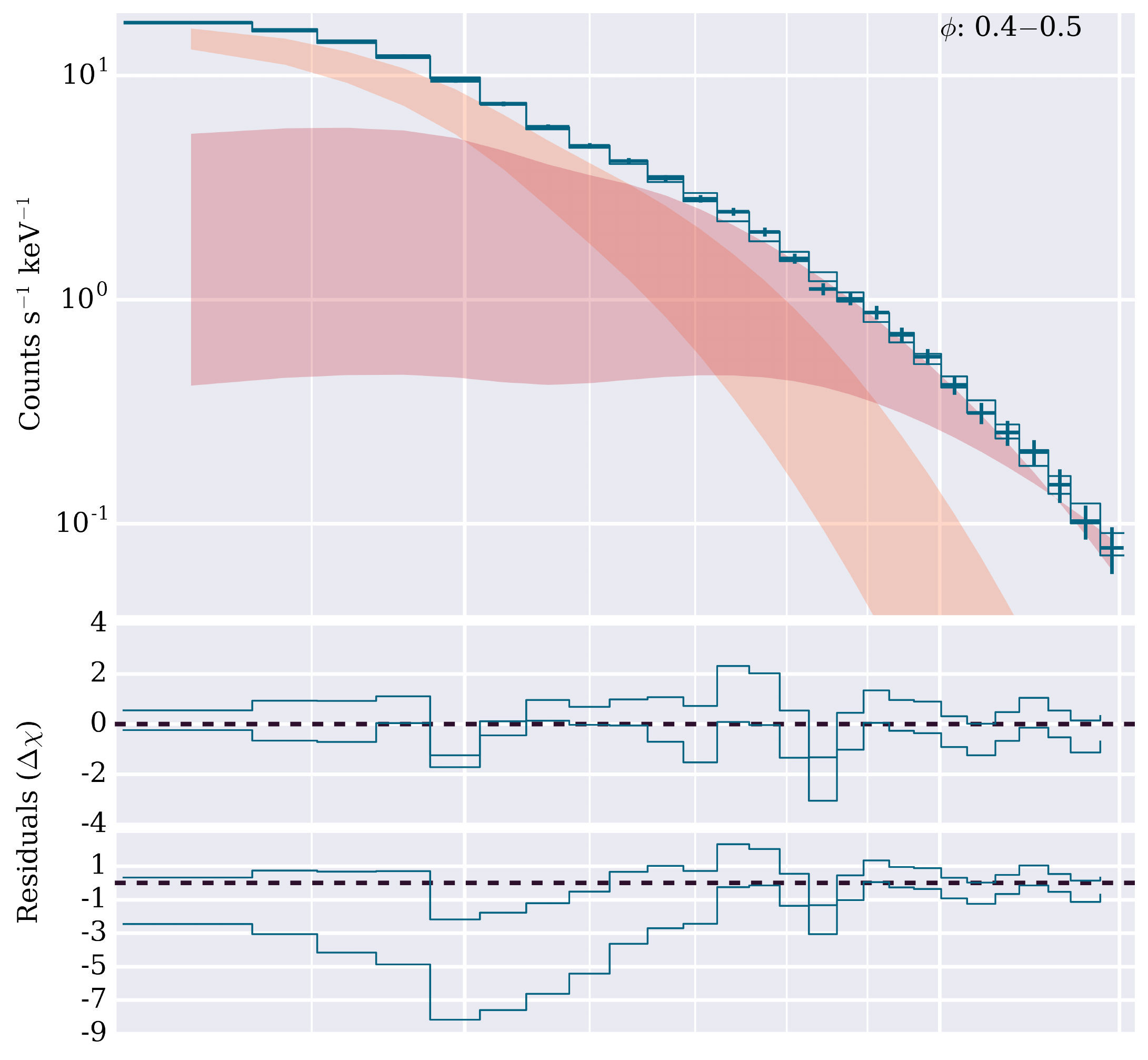}}
\vspace{-1.75\baselineskip}\\
\subfloat[]{\label{fig:J0659p61}\includegraphics[width = 0.45\textwidth]{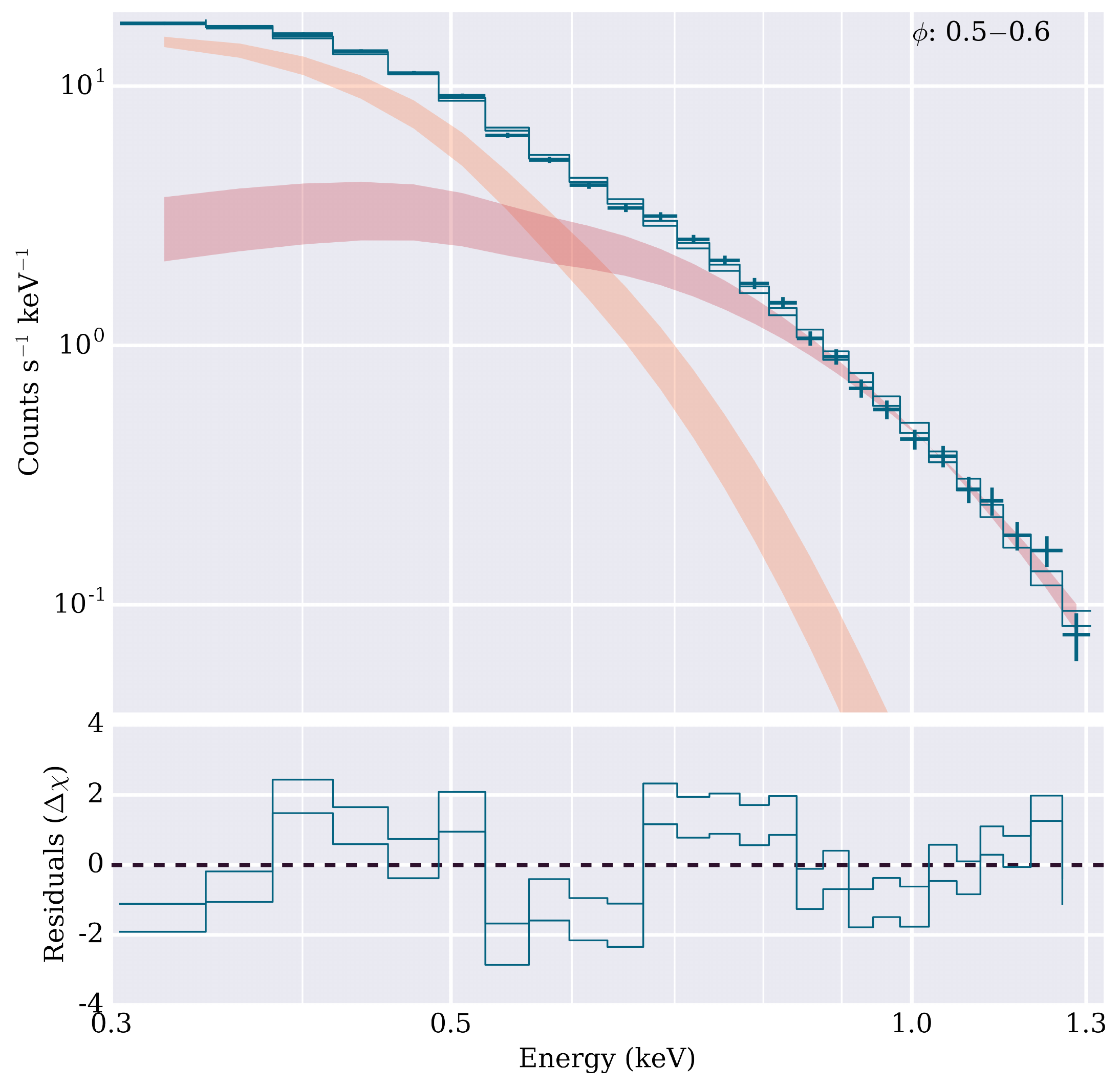}}
\subfloat[]{\label{fig:J0659p62}\includegraphics[width = 0.45\textwidth]{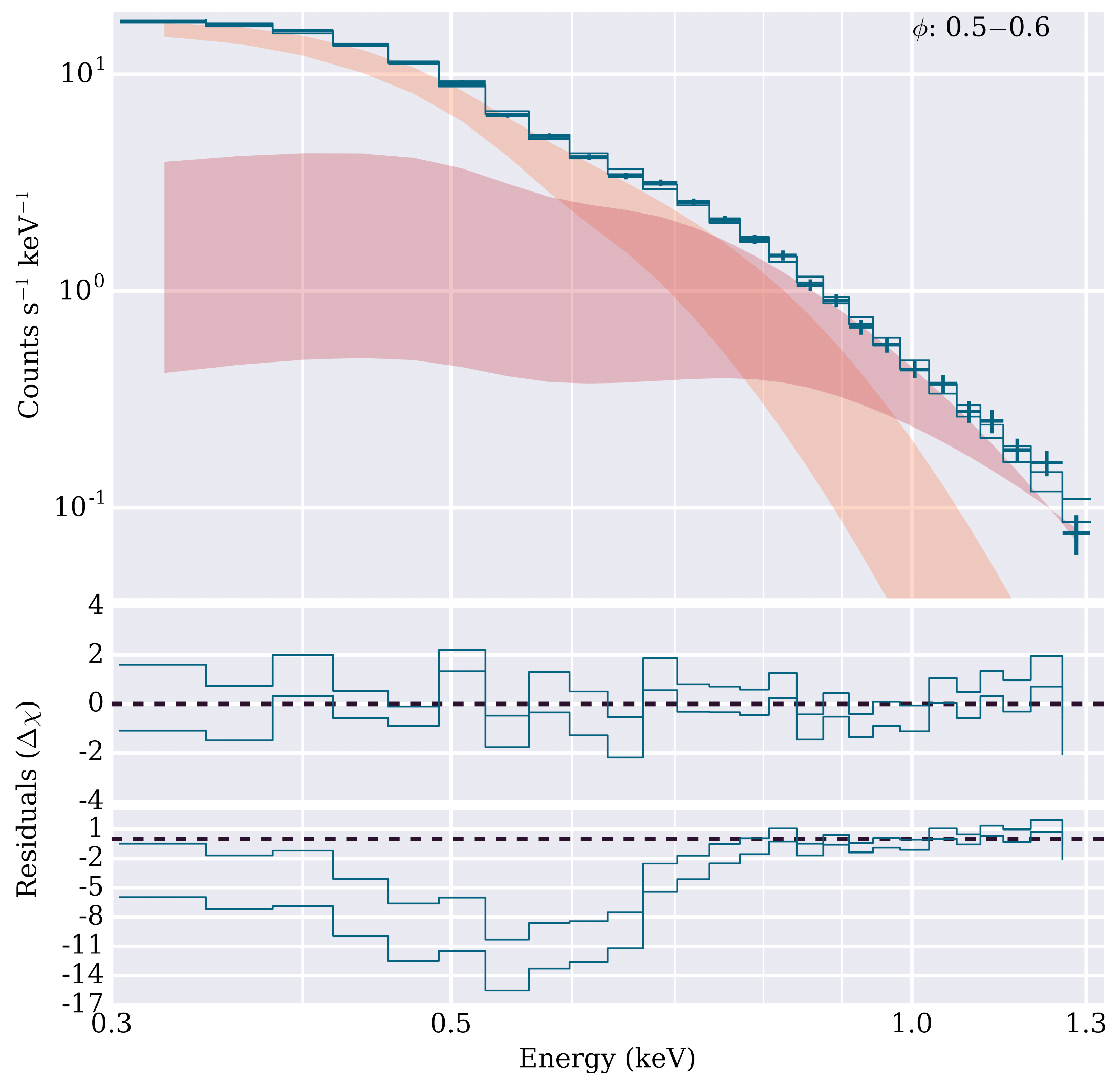}}
\caption{}
\label{J0659phspectra1}
\end{figure*}

\begin{figure*}\ContinuedFloat
\captionsetup[subfigure]{labelformat=empty}
\subfloat[]{\label{fig:J0659p71}\includegraphics[width = 0.45\textwidth]{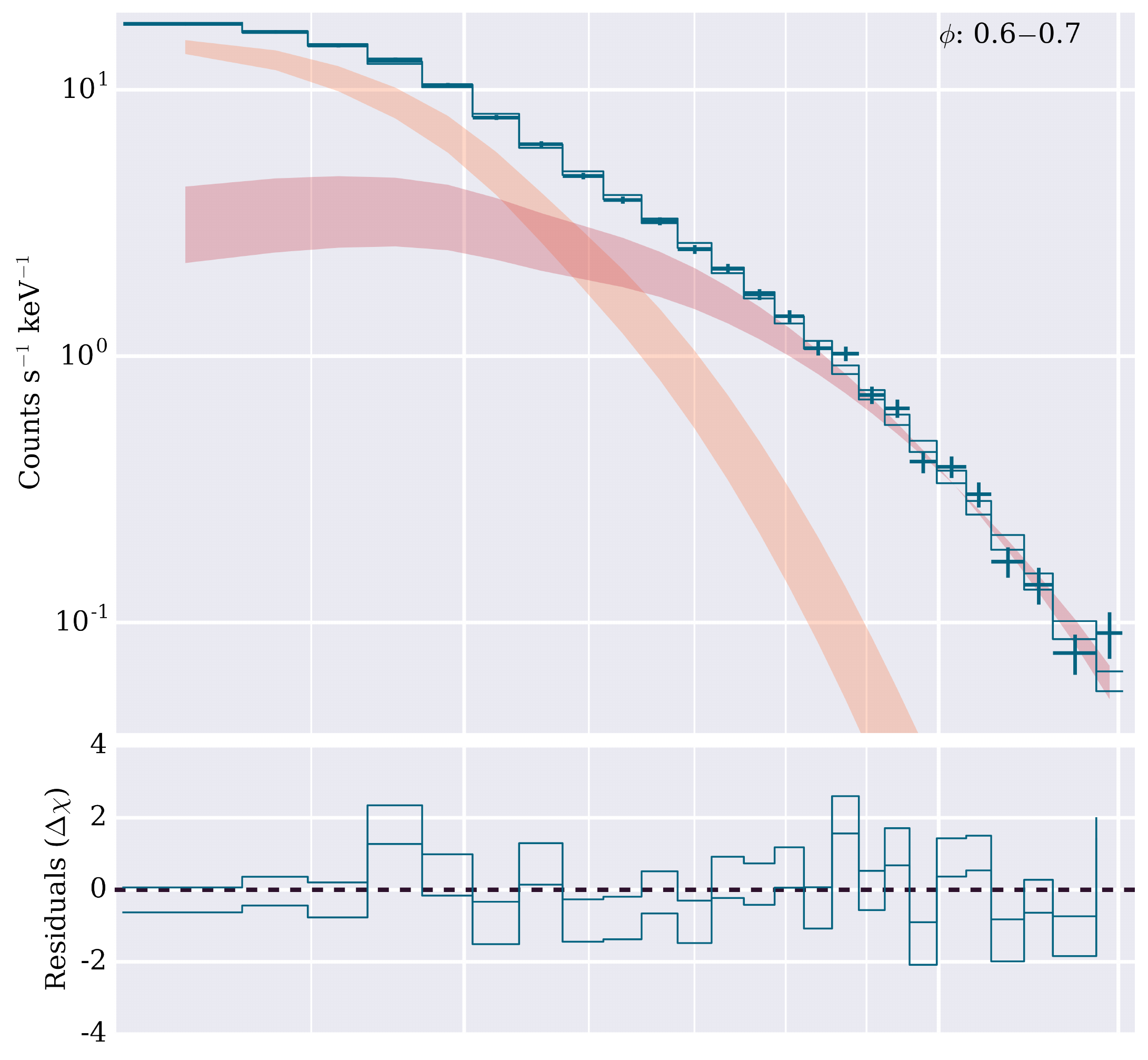}}
\subfloat[]{\label{fig:J0659p72}\includegraphics[width = 0.45\textwidth]{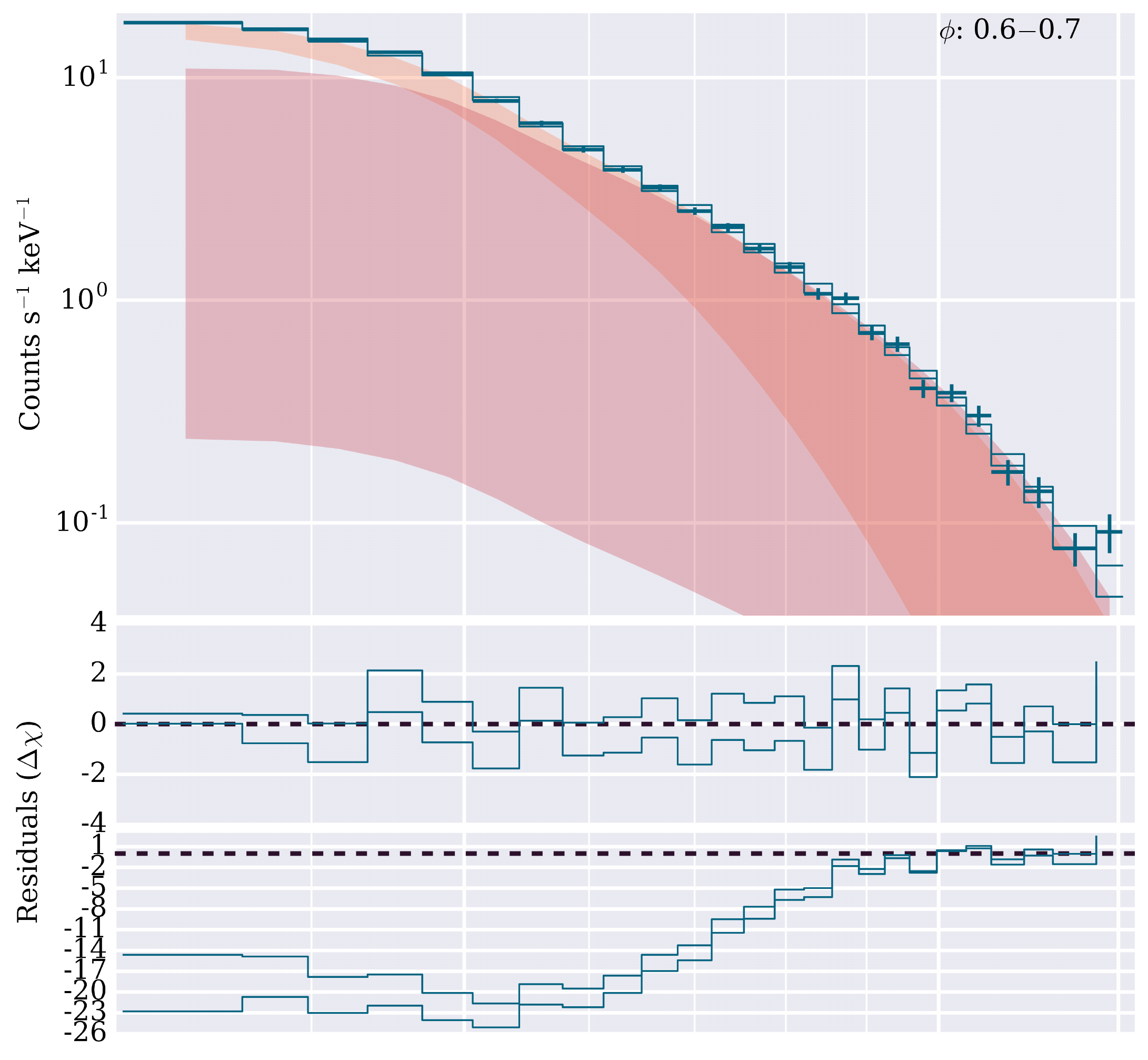}}
\vspace{-1.7\baselineskip}\\
\subfloat[]{\label{fig:J0659p81}\includegraphics[width = 0.45\textwidth]{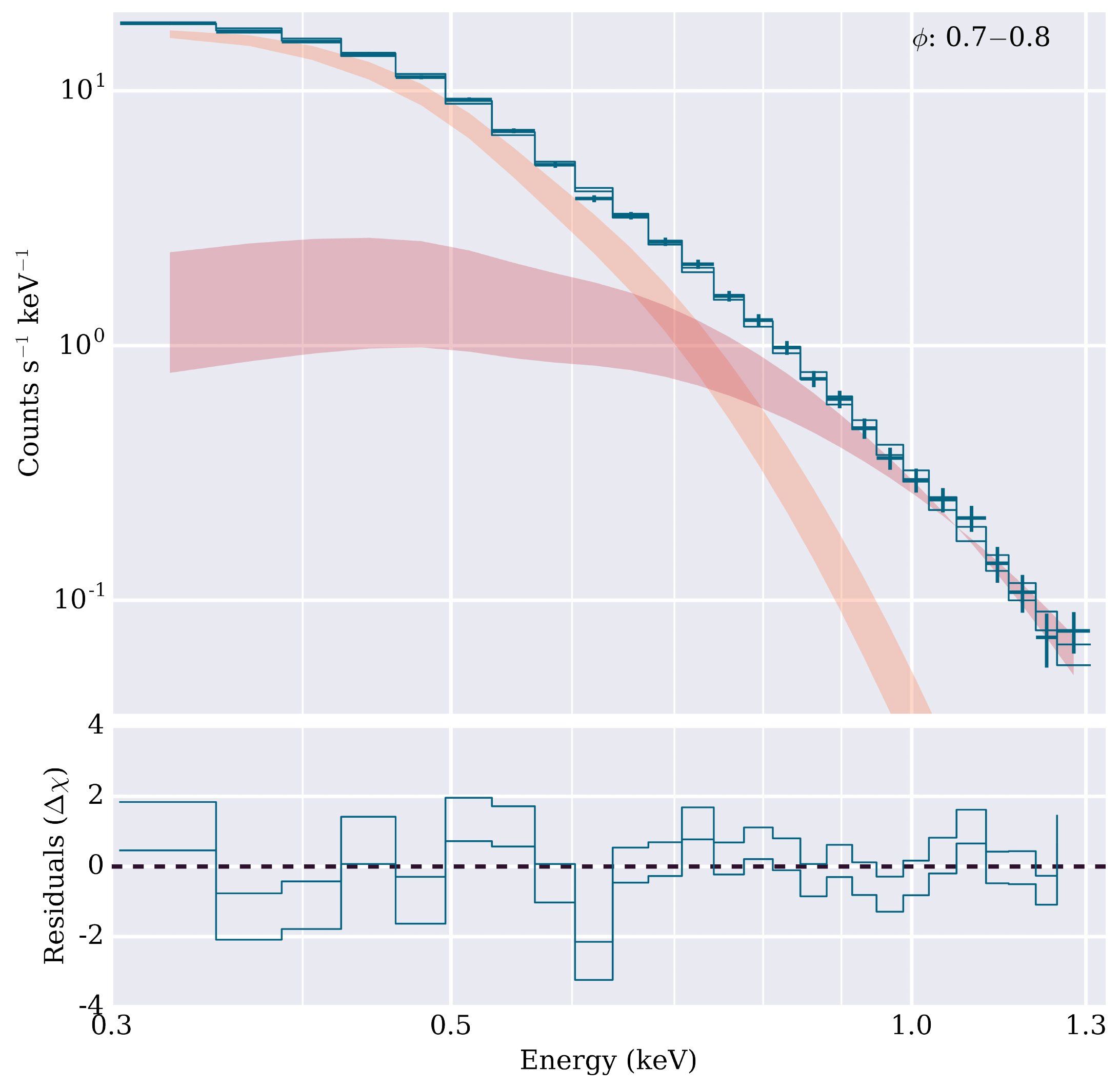}}
\subfloat[]{\label{fig:J0659p82}\includegraphics[width = 0.45\textwidth]{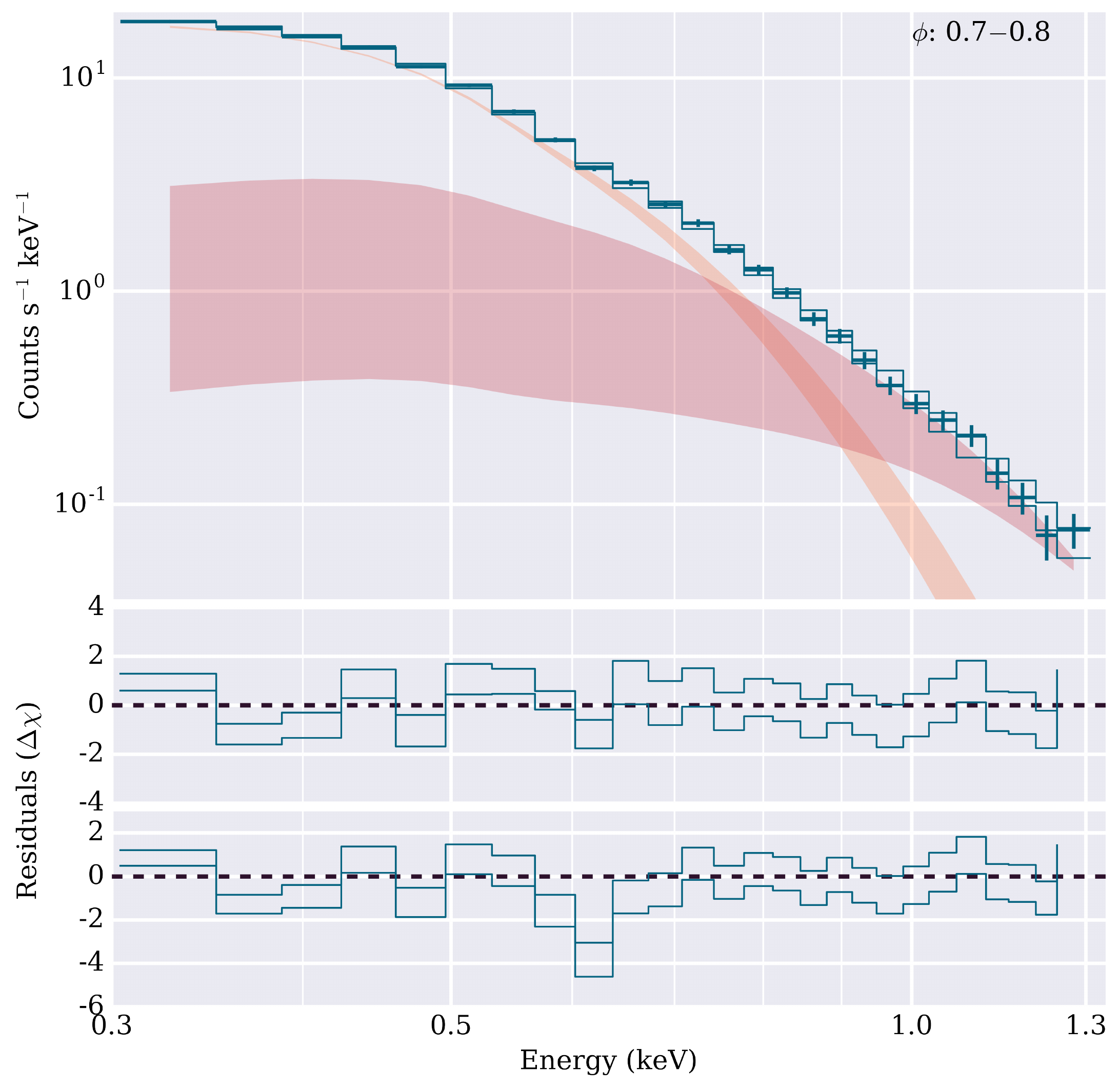}}
\vspace{-1.7\baselineskip}\\
\subfloat[]{\label{fig:J0659p91}\includegraphics[width = 0.45\textwidth]{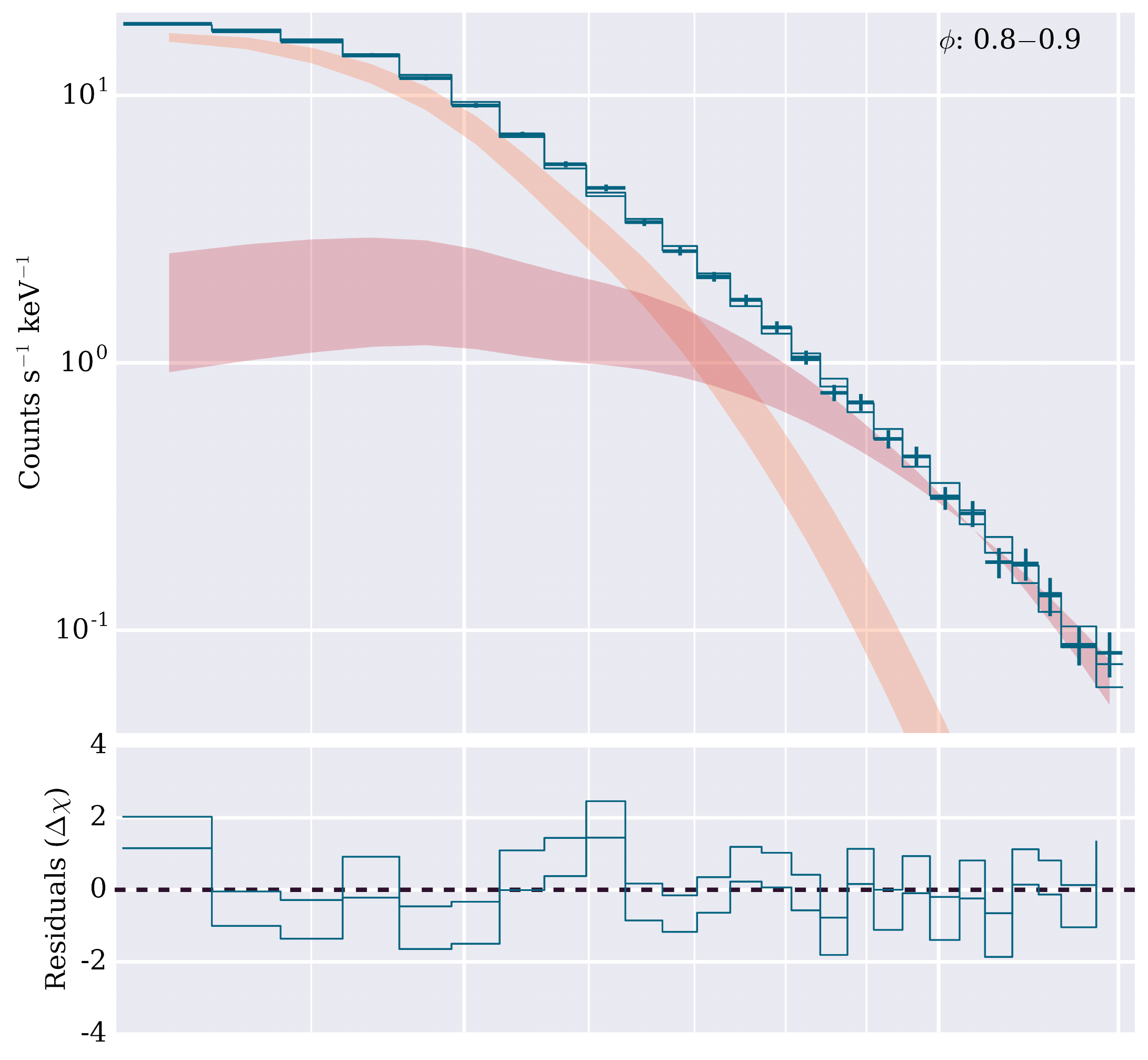}}
\subfloat[]{\label{fig:J0659p92}\includegraphics[width = 0.45\textwidth]{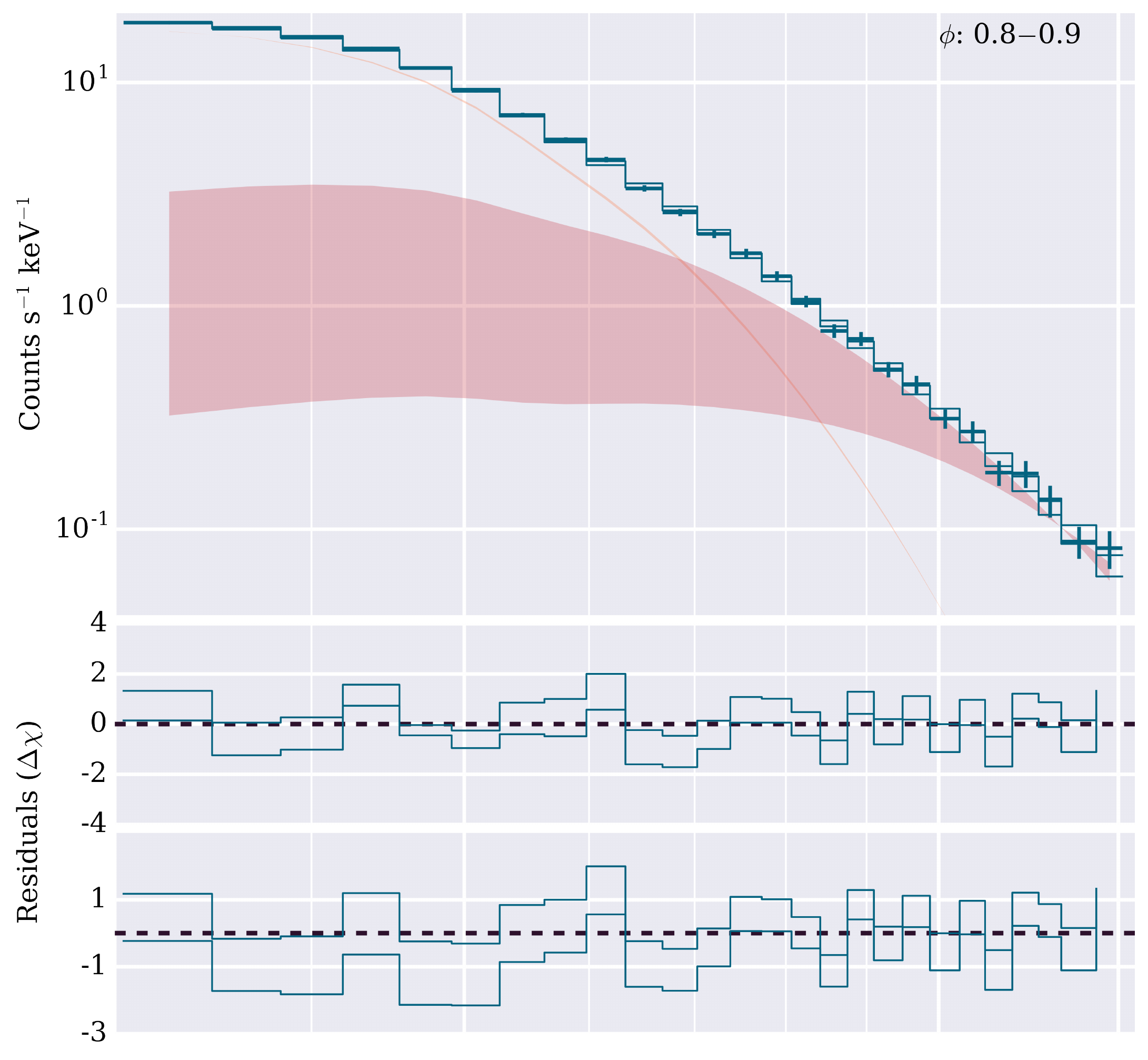}}
\caption{}
\label{J0659phspectra2}
\end{figure*}

\begin{figure*}[ht]\ContinuedFloat
\captionsetup[subfigure]{labelformat=empty}
\subfloat[]{\label{fig:J0659p101}\includegraphics[width = 0.48\textwidth]{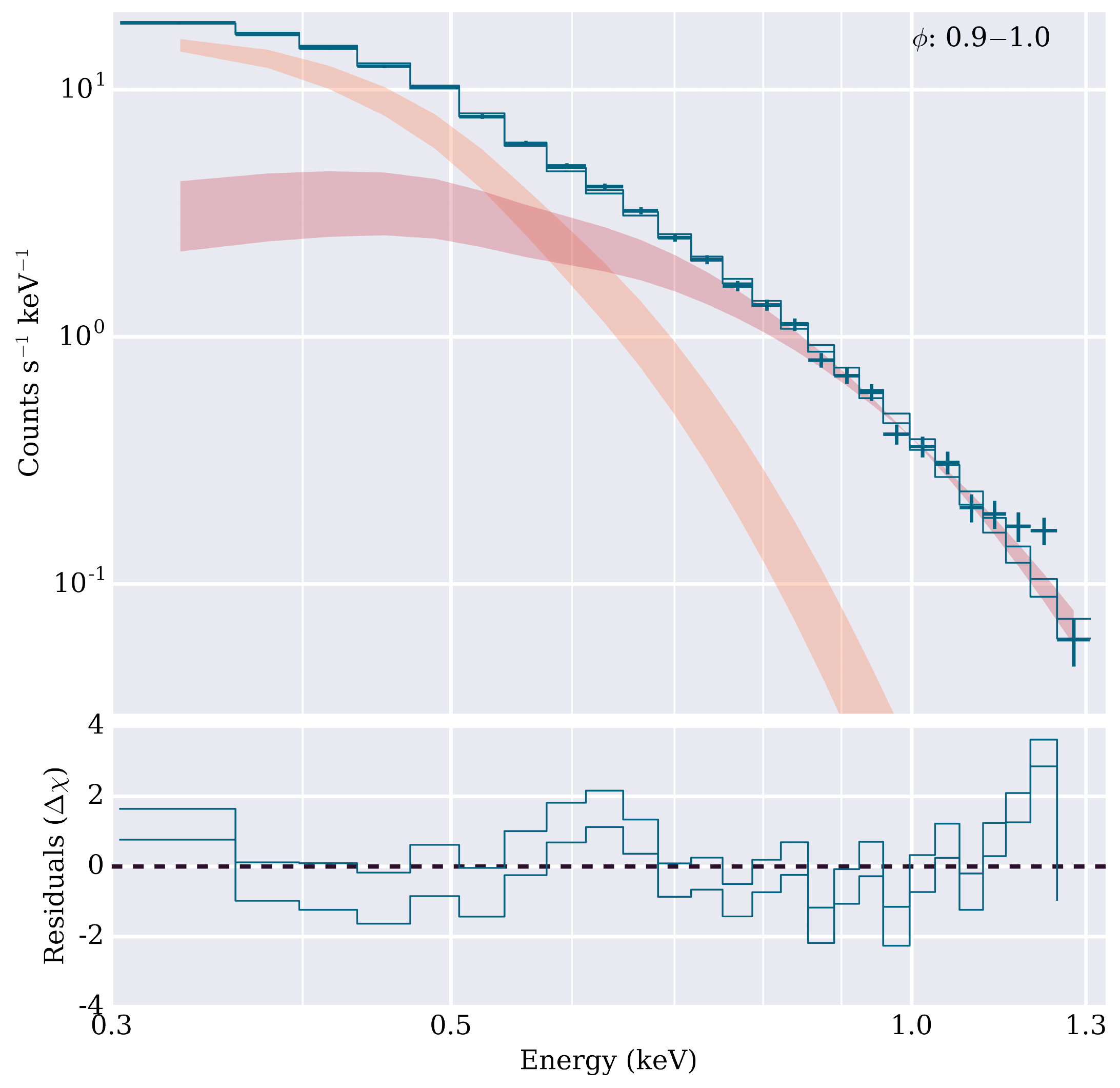}}
\subfloat[]{\label{fig:J0659p102}\includegraphics[width = 0.48\textwidth]{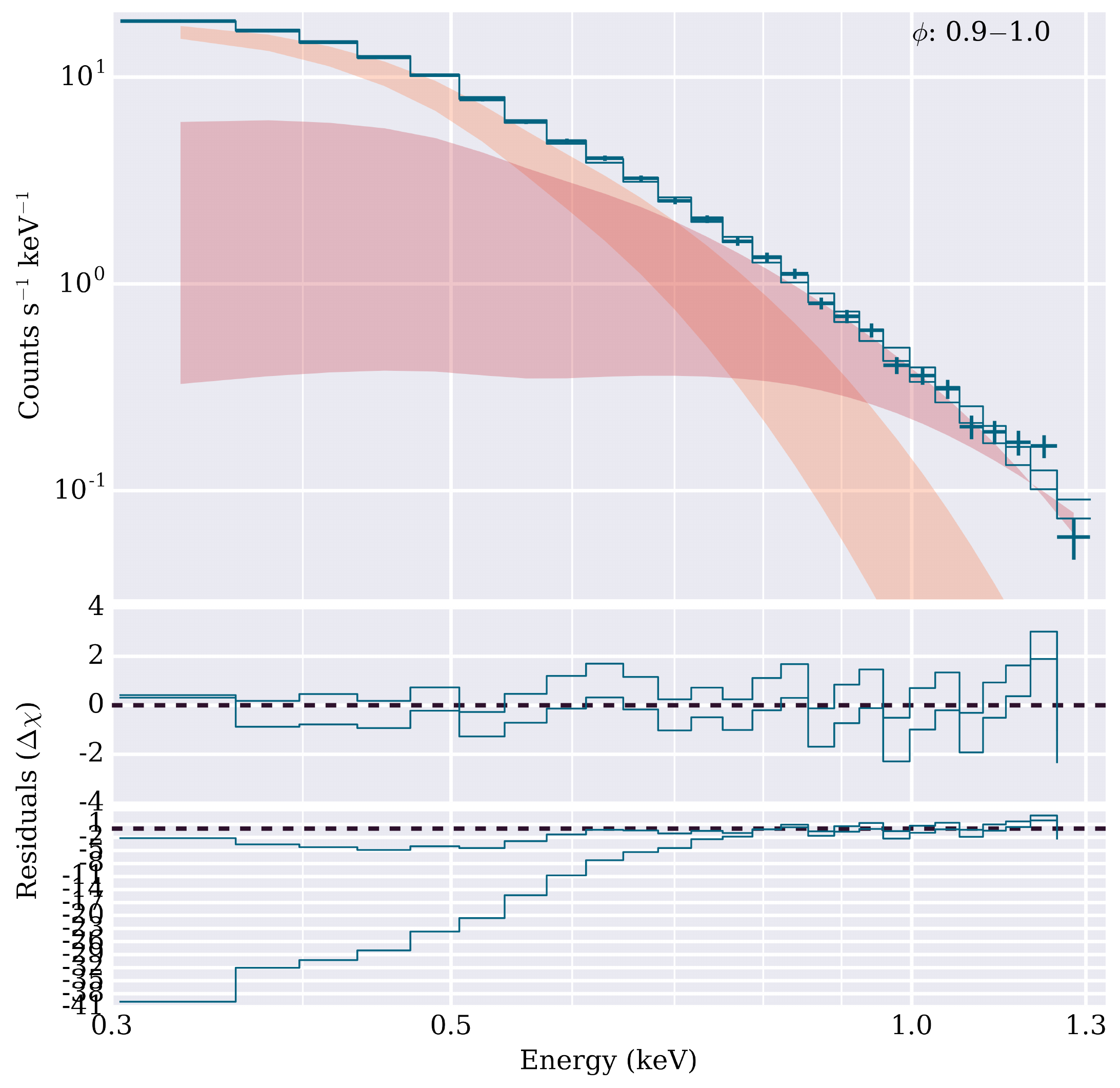}} 
\caption{Fits to phase-resolved spectra extracted for 10 equal-sized phase bins.
The models corresponding to 10 and 90 percentile of the posterior distribution are shown with solid blue lines, and the range of individual component contributions with filled regions (cold BB in orange and hot BB in red).
The panels on the left show fits with 2BB model while those on the right show fits with G2BB model.
The second set of residuals for the G2BB model is obtained by setting the strengths of the best-fit absorption lines to zero (without subsequent refitting) and is intended to show the strength of the feature.}
\label{J0659phspectra3}
\end{figure*}

\begin{figure*}
\raggedright
	\includegraphics[width=\textwidth]{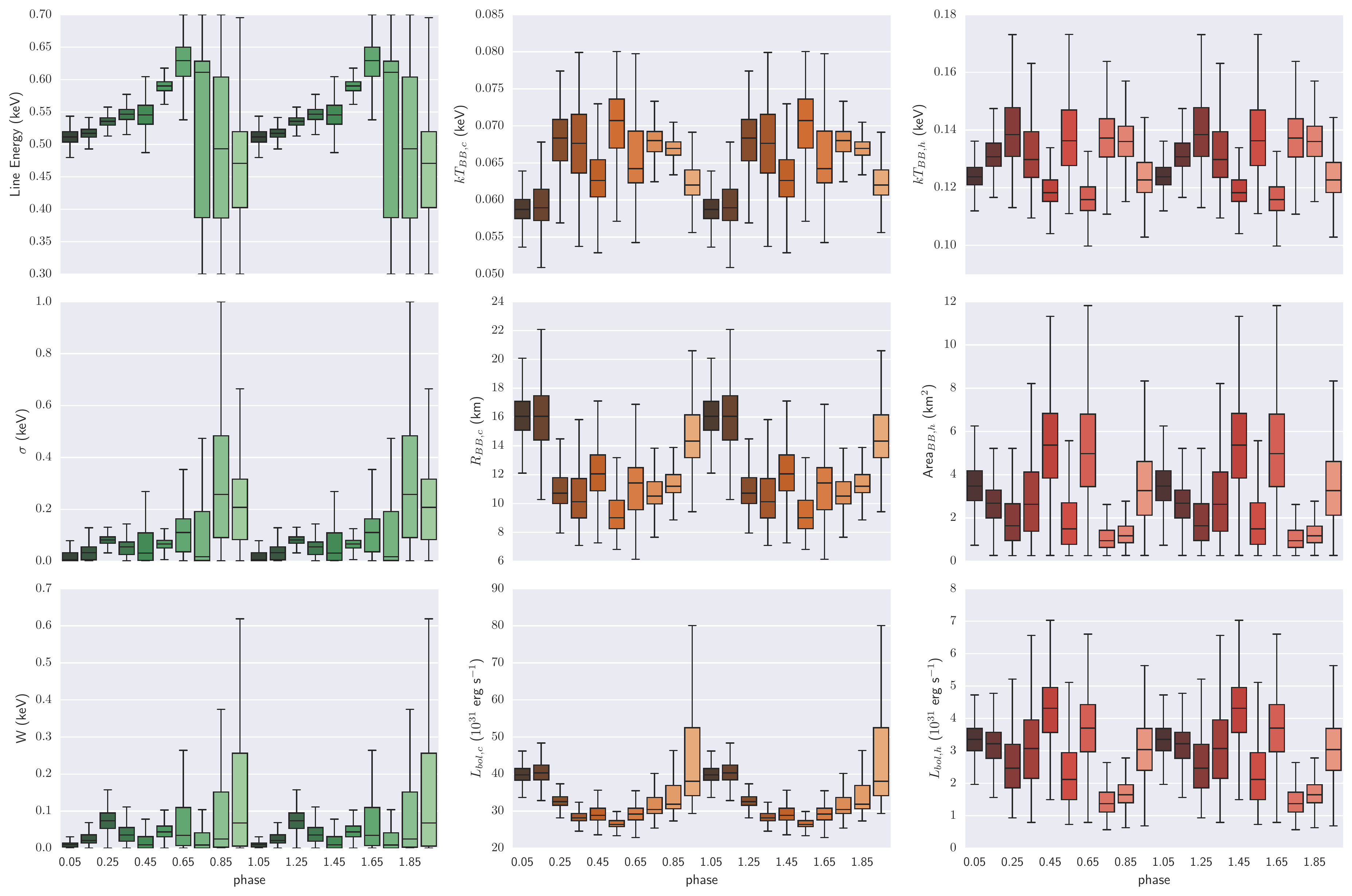}
\caption{\label{fig:Parmods}Phase variation of posterior parameter distributions for the G2BBPL fits. The boxes span the quartiles ($25-75$ percentile levels), enclosing the median values (central bar), and the whiskers show the $10-90$ percentile limits of the marginalized posterior distribution. The \texttt{Gabs} component is not required in the $0.6-1.0$ phase range, which results in the high uncertainties of \texttt{Gabs} parameters in that range.}
\end{figure*}

Adding the absorption feature at $\sim 0.5$ keV reduces the phase variability of the cold BB temperature but does not eliminate it entirely.
There is no fixed temperature value that provides good fits for all phase ranges.
This prompted us to explore the possibility that the absorption-like feature is caused by a strongly non-homogeneous temperature distribution on the NS surface.
This scenario, however, could not be modeled by simply adding more BB components.
Instead, the data favor inclusion of an absorption feature in addition to merely two BB components over a spectral model of multiple BB components.

\section{Discussion}

Absorption features reported in several isolated neutron stars of different types (e.g., \citealt{Sanwal2002}; \citealt{DeLuca2004}; \citealt{McLaughlin2007}; \citealt{Kargaltsev2012}; \citealt{Tiengo2013}; \citealt{Borghese2015}; \citealt{Rigoselli2018}) have been interpreted either as a proton (or atomic) absorption in strongly magnetized atmosphere (photosphere) or as electron cyclotron absorption in the NS magnetosphere. 

In order to produce absorption at the energy $E\approx0.54$ keV, corresponding to the most prominent absorption feature observed in the B0656 spectrum, the proton cyclotron interpretation requires a very strong magnetic field, $B_p=8.6\times10^{13}(1+z)$ G.
This field is a factor of 24 stronger than the canonical B0656's dipolar field, $B_d=4.7\times10^{12}$ G (at the NS equator for a reasonable gravitational redshift, $z=0.3$), thus requiring  presence of multipolar field components near the NS surface. Alternatively, photospheric absorption could be caused by transitions between the ground state and loosely bound states  in a once-ionized He for $B\sim5\times10^{12}$ G (these transitions have the largest oscillator strength; see \citealt{Pavlov2005}). However,  the loosely bound   states may not be available if the ions are too densely packed. If the field is stronger in some places (e.g., closer to poles or due to multipolar contributions),   transitions from the ground state to the next tightly bound state may contribute (see Figure 3 in  \citealt{Pavlov2005}).  In this scenario, the observed dependence of the absorption feature strength on rotational phase could be caused by strongly non-uniform field  or  by a non-uniform distribution of matter responsible for the absorption.

Another possibility is the electron cyclotron absorption in the NS magnetosphere at the altitude dictated by the dipolar magnetic field, $B(r)=B_{d}(1+3\cos^2\theta_{\rm B})^{1/2}(R_{\rm NS}/r)^3$, where $R_{\rm NS}$ is the NS radius, and $\theta_{\rm B}$ is magnetic co-latitude. For B0656, since the strongest absorption at $\phi=0.2-0.3$ coincides with the peak in the 0.8--1.5 keV lightcurve (see Figure 9) corresponding to the largest hot spot contribution, the absorption likely occurs
above the hot spot which is typically associated with the NS magnetic pole.  Therefore, for $E_{e,c}=11.6\;[B(r)/10^{12}~{\rm G}]$ keV to be $\approx0.54$ keV at $\theta_{\rm B}\sim 0$, the altitude of the absorbing layer has to be about $5 R_{NS}$ with the width $\Delta r\approx 7$ km (estimated from $\Delta E_{e,c}(r)/\Delta r$ with $B(r)=B_{d}R_{\rm NS}^3/r^3$) for the observed $\Delta E_{\rm e, c} = 2.3\,\sigma_{\rm gabs} = 0.23$ keV.

This absorption mechanism requires non-relativistic electrons to be present in the pulsar magnetosphere.  While rarely discussed, direct production of mildly relativistic electrons may be possible within the regions with suitable magnetic field \citep{Weise2002}.
Alternatively, non-relativistic electrons can be captured into the closed magnetic field line zone from the circumstellar medium (\citealt{Luo2007}; and references therein).
The cyclotron cooling time, relevant for non-relativistic particles, is large, which allows buildup of a large density over time.
However, since the hot BB maximum and absorption are happening in roughly the same range of rotation phases,  such interpretation would imply that the hot spots are not at the NS magnetic poles (as commonly assumed for dipolar fied) because the absorption cannot be happening in the open field line region where the $e^{+}/e^{-}$ plasma is being accelerated rapidly.
The hot spot(s) may be associated with a strong multipolar field (see, e.g., \citealt{Gourgouliatos2018}) near the NS surface rather than with the particle bombardment of (dipolar) polar caps.
Alternatively, there is the possibility of magnetic field configurations causing equatorial region of the NS to be hotter than the regions near the poles, corresponding to the dipolar external field (\citealt{Geppert2002}, \citealt{Pons2007}, \citealt{Vigano2013}).
Some indirect support for this possibility comes from the substantial offset between the radio peak and the hot spot phases (see Figure \ref{fig:stackedprofiles}). The observed optical depth, $\tau\approx 0.08$, can be achieved if the  electron density exceeds the Goldreich-Julian density by a factor of 100  for latitudes where the magnitude of the local magnetic field component along the rotation axis is about half of the strength of total local magnetic field value (see eqn. 4 from \citealt{Rajagopal1997}).

As an alternative to the absorption scenario, other possibilities are not completely ruled out by the existing data.
For instance, small deviations of the actual atmosphere spectrum from a simple 2 BB parameterization could create residuals at the crossing point if the deviations either systematically increase of decrease with photon energy.
One may expect such residuals to depend on phase, since the flux of the BB components  varies with phase.
We compared the ratio of the normalizations of the two BB components to the strength of the GABS line but did not find strong evidence for correlation.
Another possible alternative to the absorption scenario, is a more complex inhomogeneous temperature distribution on the stellar surface.
\cite{Vigano2014} demonstrated that even an ideal BB with inhomogeneous temperature distribution can mimic absorption-like features reported in spectra of at least some isolated NSs.
A much stronger support for the absorption feature scenario with underlying cyclotron mechanism is obtained when harmonics are detected in addition to the fundamental line (like, e.g., in the 1E\,1207--52 CCO: \citealt{Sanwal2002}).
Although we do not see clear evidence of cyclotron harmonics in the spectra available, they could be detected in future observations of B0656 with more sensitive instruments.

\section{Conclusions}
Our higher quality data from the 130 ks EPIC observation show phase-varying spectral feature not detected in the previous 23 ks of EPIC data \citep{DeLuca2005}.
We find that the 2BBPL fit is unsatisfactory for both phase-integrated and phase-resolved (in the phase range 0.0 to 0.6) spectra. 
Using NS atmosphere models instead of BBs or a larger number of BB components does not improve the fits and leads to physically unrealistic NS sizes.

The hard spectral tail is seen in the EPIC (from 2 to 7 keV) and {\sl NuSTAR} (from 3 to 20 keV) phase-integrated spectra.
The 2.5-7 keV EPIC-pn light curve does not show a single well-defined peak unlike the $1.5-7$ keV light curve shown in \cite{DeLuca2005} (likely, the latter is contaminated by the hot BB component).
However, the pulse profile in $3-20$ keV {\sl NuSTAR} data show a single significant peak with an intrinsic pulsed fraction of $71_{-13}^{+14}$\%. 
The slope of the X-ray PL, $\Gamma = 1.46\pm0.12$, is similar to that of the Fermi-LAT (0.1-100 GeV) PL, $\Gamma = 1.72 \pm 0.48$ \citep{2013ApJS..208...17A}.

The offset between the cold BB peak and the hot BB peak in the energy-resolved pulse profiles indicate departures from the simple, axially symmetric, dipolar magnetic field configuration.
The light curves show a highly significant difference in the pulse profile shape (with small decrease in pulsed fraction around 0.5 keV) between the energy ranges $0.3-0.7$ keV and $0.4-0.6$ keV, where only a single cold BB component is expected to dominate. 

The single soft BB description of the continuum below 0.8 keV is unsatisfactory.
The observed large residuals can be attributed to a more complex temperature distribution over the surface or to absorption feature(s).
The quality of the phase-resolved  spectral fits significantly (according to our Bayesian analysis) improves if an absorption line is added.
The central energy of the line shifts between 0.5 and 0.6 keV in the phase range $0.0-0.6$ and the line disappears (\texttt{Gabs} not required) in the phase range $0.6-1.0$.
The fits with G2BB model to the spectrum below 1.5 keV still show the variations in the $kT$ and $R$ of the soft BB component, which suggests that the single-temperature continuum model description is inaccurate below 0.8 keV.

If the absorption feature interpretation is correct, it could be attributed to cyclotron absorption in the magnetosphere requiring the presence of non-relativistic electrons in certain regions of magnetosphere \citep{Kargaltsev2012}.
Alternatively, proton cyclotron absorption in local magnetic field loops close to the NS surface with a stronger magnetic field ($B\gtrsim10^{13}$ G) could also produce such features  \citep{Tiengo2013}.
NICER observations and modeling of the temperature distribution over the neutron star surface can help to shed further light on the nature of  the B0656 soft X-ray spectrum.

\acknowledgments
We thank the \xmm helpdesk for important clarifications regarding the \xmm data analysis.
We also thank the referees for a careful reading of the manuscript and very useful remarks and suggested tests that enhance the paper.
We acknowledge NASA support through the awards NNX13AF21G and NNX16AE82G. We thank Igor Volkov for the help with simulations.  

\facilities{\xmm, CXO, {\sl NuSTAR}} 
\software{XMM-SAS, NUSTARDAS, XSPEC, PyXspec, Scipy, emcee}

\appendix
\section{Additional tests to identify systematics in data}
\subsection{Systematic error in the effective area of EPIC-pn}
As part of the \xmm calibration program\footnote{\url{http://xmm2.esac.esa.int/docs/documents/CAL-TN-0018.pdf}\label{fnxmm}}, the systematic error in the effective area of EPIC-pn is estimated by fitting a large sample of active galactic nuclei with predominantly non-thermal emission.
The statistical error of the data in the $0.4-12$ keV is subtracted from the standard deviation of the distribution of fit residuals to obtain a $3\sigma$ upper limit on the systematic error of $4\%$. 

\begin{figure*}[ht]
\captionsetup[subfigure]{labelformat=parens}
\subfloat[]{\label{fig:J0659Sa}\includegraphics[width = 0.45\textwidth]{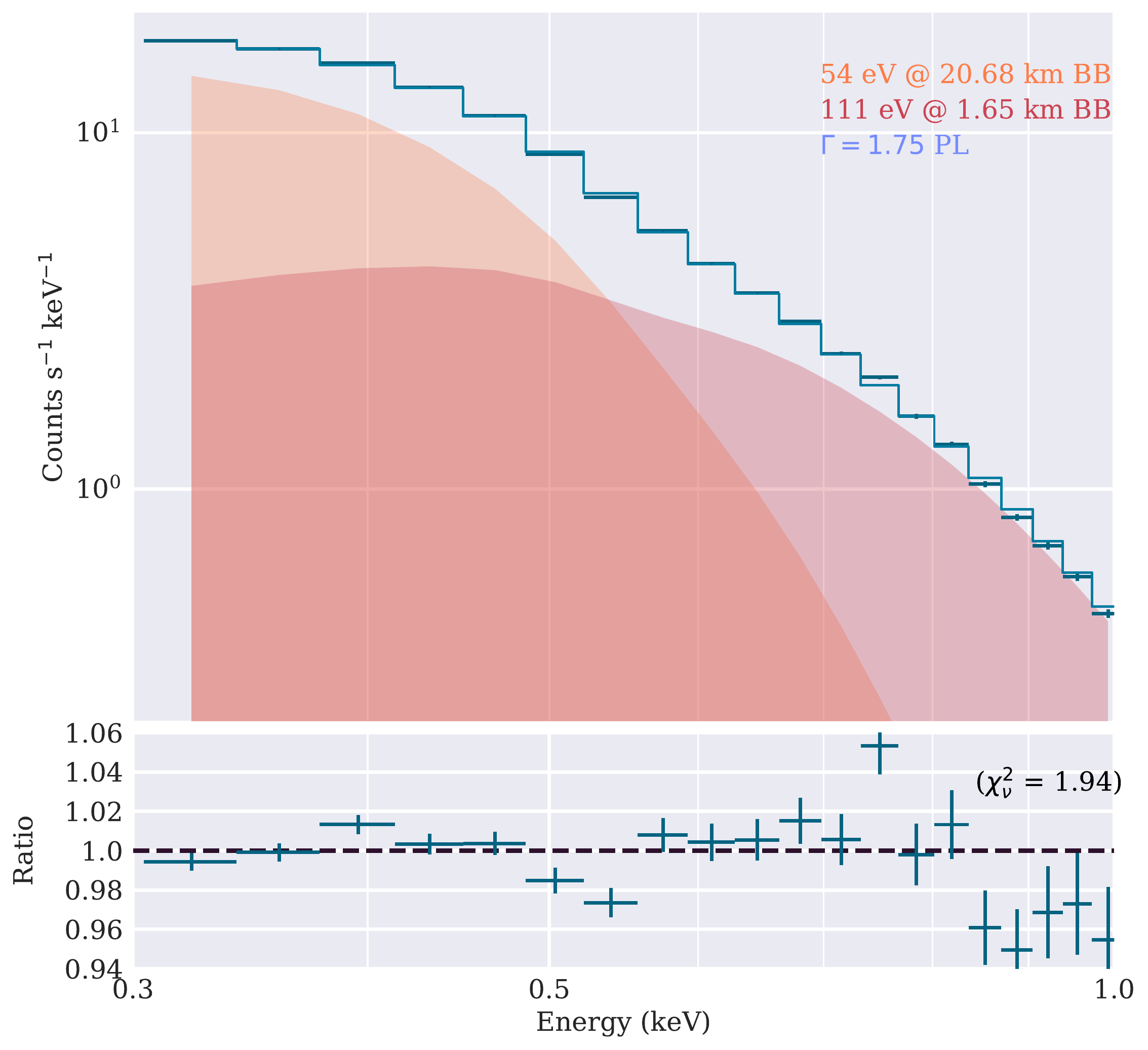}}\\
\subfloat[]{\label{fig:J0659Sb}\includegraphics[width = 0.45\textwidth]{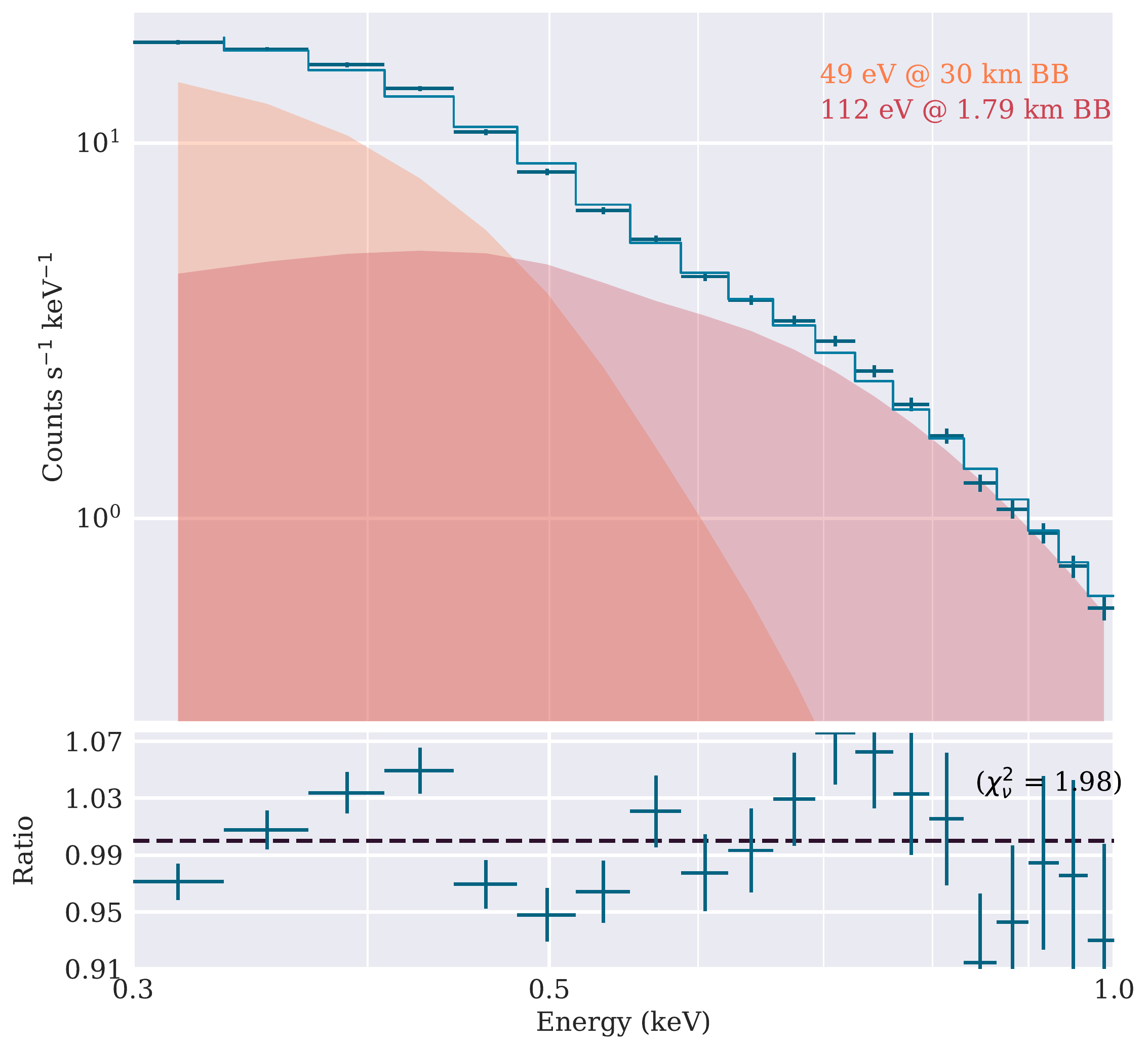}}
\subfloat[]{\label{fig:J0659Sb}\includegraphics[width = 0.45\textwidth]{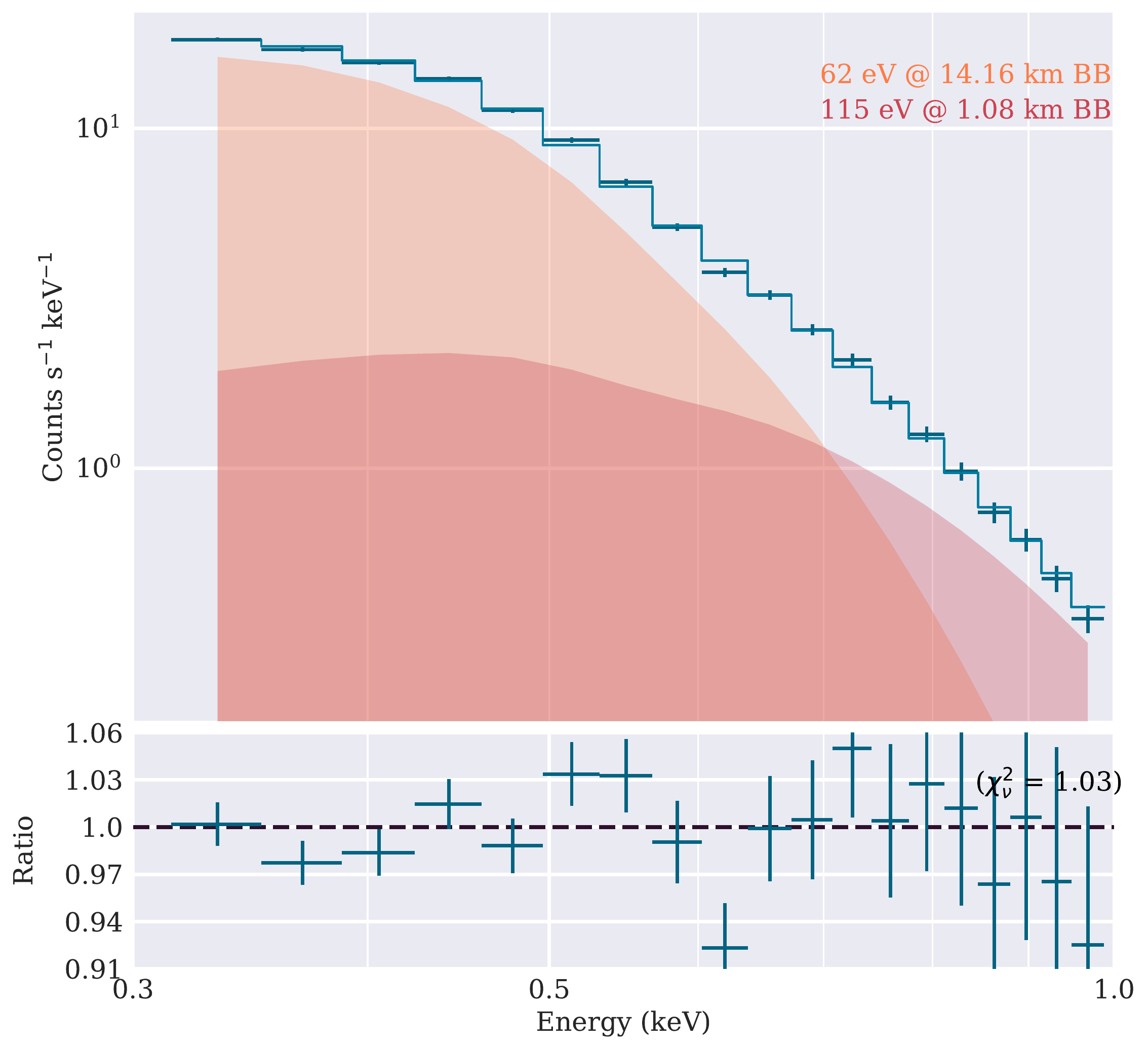}}
\caption{Best-fit model and data-to-model ratios plotted for phase-integrated spectrum (a) fit with 2BBPL model and phase-resolved spectra from phase ranges $0.2-0.3$ (b) and $0.7-0.8$ (c) fit with 2BB model.}
\label{fig:ratios}
\end{figure*}

To check for the magnitude of the residuals in our data around the absorption-like feature, we plotted the data-to-model ratio deviations from the best-fit obtained with the 2BBPL model.
For the $0.3-7$ keV phase-integrated spectrum, the data-to-model ratio deviations are between $2\%-3\%$ (Figures \ref{fig:ratios}a).
The residuals, however, are phase-dependent while any systematic residuals (related to the detector calibration imperfections) are not  expected to depend systematically on the pulsar's rotation phase.
In the $0.3-1.3$ keV phase-resolved spectra, the deviation is $\sim 5\%$ in the phase range $0.2-0.3$ (Figure \ref{fig:ratios}b) where the absorption-like feature is strongest, and within $2\%$ in the phase range $0.7-0.8$  (Figure \ref{fig:ratios}c) where there is no strong indication of any feature near 0.5 keV.
Hence, it is unlikely that a systematic error in the effective area calibration is the source of the observed feature.

\subsection{Comparisons with the calibration source RX\,J1856--3754}

We extracted spectra of the isolated NS (INS) RX\,J1856-3754 (RXJ1856) from two observations (ObsIDs 0727760301 and 0727761001, observed in SW mode, with source location on chip similar to those in the B0656 observations.).
Single BB model fits show $\sim2\%$ positive deviations in data over model ratio around 0.5 keV for one of the observations but not for the other.
Since all other INSs show obvious spectral features $< 1$ keV (see Table 1 in \citealt{Vigano2014}), it is possible that RXJ1856 has weaker yet unaccounted features and therefore is not an optimal choice for calibration in this respect.
We do see some negative (absorption) residual around 0.6 keV in these two observations.

In any case, the larger, $5\%$ deviation of data-over-model ratio seen in the phase-resolved spectra of B0656 (Figure \ref{fig:ratios}) cannot be explained easily by such instrumental systematics.

\subsection{Effect of energy scale offset}

Under normal operating conditions, the uncertainties in the gain can introduce uncertainties in line energy of $\leq12$ eV over the full energy range for EPIC-pn\textsuperscript{\ref{fnxmm}} (5 eV for EPIC-MOS).
For a rough assessment of the influence of gain uncertainties, the XSPEC model \texttt{gain} can be used to modify the response file gain by introducing a shift in energies using the offset parameter while keeping the scaling of the effective area, the slope parameter, fixed at 1.

Fitting the phase-integrated spectrum with the 2BBPL model and using the ``gain fit'' mode produces an offset of $-12$ eV.
However, similar fitting of the gain offset while fitting the 2BB model ($N_H$ fixed to the best-fit value from the phase-integrated fit) to phase-resolved spectra shows absolute offsets $<12$ eV, and they seem to vary with phase (see Table \ref{tab:gain}).
The gain offsets seem to be within the nominal value for EPIC-pn.
Moreover, an instrumental gain offset cannot depend on pulsar rotational phase (as appears to be the case; see Table \ref{tab:gain}), hence it is an unlikely explanation for the observed residuals.

\begin{deluxetable}{lccccccccccc}
\tablecaption{Fitting Gain offset\label{tab:gain}}
\tablecolumns{12}
\tablewidth{0pt}
\tablehead{
\colhead{Phase} & \colhead{$0-1$} & \colhead{$0.0-0.1$} & \colhead{$0.1-0.2$} & \colhead{$0.2-0.3$} & \colhead{$0.3-0.4$} & \colhead{$0.4-0.5$} & \colhead{$0.5-0.6$} & \colhead{$0.6-0.7$} & \colhead{$0.7-0.8$} & \colhead{$0.8-0.9$} & \colhead{$0.9-1.0$} }
\startdata
Gain offset (eV) & $-12$ & $-5.4$ & $-4.4$ & $-8.9$ & $-4.4$ & $-2.4$ & $0.6$ & $6.7$ & $0.0$ & $-6.0$ & $-3.4$ \\
\enddata
\tablecomments{2BBPL model is fit to the phase-integrated spectrum, and 2BB model is fit to the phase-resolved spectra.}
\end{deluxetable}

\subsection{Effect of small extraction aperture}
\begin{figure*}[ht]
\captionsetup[subfigure]{labelformat=empty}
\subfloat[]{\label{fig:J0659Sb}\includegraphics[width = 0.5\textwidth]{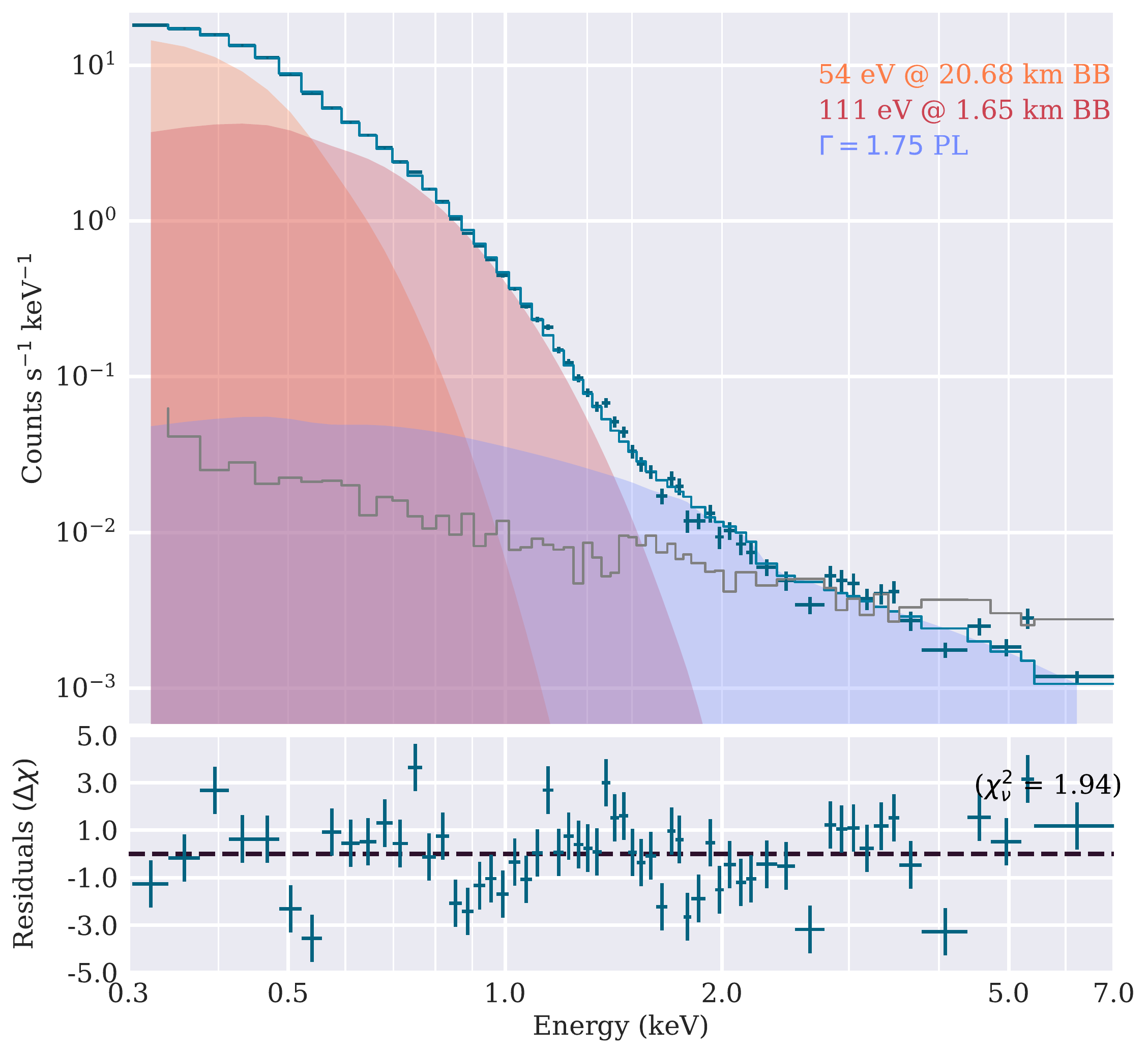}}
\subfloat[]{\label{fig:J0659Sb}\includegraphics[width = 0.5\textwidth]{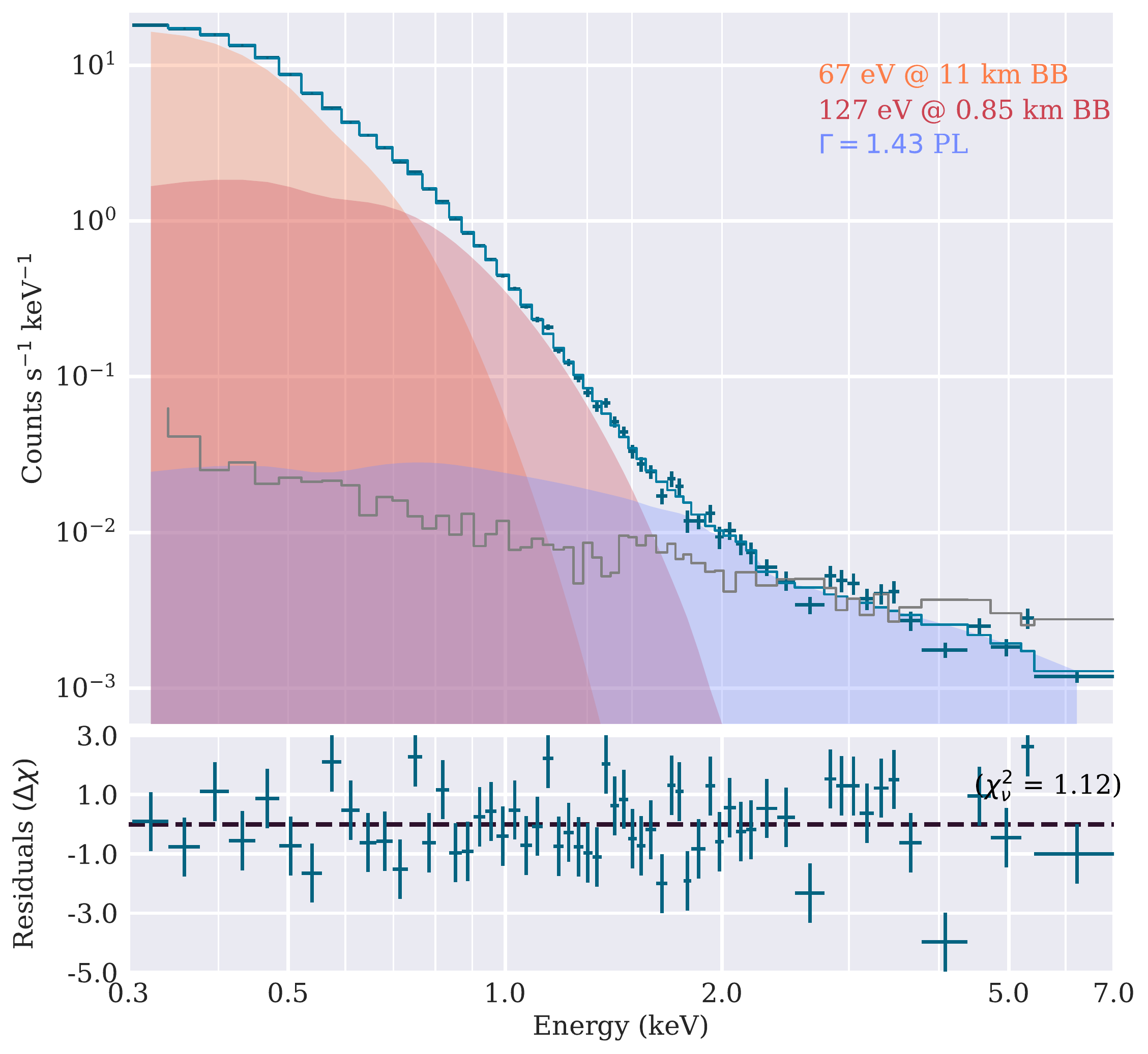}}
\caption{2BBPL (left panel) and G2BBPL (right panel) models fit to phase-integrated spectra extracted from $37\farcs5$ radius circular aperture as shown in Figures \ref{fig:fig1_cmap}.
The background spectrum is overplotted in grey using a step plot.
This figure should be compared with Figure \ref{J0659spectra} which shows the corresponding fits for an extraction radius of $15\arcsec$.}
\label{J0659spectra37p5}
\end{figure*}

We investigate the effect of small versus large extraction aperture on the spectral fit parameters and the observed residuals.
We fit the phase-integrated spectrum extracted from a large $37\farcs5$ radius region\footnote{This is highest S/N extraction region in the $0.3-7$ keV range within the chip boundary} with the same 2BBPL and G2BBPL models that we used for the small extraction area of $15\arcsec$ radius (Figure \ref{J0659spectra37p5}).
The fits to the spectra from the small and large extraction regions match well at low energies, where the S/N is high and the absorption-like feature is seen (compare Figures \ref{J0659spectra} and \ref{J0659spectra37p5}).
The deviations at higher energies ($\gtrsim2$ keV), as seen from the spread in residuals, are due to the low S/N in the spectral bins and high background contribution in the spectrum extracted from the larger aperture.

Comparing the parameters obtained from best fit models using the GBBPL model to the spectra extracted from $15\arcsec$ and $37\farcs5$ radius regions (Table \ref{tab:regions}), we see no statistically significant effect on the fit parameters.

\begin{deluxetable}{lcc}[h]
\tablecaption{The comparison of the effect of small and large extraction aperture on phase-integrated spectral fits.\label{tab:regions}}
\tablecolumns{3}
\tablewidth{0pt}
\tablehead{
\colhead{Parameter} & \colhead{$15\arcsec$ region} & \colhead{$37\farcs5$ region} \\
\colhead{} & \colhead{$0.3-7$ keV} & \colhead{$0.3-7$ keV}}
\startdata
$N_{H}$ ($10^{20}$ cm$^{-2}$)   & $3.30^{+2.21}_{-1.24}$ & $2.90^{+1.76}_{-0.96}$ \\ 
$E_c$ (keV) & $0.54^{+0.03}_{-0.06}$ & $0.55^{+0.02}_{-0.04}$ \\ 
$\sigma$ (eV) & $106^{+52}_{-36}$ & $97^{+40}_{-26}$ \\ 
s & $0.06^{+0.18}_{-0.04}$ & $0.06^{+0.11}_{-0.03}$ \\ 
$kT_{BB}$ (eV) & $65^{+ 5}_{- 6}$ & $67^{+ 4}_{- 4}$ \\ 
BB norm ($10^5$) & $1.95^{+4.55}_{-0.86}$ & $1.53^{+2.22}_{-0.53}$ \\ 
$R_{BB}$ (km) & $13^{+10}_{- 3}$ & $11^{+ 6}_{- 2}$  \\ 
$kT_{BB}$ (eV) & $123^{+ 8}_{- 6}$ & $127^{+ 8}_{- 6}$ \\ 
BB norm ($10^3$) & $1.19^{+0.79}_{-0.55}$ & $0.87^{+0.54}_{-0.36}$ \\ 
$R_{BB}$ (km) & $0.99^{+0.29}_{-0.26}$ & $0.85^{+0.23}_{-0.20}$  \\ 
$\Gamma$ & $1.74^{+0.20}_{-0.20}$ & $1.43^{+0.11}_{-0.22}$ \\ 
PL norm ($N_{-5}$) & $2.71^{+0.64}_{-0.56}$ & $2.36^{+0.63}_{-0.53}$ \\ 
$\chi_\nu^2$ & 0.73 & 1.12 \\ 
\enddata
\tablecomments{Description of parameters same as in Table \ref{tbl-2_intspectra}.}
\end{deluxetable}

\newpage
\bibliographystyle{yahapj}
\bibliography{references}

\end{document}